\documentclass[reprint,superscriptaddress,showpacs,amsmath,amssymb,aps,prb]{revtex4-1}

\usepackage{graphicx}  
\usepackage{dcolumn}   
\usepackage{bm}        
\usepackage{amssymb}   
\usepackage{mathrsfs} 
\usepackage{amsthm}
\usepackage{mathtools}
\usepackage{framed}
\usepackage{bbm}     
\usepackage[colorlinks=True,urlcolor=blue,linkcolor=blue,citecolor=blue]{hyperref}
\usepackage{xcolor}
\usepackage{multirow}
\usepackage{setspace}
\usepackage{CJKutf8}
\usepackage{color,soul}

\usepackage[utf8]{inputenc}
\usepackage[T1]{fontenc}
\usepackage{lmodern}
\usepackage[toc,page]{appendix}
\usepackage{tikz}
\usetikzlibrary{arrows,arrows.meta,decorations.markings,decorations.pathreplacing,calc}
\tikzset{->-/.style={decoration={markings,mark=at position #1 with {\arrow{Stealth}}},postaction={decorate}},->-/.default=0.6}

\theoremstyle{definition}

\def\v#1{\mathbf{#1}}

\newcommand{\be}{\begin{equation}}
\newcommand{\ee}{\end{equation}}

\newcommand{\Ac}           {\mathcal{A}}
\newcommand{\Ab}           {\mathbf{A}}

\newcommand\Tb            {\mathbb{T}}

\newcommand{\Zb}           {\mathbb{Z}}

\begin{document}

\begin{CJK*}{UTF8}{}

\title{A lattice realization of general three-dimensional topological order}

\author{Wenjie Xi}
\thanks{These two authors contributed equally to this work.}
\affiliation{Shenzhen Key Laboratory of for Advanced Quantum Functional Materials and Devices, Southern University of Science and Technology, Shenzhen 518055, China}
\affiliation{Institute for Quantum Science and Engineering and Department of Physics, Southern University of Science and Technology, Shenzhen 518055, China}

\author{Ya-Lei Lu}
\thanks{These two authors contributed equally to this work.}
\affiliation{Key Laboratory for Magnetism and Magnetic Materials of MOE, Lanzhou University, 730000 Lisanzhou, China}
\affiliation{Shenzhen Key Laboratory of for Advanced Quantum Functional Materials and Devices, Southern University of Science and Technology, Shenzhen 518055, China}
\affiliation{Institute for Quantum Science and Engineering and Department of Physics, Southern University of Science and Technology, Shenzhen 518055, China}

\author{Tian Lan}
\affiliation{Department of Physics, The Chinese University of Hong Kong, Shatin, New Territories, Hong Kong, China}

\author{Wei-Qiang Chen}
\email{chenwq@sustech.edu.cn}
\affiliation{Shenzhen Key Laboratory of for Advanced Quantum Functional Materials and Devices, Southern University of Science and Technology, Shenzhen 518055, China}
\affiliation{Institute for Quantum Science and Engineering and Department of Physics, Southern University of Science and Technology, Shenzhen 518055, China}

\begin{abstract}
Topological orders are a class of phases of matter that beyond the Landau symmetry breaking paradigm.  The two (spatial) dimensional (2d) topological orders have been thoroughly studied.  It is known that they can be fully classified by a unitary modular tensor category (UMTC) and a chiral central charge c.  And a class of 2d topological orders whose boundary are gappable can be systematically constructed by Levin-Wen model whose ground states are string-net condensed states.  
Previously, the three spatial dimensional topological orders have been classified based on their canonical boundary described by some special unitary fusion 2-category, $2\mathcal{V}ec_G^\omega$ or an EF 2-category. However, a lattice realization of a three spatial dimensional topological orders with both canonical boundary and arbitrary boundaries are still lacking. 
In this paper, we construct a 3d membrane-net model based on spherical fusion 2-category, which can be used to systematically study all general 3d topological order with gapped boundary.  The partition function and lattice Hamiltonian of the membrane-net model is constructed based on state sum of spherical fusion 2-category.  We also construct the 3d tube algebra of the membrane-net model to study excitations in the model. We conjecture all intrinsic excitations in membrane-net model have a one-to-one correspondence with the irreducible central idempotents (ICI) of the 3d tube algebra.
We also provide a universal framework to study mutual statistics of all excitations in 3d topological order through 3d tube algebra. Our approach can be straightforwardly generalized to arbitrary dimension.
\end{abstract}
\maketitle
\end{CJK*}

\section{Introduction}
It is long believed that Landau's spontaneous symmetry breaking theory\cite{Landau1937} can describe all phases and phase transitions. However, with the discovery of fractional quantum hall effect\cite{Tsui1982}, it is realized that quantum states with long range entanglement can give rise to a class of non-trivial quantum phases beyond Landau's theory, the topological orders (TO)\cite{Wen1989,Wen1990,Kitaev2003,Wen2005}. Besides fundamental physical importance, the nontrivial statistics of the excitations in a topological order\cite{Kitaev2003} make them promising candidates for universal quantum computing.

The two (spatial) dimensional topological orders have been thoroughly studied. Mathematically, a 2d topological order is fully classified by a unitary modular tensor category (UMTC) and a chiral central charge $c$.  Physically, the most important observables of a 2d topological order are the nontrivial statistics of the anyonic excitations, which is 
characterized by the $S,\ T$ matrices.  The string-net model (also known as the Levin-Wen model)\cite{Wen2005} provides a general framework to construct non-chiral 2d topological orders, i.e. the topological orders that can have gapped boundaries.  In brief, one can construct a Levin-Wen model from an input unitary fusion 1-category, whose partition function is just the state-sum of the unitary fusion 1-category (known as the Turaev-Viro topological quantum field theory)\cite{TURAEV1992,Barrett96}.  Such a model gives a 2d topological order described (in mathematics) by the Drinfeld center (i.e. a UMTC) \cite{Joyal1991} of the input unitary fusion 1-category and the zero central charge \cite{KK1104.5047}.
It is known that the excitations in the Levin-Wen model have a one-to-one correspondence with the ICIs of the tube algebra\cite{Bultinck2017} or modules of the $Q$-algebra\cite{Lan2014} constructed with the fusion 1-category. And information of mutual statistics of these excitations is also encoded in ICIs of the tube algebra.

The Levin-Wen construction exhibits the boundary-bulk duality~\cite{KK1104.5047}, i.e., boundary uniquely determines the bulk, a principle that we believe to hold for topological phases in all dimensions. More precisely, the unitary fusion 1-category is the data that describes a one dimensional gapped boundary, which hosts some point-like excitations corresponding to the objects in the 1-category. By inputting a fusion 1-category into the Levin-Wen model, or equivalently the Turaev-Viro state sum, one obtains a unique two dimensional bulk described by the Drinfeld center of the input fusion 1-category. The similar procedure should also work for higher dimensional topological orders, as long as we know how to describe its boundary and construct a proper state sum using the boundary data. 

There have been many partial studies on three dimensional topological orders from different aspects. Walker and Wang\cite{walker2011} construct lattice models of a class of 3d topological orders whose partition function is the Crane-Yetter invariant\cite{Crane1993}. Furthermore, mutual statistics of the excitations, such as loop-loop braiding, loop-particle braiding and three-loop braiding, in several 3d topological orders have been studied by different approaches\cite{Wang2014,Jiang2014,JuvenWang2015,Ye2021}. 
However, we aim to provide a universal framework to construct general
3d topological order and study the mutual statistics of all excitations in them. The following recent progresses make the realization of such a framework possible:
\begin{enumerate}
    \item Lan, Kong and Wen proposed a classification of three dimensional topological orders~\cite{LKW1704.04221,LW1801.08530}, the mathematical part of which was later confirmed by Johnson-Freyd~\cite{Joh2003.06663}. They showed that it is possible to condense half of the excitations in a three dimensional topological order, to create a canonical gapped two dimensional boundary, which is described by some special unitary fusion 2-category, $2\mathcal{V}ec_G^\omega$ or an EF 2-category~\cite{LW1801.08530}.
    In particular, it means that three dimensional topological orders can all have gapped boundary, thus a Levin-Wen like construction is able to include all three dimensional topological orders. However, a detailed analysis of statistics data is lacking in these works.
    \item Besides the canonical gapped boundary, a three dimensional topological order may have other gapped boundaries. Comparing to a one dimensional gapped boundary, a two dimensional gapped boundary can host both string-like excitations and point-like excitations, thus has to be described by a higher dimensional, unitary fusion 2-category. The objects in the 2-category correspond to the string-like excitations and the 1-morphisms correspond to point-like defects on the strings (point-like excitations are just defects on the trivial string). The detailed mathematical structures of unitary fusion 2-categories have been clarified by recent works~\cite{douglas2018,johnson2020}, enabling us to explicitly construct the 3d model. 
\end{enumerate}

In this paper, we construct a membrane-net model to systematically study all 3d topological orders with a generic, arbitrary gapped boundary described by an arbitrary unitary fusion 2-category. The model Hamiltonian is derived from the 3+1D partition function, which is taken as the state sum of the spherical fusion 2-category \cite{douglas2018}.   Instead of solving the Hamiltonian directly, we study the excitations in the membrane-net model with the 3d tube algebra\cite{Bullivant2019,Bullivant2020} constructed with the input spherical fusion 2-category.  
We also provide a universal framework to study mutual statistics of all excitations in the 3d topological order through 3d tube algebra. We demonstrate by a few examples that our result is consistent with known results.
Finally, we propose that our construction and study can be generalized to higher dimensional topological orders.
By the results of Ref.~\onlinecite{LKW1704.04221,LW1801.08530,Joh2003.06663}, the constructed membrane-net model should always admit also the canonical boundary described by either $2\mathcal{V}ec_G^\omega$ or an EF 2-category, thus we provide an explicit approach to demonstrate the Morita equivalence between any fusion 2-category and the canonical ones. The study of the gapped boundary conditions of the membrane-net model is left for future work.

\section{Construction of the Membrane-net Model}

\subsection{Partition function of Levin-Wen model}

We first briefly review the way to construct Levin-Wen Hamiltonian from the partition function of a Turaev-Viro TQFT.  In this subsection, we will introduce the partition function of Turaev-Viro TQFT, while the derivation of the Levin-Wen Hamiltonian will be introduced in the next subsection.  

We consider a triangulation $\mathcal{K}$ of a closed oriented three-dimensional manifold $\mathcal{M}$, i.e. split the space-time into tetrahedrons.  Given an order of vertices on $\mathcal{K}$, and a spherical fusion 1-category $\mathcal{C}$, the partition function can be constructed as the following. We assign one object in $\mathcal{C}$ to each 1-simplex, or edge, $\bigtriangleup_1$ in $\mathcal{K}$, and one 1-morphism in $\mathcal{C}$ to each 2-simplex, or triangle, $\bigtriangleup_2$ in $\mathcal{K}$.  Please note that the 1-morphism must be in the hom space, which is just a vector space in fusion 1-category, determined by the three objects on the edges of the triangle.  The Hilbert space of Levin-Wen model is spanned by these hom spaces\cite{Wen2005,kong2012}.
With all these data, we may assign a F symbol (6j symbol) to each 3-simplex, or tetrahedron, $\bigtriangleup_3$.  Then the partition function of Turaev-Viro TQFT \cite{TURAEV1992,Barrett96} can be written as a product of the F symbols
\begin{align}
Z_{\mathcal{M},\mathcal{C}} = &\frac{1}{|D_\mathcal{C}|^{N_v}}
\sum_{\{A\} \in \mathcal{O}bj(\mathcal{C})}
\prod_{A}d_{A} \prod_{\rm{\bigtriangleup_3\in \mathcal{K}}} F_{\bigtriangleup_3}^{\epsilon(\bigtriangleup_3)},
\label{CLW TQFT}
\end{align}
where $N_v$ is the total number of vertices, $D_\mathcal{C}$ is the dimension of $\mathcal{C}$, $\mathcal{O}bj(\mathcal{C})$ is the class of objects in category $\mathcal{C}$, $\{A\}$ is a configuration of the objects assigned to the 1-simplex in $\mathcal{K}$,
and $d_{A}$ is quantum dimension of the object $A$.
The F symbol $F_{\bigtriangleup_3}$ assigned to a tetrahedron $\bigtriangleup_3$ is just the associator of the fusion category, which is determined by the six objects on the edges of the tetrahedron. $\epsilon(\bigtriangleup_3) = \pm $ determined by the orientation of the 3-simplex.
For a given closed manifold $\mathcal{M}$ and a spherical fusion 1-category $\mathcal{C}$, the partition function $Z_{\mathcal{M},\mathcal{C}}$ is just a number . It has been proved that this number is a topological invariant, i.e. it is independent of the choice of triangulation $\mathcal{K}$ and order of vertices. 

However, the partition function of a TQFT is more than just a number.  If one consider an open manifold $\mathcal{M}$ of dimension $n$, one can divide its boundary of dimension $n-1$, denoted as $\partial \mathcal{M}$, into two pieces $\partial \mathcal{M}_1$ and $\partial \mathcal{M}_2$ such that $\partial \mathcal{M}=\overline{\partial \mathcal{M}_1} \cup \partial \mathcal{M}_2$.
Then $\mathcal{M}$ can be regarded as a cobordism between $\partial \mathcal{M}_1$ and $\partial \mathcal{M}_2$.
In Atiyah's definition of TQFT\cite{atiyah1988,lurie2009}, an $nD$ TQFT $Z$ determines a linear map for any pairs of $(\mathcal{M}, \partial \mathcal{M})$
\begin{align}
Z[\mathcal{M},\partial \mathcal{M}]: \mathcal{H}[\partial \mathcal{M}_1] \rightarrow \mathcal{H}[\partial \mathcal{M}_2],
\end{align}
where $\mathcal{H}[\mathcal{N}]$ is a vector space assigned to the $(n-1)D$ oriented manifold $\mathcal{N}$.

For the 2+1D TQFT corresponds to the Levin-Wen model constructed from a spherical fusion 1-category $\mathcal{C}$, the vector space on the boundary is given by
\begin{align}
\label{eq:hilbertspace}
    \mathcal{H}_{\mathcal{C}}[ \mathcal{N}] = \bigotimes_{\bigtriangleup_2 \in \mathcal{N}} V_{\bigtriangleup_2},
\end{align}
where $\mathcal{N} = \partial \mathcal{M}_1, \partial \mathcal{M}_2$ corresponds to the two pieces respectively, 
$V_{\bigtriangleup_2} = V^{\pm}(ijk)$ is the vector space corresponds to the triangle $\bigtriangleup_2$ in $\mathcal{N}$, $i,j,k$ are the indices of the vertices of $\bigtriangleup_2$ with $i<j<k$.  $V^{\pm}(ijk)$ are defined as
\begin{align}
\label{def_v1}
V^+(ijk)\equiv \hom_{\mathcal{C}}(A_{ij} \boxtimes A_{jk}, A_{ik})\nonumber\\
V^-(ijk)\equiv \hom_{\mathcal{C}}(A_{ik}, A_{ij} \boxtimes A_{jk}),
\end{align}
where $\boxtimes$ denotes the fusion of objects, and $\hom_{\mathcal{C}}(A, B)$ is the hom space from object $A$ to object $B$ in fusion category $\mathcal{C}$.  The vector space $V^{-}(ijk)$ is nothing but time reversal of the vector space $V^+(ijk)$. The $\pm$ sign is chosen according to the orientation of the triangles $\bigtriangleup_2$.

Though the dimension of $V^{\pm}(ijk)$ could be any non-negative integer, $V^{\pm}(ijk)$ is a one-dimensional vector space
in most of the cases that has been studied.  Thus, for simplicity, we will consider only the case where all of the $V^{\pm}(ijk)$ are one dimensional in the following.  Then a set of orthonormalized basis of the Hilbert space $\mathcal{H}_{\mathcal{C}}[ \mathcal{N}]$ is given by
$\otimes_{\bigtriangleup_2 \in \mathcal{N}} \mid V_{\bigtriangleup_2} \rangle$, where $\mid V_{\bigtriangleup_2} \rangle$ is the unit vector in vector space $V_{\bigtriangleup_2}$.
And the partition function of Levin-Wen model on an open manifold $\mathcal{M}$ reads
\begin{widetext}
\begin{align}
\label{OLW TQFT}
Z_{(\mathcal{M},\partial \mathcal{M}),\mathcal{C}} = &\frac{1}{|D_\mathcal{C}|^{N_v-\frac{1}{2}N_{\partial \mathcal{M}}}}\sum_{\{A\} \in \mathcal{O}bj(\mathcal{C})}
\prod_{\bigtriangleup_1\in \mathcal{K}^{in}}d_{{\bigtriangleup_1}}
\prod_{\rm{\bigtriangleup_3\in \mathcal{K}}} F_{\bigtriangleup_3}^{\epsilon(\bigtriangleup_3)} \quad
\left\{  \bigotimes_{\bigtriangleup_2 \in \partial \mathcal{M}_2} \mid V_{\bigtriangleup_2} \rangle\right\}\quad
\left\{ \bigotimes_{\bigtriangleup_2 \in \partial \mathcal{M}_1} \langle V_{\bigtriangleup_2}  \mid \right\},
\end{align}
\end{widetext}
where $N_{\partial \mathcal{M}}$ is number of vertices on boundary of $\mathcal{M}$. $\prod_{\bigtriangleup_1\in \mathcal{K}^{in}}$ means that the product is only over all internal edges $\bigtriangleup_1\in\mathcal{K}^{in}$. The conventional partition function $Z_{\mathcal{M},\mathcal{C}}$, which is just a number, of a closed manifold $\mathcal{M}$ is a special case in this framework, which corresponds to a map from $\mathbb{C}$ to $\mathbb{C}$.
The partition function~\eqref{OLW TQFT} provides a canonical approach to construct the lattice model of Levin-Wen. To write down partition function and F symbol explicitly, one must choose a set of basis in $V_{\bigtriangleup_2}$ and orientations of all simplexes in $\mathcal{K}$. We will discuss our choice when we need to do some explicit calculations in the following sections.

\subsection{Hamiltonian of Levin-Wen model}

The Levin-Wen model for a given spherical fusion category $\mathcal{C}$ is defined on a two-dimensional honeycomb lattice, where each edge is assigned an object in $\mathcal{C}$.  The Hamiltonian of Levin-Wen model reads
\begin{align}
H=-\sum_{\v I} Q_{\v I}-\sum_{\v P} B_{\v P},
\end{align}
where ${\v I}$ and ${\v P}$ are the indices of vertices and plaquettes of the honeycomb lattice respectively.  
The eigenvalues of the operator $Q_{\v I}$ is denoted as $\delta_{IJK}$ and given by
\begin{align}
\delta_{IJK}=\left \{
\begin{aligned}
&1,\quad \mbox{if ${I, J, K}$ are allowed}\\
&0, \quad \mbox{otherwise}
\end{aligned}
\right.,
\end{align}
where $I, J, K$ are the objects on the three edges connect to the vertex $\v I$.  They are allowed if and only if the hom space $\hom_{\mathcal{C}}(I\boxtimes J, K)$ is not empty.

The $B_{\v P}$ operator is given by
\begin{widetext}
\begin{align}
\label{Bp}
B_{\v p} = \frac{1}{D_{\mathcal{C}}}
\sum_{\substack{ABCDEF\\GHIJKL\\G'H'I'J'K'L'S}}d_S
F^{BG^*H}_{S^*H'G'^*}
F^{CH^*I}_{S^*I'H'^*}
F^{DI^*J}_{S^*J'I'^*}
F^{EJ^*K}_{S^*K'J'^*}
F^{FK^*L}_{S^*L'K'^*}
F^{AL^*G}_{S^*G'L'^*}
\Biggl|
\begin{array}{c}
\begin{tikzpicture}[scale=0.5]
\draw[->-] (180:2.5)--(180:1.5) node[midway,below] {$A$};
\draw[->-] (120:1.5)--(180:1.5) node[midway,left] {$G'$};
\draw[->-] (120:2.5)--(120:1.5) node[midway,left] {$B$};
\draw[->-] (60:1.5)--(120:1.5) node[midway,above] {$H'$};
\draw[->-] (60:2.5)--(60:1.5) node[midway,right] {$C$};
\draw[->-] (0:2.5)--(0:1.5) node[midway,below] {$D$};
\draw[->-] (-60:2.5)--(-60:1.5) node[midway,right] {$E$};
\draw[->-] (-120:2.5)--(-120:1.5) node[midway,left] {$F$};
\draw[->-] (0:1.5)--(60:1.5) node[midway,right] {$I'$};
\draw[->-] (-60:1.5)--(0:1.5) node[midway,left] {$J'$};
\draw[->-] (-120:1.5)--(-60:1.5) node[midway,below] {$K'$};
\draw[->-] (180:1.5)--(-120:1.5) node[midway,left] {$L'$};
\end{tikzpicture}
\end{array} 
\Biggr> 
\Biggl<
\begin{array}{c}
\begin{tikzpicture}[scale=0.5]
\draw[->-] (180:2.5)--(180:1.5) node[midway,below] {$A$};
\draw[->-] (120:1.5)--(180:1.5) node[midway,left] {$G$};
\draw[->-] (120:2.5)--(120:1.5) node[midway,left] {$B$};
\draw[->-] (60:1.5)--(120:1.5) node[midway,above] {$H$};
\draw[->-] (60:2.5)--(60:1.5) node[midway,right] {$C$};
\draw[->-] (0:2.5)--(0:1.5) node[midway,below] {$D$};
\draw[->-] (-60:2.5)--(-60:1.5) node[midway,right] {$E$};
\draw[->-] (-120:2.5)--(-120:1.5) node[midway,left] {$F$};
\draw[->-] (0:1.5)--(60:1.5) node[midway,right] {$I$};
\draw[->-] (-60:1.5)--(0:1.5) node[midway,left] {$J$};
\draw[->-] (-120:1.5)--(-60:1.5) node[midway,below] {$K$};
\draw[->-] (180:1.5)--(-120:1.5) node[midway,left] {$L$};
\end{tikzpicture}
\end{array} 
\Biggl|,
\end{align}
\end{widetext}
where the labels in the formula correspond to the objects on the edges connect to the plaquettes $\v P$ (see Ref.~\onlinecite{Wen2005} for details).

\begin{figure}[htbp]
\centering
\begin{tikzpicture}
\coordinate[label=right:$0$] (0) at (0,0) ;
\coordinate[label=above:$1$] (1) at (-2.6,1.5) ;
\coordinate[label=left:$2$] (2) at (-2.6,-1.5) ;
\coordinate[label=below:$3$] (3) at (0,-3) ;
\coordinate[label=right:$4$] (4) at (2.6,-1.5) ;
\coordinate[label=right:$5$] (5) at (2.6,1.5) ;
\coordinate[label=right:$6$] (6) at (0,3) ;
\draw[thick,dashed,red] (0)--(1);
\draw[thick,dashed,red] (0)--(2);
\draw[thick,dashed,red] (0)--(3);
\draw[thick,dashed,red] (0)--(4);
\draw[thick,dashed,red] (0)--(5);
\draw[thick,dashed,red] (0)--(6);
\draw[thick,dashed,red] (6)--(1);
\draw[thick,dashed,red] (1)--(2);
\draw[thick,dashed,red] (2)--(3);
\draw[thick,dashed,red] (3)--(4);
\draw[thick,dashed,red] (4)--(5);
\draw[thick,dashed,red] (5)--(6);

\draw[->-,very thick] (180:2.8)--(180:1.5) node[midway,below] {$A$};
\draw[->-,very thick] (120:1.5)--(180:1.5) node[midway,left] {$G$};
\draw[->-,very thick] (120:2.8)--(120:1.5) node[midway,left] {$B$};
\draw[->-,very thick] (60:1.5)--(120:1.5) node[midway,above] {$H$};
\draw[->-,very thick] (60:2.8)--(60:1.5) node[midway,right] {$C$};
\draw[->-,very thick] (0:2.8)--(0:1.5) node[midway,below] {$D$};
\draw[->-,very thick] (-60:2.8)--(-60:1.5) node[midway,right] {$E$};
\draw[->-,very thick] (-120:2.8)--(-120:1.5) node[midway,left] {$F$};
\draw[->-,very thick] (0:1.5)--(60:1.5) node[midway,right] {$I$};
\draw[->-,very thick] (-60:1.5)--(0:1.5) node[midway,left] {$J$};
\draw[->-,very thick] (-120:1.5)--(-60:1.5) node[midway,above] {$K$};
\draw[->-,very thick] (180:1.5)--(-120:1.5) node[midway,left] {$L$};

\end{tikzpicture}
\caption{Duality between triangular lattice and honeycomb lattice.}
\label{2d duality}
\end{figure}
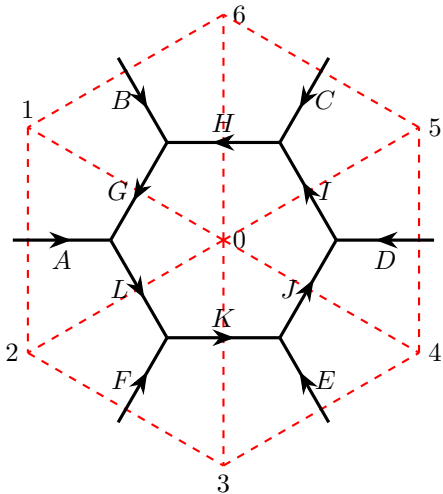

In the following, we will show how to derive the Levin-Wen Hamiltonian from the partition function discussed in previous subsection.  We start from the underlying lattice of the Levin-Wen model.  The two-dimensional honeycomb lattice can be considered as the dual lattice of the 2d boundary of a special triangulation $\mathcal{K}$ of the 2+1D space-time, as shown in Fig.~\ref{2d duality}.  According to the figure, the vertices in $\mathcal{K}$, which are labeled by numbers, are mapped to the plaquettes in the honeycomb lattice; the 1-simplexes or edges in $\mathcal{K}$ are mapped to the edges in honeycomb lattice and labeled by capital letters; while the 2-simplexes of $\mathcal{K}$ are mapped to the vertices of honeycomb lattice.

Under this transformation, the objects assigned to 1-simplexes in $\mathcal{K}$ are now assigned to the edges of honeycomb lattice, which is same as the Levin-Wen model.  Then the 1-morphism and the hom space assigned to the 2-simplexes in $\mathcal{K}$ are now assigned to the vertices of honeycomb lattice.  Since the Hilbert space in the partition function approach is spanned by the tensor product of the hom spaces, a non-vanishing Hilbert space requires non-vanishing hom space on each vertex on the honeycomb lattice.  This constraint is just the flatness condition in Levin-Wen model, which leads to the $Q_{\v I}$ operator in the model.

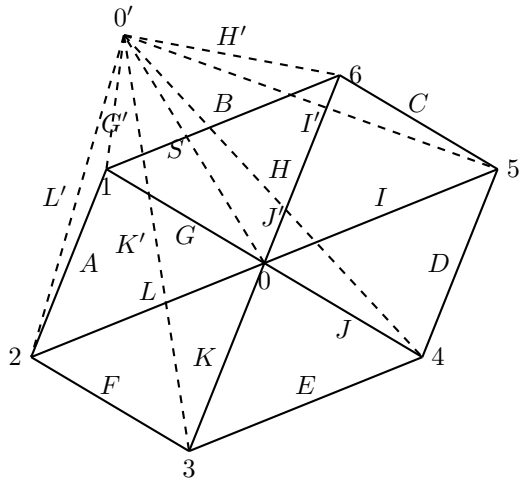
\begin{figure}[htbp]
\centering
\begin{tikzpicture}
\coordinate[label=below:$0$] (0) at (0,0,0) ;
\coordinate[label=below:$1$] (1) at (-2.6,0.75,-1.3) ;
\coordinate[label=left:$2$] (2) at (-2.6,-0.75,1.3) ;
\coordinate[label=below:$3$] (3) at (0,-1.5,2.6) ;
\coordinate[label=right:$4$] (4) at (2.6,-0.75,1.3) ;
\coordinate[label=right:$5$] (5) at (2.6,0.75,-1.3) ;
\coordinate[label=right:$6$] (6) at (0,1.5,-2.6);
\coordinate[label=above:$0'$] (0') at (-1,3.9,2.25);
\draw[thick] (0)--(1) node[midway,below] {$G$};
\draw[thick] (0)--(2) node[midway,above] {$L$};
\draw[thick] (0)--(3) node[midway,left] {$K$};
\draw[thick] (0)--(4) node[midway,below] {$J$};
\draw[thick] (0)--(5) node[midway,above] {$I$};
\draw[thick] (0)--(6) node[midway,left] {$H$};
\draw[thick] (6)--(1) node[midway,above] {$B$};
\draw[thick] (1)--(2) node[midway,right] {$A$};
\draw[thick] (2)--(3) node[midway,above] {$F$};
\draw[thick] (3)--(4) node[midway,above] {$E$};
\draw[thick] (4)--(5) node[midway,left] {$D$};
\draw[thick] (5)--(6) node[midway,above] {$C$};
\draw[thick,dashed] (0)--(0') node[midway,left] {$S$};
\draw[thick,dashed] (0')--(1) node[midway,below] {$G'$};
\draw[thick,dashed] (0')--(2) node[midway,left] {$L'$};
\draw[thick,dashed] (0')--(3) node[midway,left] {$K'$};
\draw[thick,dashed] (0')--(4) node[midway,below] {$J'$};
\draw[thick,dashed] (0')--(5) node[midway,below] {$I'$};
\draw[thick,dashed] (0')--(6) node[midway,above] {$H'$};
\end{tikzpicture}
\caption{Evolution of the Levin-Wen model's Hamiltonian}
\label{LW evolution}
\end{figure}

The $B_{\v P}$ operators can be derived from the partition function of the triangulation shown in Fig.~\ref{LW evolution}.  We can divide the triangulation on the boundary of the space-time into two pieces, $\partial \mathcal{M}_1$ consists of the six triangles 012, 023, 034, 045, 056, and 061, and $\partial \mathcal{M}_2$ consists of the six triangles, 0'12, 0'23, 0'34, 0'45, 0'56, and 0'61.  According to
eqn.~\eqref{OLW TQFT}, we have a linear map between the vector space $V^{+}(012)\otimes V^{+}(023)\otimes  V^{+}(034)\otimes  V^{+}(045)\otimes  V^{+}(056)\otimes V^{+}(061)$ and the vector space $V^{+}(0'12) \otimes V^{+}(0'23) \otimes  V^{+}(0'34) \otimes V^{+}(0'45) \otimes V^{+}(0'56) \otimes  V^{+}(0'61)$.  And the F symbols in eqn.~\eqref{OLW TQFT} correspond to the six tetrahedrons, 00'12, 00'23, 00'34, 00'45, 00'56, and 00'61.
Because there are seven vertices in each of the two boundary pieces and eight vertices in bulk, we have $N_v-\frac{1}{2}N_{\partial \mathcal{M}}=8-7=1$.  Therefore, if we move to the dual lattice, one can find that the partition function discussed above is just the $B_{\v P}$ in Levin-Wen model\cite{Wen2005}.  Physically, the partition function could be understood as the "evolution" of a string-net configuration to another string-net configuration by creating an $s$-string, corresponds to the object on 00' edge in Fig.~\ref{LW evolution}, from vacuum in the plaquettes 0 in the honeycomb lattice.

To this end, we have provided a way to construct the Levin-Wen model from the partition function of 2+1D lattice TQFT for a given spherical fusion category $\mathcal{C}$.  The Hilbert space is spanned by 1-morphism in $\mathcal{C}$ assigned to 2-simplexes on a triangulation $\mathcal{K}$ of the 2+1D space-time. The lattice of Levin-Wen model is dual to the lattice on the boundary of $\mathcal{K}$. $Q_{\v I}$ operator is determined by fusion rule of objects in $\mathcal{C}$. $B_{\v p}$ operator is the partition function which corresponds to the evolution of string-net configuration. In the following, we will generalize all of them to 3+1D case and construct a lattice model that can realize all three-dimensional topological orders. We will refer to this model as membrane-net model in the context.

\subsection{Partition function of three dimensional topological order}

We will follow the above approach to construct the membrane-net model that describes a 3d topological order.  In this subsection, we will discuss the partition function of a 3d topological order.  Then we will derive the model Hamiltonian from the partition function in the next subsection.

In the 3+1D case, a lattice TQFT can be constructed from a spherical fusion 2-category as the following.  We consider a given spherical fusion 2-category $\mathcal{B}$ (see more details in appendix \ref{F2CAT}), a closed oriented four-dimensional manifold $\mathcal{M}$ with a triangulation $\mathcal{K}$, and an order of vertex on $\mathcal{K}$.  Similar to the 2+1D case,  we can assign one object in $\mathcal{B}$ to each 1-simplex $\bigtriangleup_1$ of $\mathcal{K}$, and one 1-morphism in $\hom_{\mathcal{B}}(A\otimes B, C)$ or $\hom_{\mathcal{B}}(C, A\otimes B)$ (depends on the orientation of the 2-simplex) to each 2-simplex $\bigtriangleup_2$ of $\mathcal{K}$, where $A, B, C$ are the objects assigned to the edges of the 2-simplex, $\hom_{\mathcal{B}}(X, Y)$ is the hom space or the set of the 1-morphisms from object $X$ to object $Y$ of $\mathcal{B}$.

In a 2-category, besides of the objects and 1-morphisms, there are another class of data, the 2-morphisms or the associate states.  The set of the 2-morphism, or the associator state space, between two 1-morphisms $a, b \in \hom_{\mathcal{B}}(X, Y)$ are denoted as $\hom_{\mathcal{B}}(a, b)$.  For a fusion 2-category, the associator state spaces are just vector spaces\cite{douglas2018}. Similar to the 2+1D case, we assign a vector space $V_{\bigtriangleup_3} = V^{\pm}(ijkl)$ to each 3-simplex $\bigtriangleup_3$ of $\mathcal{K}$, which are defined as
\begin{align}
\label{def_v2}
V^+(ijkl)=V^+(0123)\equiv \hom_{\mathcal{B}}(i_{012} \otimes j_{023}, k_{013} \otimes l_{123})\nonumber\\
V^-(ijkl)=V^-(0123)\equiv \hom_{\mathcal{B}}(k_{013} \otimes l_{123}, i_{012} \otimes j_{023}),
\end{align}
where $0, 1, 2, 3$ label the four vertices of the 3-simplex, $i,j,k,l$ labels the four 1-morphisms, 
the sign in the superscript is determined by the orientation of the tetrahedron.
These $V_{\bigtriangleup_3}$ can be used to span the Hilbert space of membrane-net model, just like what we did with the $V_{\bigtriangleup_2}$ in the 2+1D case.

With all these data, the partition function on a closed 4D manifold $\mathcal{M}$ can be written as the product of the 10j symbols, $\Gamma$, filling in the 4-simplex $\bigtriangleup_4$
\begin{align}
Z_{\mathcal{M},\mathcal{B}} = &\frac{1}{|D_\mathcal{B}|^{N_v}}\sum_{\substack{ \{A\} \in \mathcal{B}\\ \{a\} \in \mathcal{B}}}
\prod_{\substack{\bigtriangleup_1\in \mathcal{K}\\\bigtriangleup_2\in \mathcal{K}}}f_{\bigtriangleup_1, \bigtriangleup_2}
\prod_{\rm{\bigtriangleup_4\in \mathcal{K}}} \Gamma_{\bigtriangleup_4}^{\epsilon(\bigtriangleup_4)},
\label{CMN TQFT}
\end{align}
where $D_\mathcal{B}$ is the dimension of $\mathcal{B}$, $\{A\}$ and $\{a\}$ are the configurations of the objects and 1-morphisms respectively.  
$\Gamma_{\bigtriangleup_4}$ is the 10j symbol that defines the pentagonator structure of the fusion 2-category\cite{douglas2018}.  It actually has 20 labels because it is determined by the 10 objects and 10 1-morphisms assigned to the 1- and 2-simplexes in each  4-simplex $\bigtriangleup_4$.  $\epsilon(\bigtriangleup_4) = \pm $ is determined by the orientation of the 4-simplex.
$\prod_{\substack{\bigtriangleup_1\in \mathcal{K}\\\bigtriangleup_2\in \mathcal{K}}}f_{\bigtriangleup_1, \bigtriangleup_2}$ is the normalization factor (similar to the term $\prod_{\bigtriangleup_1 \in \mathcal{K}}d_{{\bigtriangleup_1}}$ in two-dimensional case) and given by
\begin{align*}
\prod_{\bigtriangleup_1\in \mathcal{K}}
(d_{A_{\bigtriangleup_1}} d_{\mathrm{End}_{\mathcal{B}}(A_{\bigtriangleup_1})} \mathrm{N}(A_{\bigtriangleup_1}))^{-1}
\prod_{\bigtriangleup_2\in \mathcal{K}}
d_{{\bigtriangleup_2}},
\end{align*}
where $A_{\bigtriangleup_1}$ is the object assigned to the 1-simplex ${\bigtriangleup_1}$, $d_{A}$ is the quantum dimension of $A$. $D_{ \mathrm{End}_{\mathcal{B}}(A)}$ is the dimension of $\mathrm{End}_{\mathcal{B}}(A) \equiv \hom_{\mathcal{B}}(A, A)$, which is actually a fusion 1-category.  $\mathrm{N}(A_{\bigtriangleup_1})$ is the number of equivalence classes of simple objects in the connected component of the object $A_{\bigtriangleup_1}$. $d_{{\bigtriangleup_2}}$ is the quantum dimension of the 1-morphism assigned to the 2-simple ${\bigtriangleup_2}$ (see Ref.~\onlinecite{douglas2018} for more details.)

For an open manifold $\mathcal{M}$ and its boundary $\partial \mathcal{M}$, as discussed in previous subsections, the partition function $Z_{(\mathcal{M},\partial \mathcal{M}),\mathcal{B}}$ is a linear map from $\mathcal{H}_{\mathcal{B}}[\partial \mathcal{M}_1]$ to $\mathcal{H}_{\mathcal{B}}[\partial \mathcal{M}_2]$ and reads
\begin{widetext}
\begin{align}
Z_{(\mathcal{M},\partial \mathcal{M}),\mathcal{B}} = &\frac{1}{|D_\mathcal{B}|^{N_v-\frac{1}{2}N_{\partial \mathcal{M}}}}\sum_{\substack{\{A\} \in \mathcal{B}\\ \{a\} \in \mathcal{B}}}
\prod_{\substack{\bigtriangleup_2\in \mathcal{K}^{in}\\\bigtriangleup_1\in \mathcal{K}^{in}}}f_{\bigtriangleup_1, \bigtriangleup_2}
\prod_{\rm{\bigtriangleup_4\in \mathcal{K}}} \Gamma_{\bigtriangleup_4}^{\epsilon(\bigtriangleup_4)}
\left\{ \bigotimes_{\bigtriangleup_3 \in \partial \mathcal{M}_2} \mid V_{\bigtriangleup_3} \rangle
\right\}\quad
\left\{ \bigotimes_{\bigtriangleup_3 \in \partial \mathcal{M}_1} \langle V_{\bigtriangleup_3}  \mid  \right\}.
\label{OMN TQFT}
\end{align}
\end{widetext}
Here, we assume that all of the $V_{\bigtriangleup_3}$ are one-dimensional for simplicity.  It can be easily generalized to the case with higher dimensional $V_{\bigtriangleup_3}$.  $\mid V_{\bigtriangleup_3} \rangle$ is the unit vector in $V_{\bigtriangleup_3}$.  
The Hilbert space $\mathcal{H}_{\mathcal{B}}[\partial \mathcal{M}_1]$
is spanned by $\otimes_{\bigtriangleup_3 \in \partial \mathcal{M}_1}\mid V_{\bigtriangleup_3} \rangle$, and so does $\mathcal{H}_{\mathcal{B}}[\partial \mathcal{M}_2]$.

\subsection{Hamiltonian of membrane-net model}

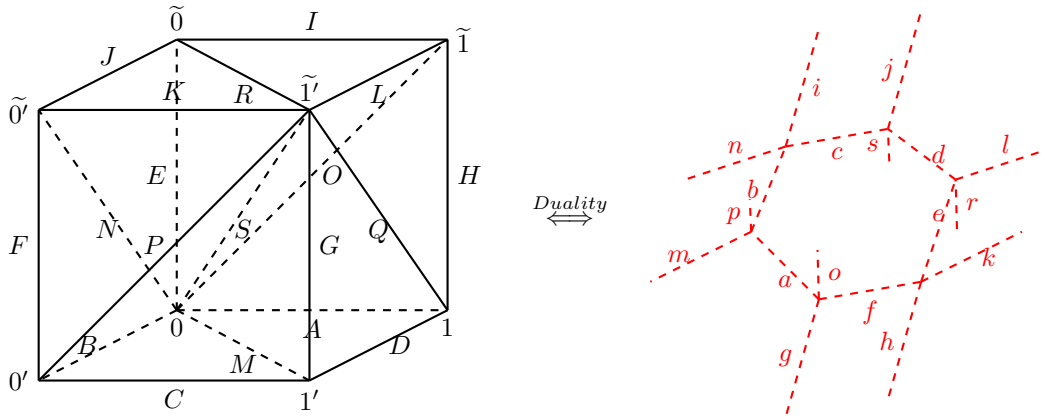
\begin{figure*}[htbp]

$$
\begin{array}{cc}
\begin{array}{c}
\begin{tikzpicture}[scale=0.9]

\coordinate[label=below:$0$] (0) at (0,0,0) ; 
\coordinate[label=below:$1$] (1) at (4,0,0) ; 
\coordinate[label=left:$0'$] (0') at (-0.5,0.5,4) ; 1
\coordinate[label=below:$1'$] (1') at (3.5,0.5,4) ;  
\coordinate[label=above:$\widetilde{0}$] (0t) at (0,4,0); 
\coordinate[label=right:$\widetilde{1}$] (1t) at (4,4,0); 
\coordinate[label=left:$\widetilde{0'}$] (0't) at (-0.5,4.5,4); 
\coordinate[label=above:$\widetilde{1'}$] (1't) at (3.5,4.5,4); 
\coordinate[label=below:$ $] (a) at (6.5/4,5.5/4,3);
\coordinate[label=below:$ $] (b) at (2.5/4,9.5/4,3);
\coordinate[label=below:$ $] (c) at (3/4,13/4,2);
\coordinate[label=below:$ $] (d) at (11/4,5/4,2);
\coordinate[label=below:$ $] (e) at (11.5/4,9.5/4,1);
\coordinate[label=below:$ $] (f) at (7.5/4,12.5/4,1);
\coordinate[label=below:$ $] (cb) at (3/4,13/4-4,2);
\coordinate[label=below:$ $] (fb) at (7.5/4,12.5/4-4,1);
\coordinate[label=below:$ $] (au) at (6.5/4,5.5/4+4,3);
\coordinate[label=below:$ $] (du) at (11/4,5/4+4,2);
\coordinate[label=below:$ $] (cr) at (3/4+4,13/4,2);
\coordinate[label=below:$ $] (br) at (2.5/4+4,9.5/4,3);
\coordinate[label=below:$ $] (dl) at (11/4-4,5/4,2);
\coordinate[label=below:$ $] (el) at (11.5/4-4,9.5/4,1);
\coordinate[label=below:$ $] (eif) at (11.5/4-0.5,9.5/4+0.5,1+4);
\coordinate[label=below:$ $] (fif) at (7.5/4-0.5,12.5/4+0.5,1+4);
\coordinate[label=below:$ $] (abh) at (6.5/4+0.5,5.5/4-0.5,3-4);
\coordinate[label=below:$ $] (bbh) at (2.5/4+0.5,9.5/4-0.5,3-4);
\draw[thick,dashed] (0)--(1) node[midway,below] {$A$};
\draw[thick,dashed] (0)--(0') node[midway,left] {$B$};
\draw[thick] (0')--(1') node[midway,below] {$C$};
\draw[thick] (1)--(1') node[midway,right] {$D$};
\draw[thick,dashed] (0)--(0t) node[midway,left] {$E$};
\draw[thick] (0')--(0't) node[midway,left] {$F$};
\draw[thick] (1')--(1't) node[midway,right] {$G$};
\draw[thick] (1)--(1t) node[midway,right] {$H$};
\draw[thick] (0t)--(1t) node[midway,above] {$I$};
\draw[thick] (0t)--(0't) node[midway,above] {$J$};
\draw[thick] (0't)--(1't) node[midway,above] {$K$};
\draw[thick] (1t)--(1't) node[midway,below] {$L$};
\draw[thick,dashed] (0)--(1') node[midway,below] {$M$};
\draw[thick,dashed] (0)--(0't) node[midway,below] {$N$};
\draw[thick,dashed] (0)--(1t) node[midway,right] {$O$};
\draw[thick] (0')--(1't) node[midway,left] {$P$};
\draw[thick] (1)--(1't) node[midway,below] {$Q$};
\draw[thick] (0t)--(1't) node[midway,below] {$R$};
\draw[thick,dashed] (0)--(1't) node[midway,below] {$S$};
\end{tikzpicture}
\end{array}
\quad \stackrel{Duality}{\Longleftrightarrow} \quad
\begin{array}{c}
\begin{tikzpicture} [scale=0.9]
\draw[thick,dashed,red] (a)--(b) node[midway,below] {$a$};
\draw[thick,dashed,red] (b)--(c) node[midway,left] {$b$};
\draw[thick,dashed,red] (a)--(d) node[midway,below] {$f$};
\draw[thick,dashed,red] (d)--(e) node[midway,above] {$e$};
\draw[thick,dashed,red] (e)--(f) node[midway,right] {$d$};
\draw[thick,dashed,red] (c)--(f) node[midway,below] {$c$};
\draw[thick,dashed,red] (a)--(cb) node[midway,left] {$g$};
\draw[thick,dashed,red] (d)--(fb) node[midway,left] {$h$};
\draw[thick,dashed,red] (c)--(au) node[midway,right] {$i$};
\draw[thick,dashed,red] (f)--(du) node[midway,left] {$j$};
\draw[thick,dashed,red] (d)--(br) node[midway,right] {$k$};
\draw[thick,dashed,red] (e)--(cr) node[midway,above] {$l$};
\draw[thick,dashed,red] (b)--(dl) node[midway,left] {$m$};
\draw[thick,dashed,red] (c)--(el) node[midway,above] {$n$};
\draw[thick,dashed,red] (a)--(eif) node[midway,right] {$o$};
\draw[thick,dashed,red] (b)--(fif) node[midway,left] {$p$};
\draw[thick,dashed,red] (abh)--(e) node[midway,right] {$r$};
\draw[thick,dashed,red] (bbh)--(f) node[midway,left] {$s$};
\end{tikzpicture}
\end{array}
\end{array}
$$

\caption{Duality between tetrahedron lattice (in black) and membrane-net lattice (in red).}
\label{3d dual}
\end{figure*}

Similar to the Levin-Wen model, our membrane-net model is constructed on the dual lattice of the boundary of the triangulation $\mathcal{K}$ of the partition function discussed above.  For simplicity, we consider a special triangulation, whose boundary is the tetrahedron lattice.  We have depicted the tetrahedron lattice and its dual lattice in Fig.~\ref{3d dual} with black lines and red lines respectively.  In the transformation, the 3-simplexes (tetrahedrons) of the tetrahedron lattice are mapped to the vertices of membrane-net lattice, the 2-simplexes (triangles) of the tetrahedron lattice are mapped to the edges of membrane-net lattice, the 1-simplexes (edges) of the tetrahedron lattice are mapped to the faces of the membrane-net lattice, and the vertices of the tetrahedron lattice are mapped to the cells of the membrane-net lattice.

Then we can see how the datas are assigned in the membrane-net lattice.
The 1-morphisms assigned to 2-simplexes of $\mathcal{K}$ are assigned to the edges of membrane-net model.  For example, as shown in Fig.~\ref{3d dual}, the 1-morphism $g$ on the 2-simplex $(00'1')$ of the tetrahedron lattice is assigned to the edge of the membrane-net lattice that penetrates the triangles $(00'1')$.  Similarly, the objects assigned to 1-simplexes of $\mathcal{K}$ are assigned to the faces of the membrane-net model. For example, the object $S$ on the 1-simplex $(0\widetilde{1'})$ of the tetrahedron lattice is assigned to the hexagon enclosed by the red lines in the center of the cube.

In previous subsection, we have mentioned that a vector space depends on the four 1-morphisms on the surface of a tetrahedron has been assigned to that tetrahedron.  In the dual picture, the vector spaces can be understood as being assigned to the corresponding vertices on the membrane-net lattice, and can be denoted as $V^{ijkl}_{\v V}$, where ${\v V}$ labels the vertex, and $ijkl$ labels the 1-morphisms on the edge connect to vertex $\v V$ here.  The Hilbert space is spanned by these vector spaces, i.e.
\begin{align}
    \mathcal{H} = \bigoplus_{\{a\} \in \partial \mathcal{M}} \bigotimes_{\v V} V_{\v V},
\end{align}
where we sum over all possible configurations of 1-morphisms on the edges.

Now, we can construct the Hamiltonian of membrane-net model based on the partition function~\eqref{OMN TQFT}. The Hamiltonian can be written as:
\begin{align}
H=-\sum_{\v V} Q_{\v V}-\sum_{\v E} Q_{\v E}-\sum_{\v C} B_{\v C},
\end{align}
where ${\v V}$, ${\v E}$, ${\v F}$, ${\v C}$ are indices of vertices, edges, faces and cells respectively.

The operator $Q_{\v V}$ acts on each vertex of membrane-net model tracks the constraints of a nonempty $V^{ijkl}_{\v V}$.  $V^{ijkl}_{\v V}$ is nonempty if and only if the associator state space $\hom_{\mathcal{B}}(i \otimes j, k \otimes l)$ is not empty.  Thus the eigenvalue of $Q_{\v V}$ is given by $\delta_{ijlk}$ defined as
\begin{align}
\delta_{ijkl}=\left \{
\begin{aligned}
&1,\quad \mbox{if $\hom_{\mathcal{B}}(i \otimes j, k \otimes l)$ is not empty}\\
&0, \quad \mbox{otherwise}
\end{aligned}
\right.
\end{align}

The operator $Q_{\v E}$ acts on one edge, which is shared by 3 faces, tracks the constraints that the 1-morphism on the triangles in the tetrahedron lattice must be in the space determined by the objects on the edges of the triangle.  Thus, 
the eigenvalue of $Q_{\v E}$ is $\delta_{ABC,f}$ defined as
\begin{align}
\delta_{ABC,f}=\left \{
\begin{aligned}
&1,\quad f \in \hom_{\mathcal{B}}(A \otimes B, C) \\
&0, \quad \mbox{otherwise}
\end{aligned}
\right.
\end{align}
The combinations of $Q_{\v V}$ and $Q_{\v E}$ corresponds to the flatness condition of the membrane-net model. 

The operator $B_{\v C}$ acting on each cell of membrane-net lattice states describes the effect of creating a closed membrane in the cell ${\v C}$ from the vacuum and fusing it into the current membrane-net configuration($\{A\}$ and $\{a\}$) to create a new membrane-net configuration($\{A'\}$ and $\{a'\}$). 
This operator is much more complicated then $Q_{\v V}$ and $Q_{\v E}$. However, we can have a better understanding of this operator on dual lattice of the membrane-net model-the tetrahedron lattice and $B_{\v C}$ can be written as:
\begin{widetext}
\begin{align}
B_{\v C} = &\frac{1}{|D_\mathcal{B}|} \sum_{\substack{\{A\}, \{A'\} \in \mathcal{B} \\ \{a\}, \{a'\} \in \mathcal{B}}}
\prod_{\substack{\bigtriangleup_2\in \mathcal{K}^{in}\\\bigtriangleup_1\in \mathcal{K}^{in}}}f_{\bigtriangleup_1, \bigtriangleup_2}
\prod_{\rm{\bigtriangleup_4\in \mathcal{K}}} \Gamma_{\bigtriangleup_4}^{\epsilon(\bigtriangleup_4)}
\left\{ \bigotimes_{\bigtriangleup_3 \in \partial \mathcal{M}_2} \mid V^{ \{a'_{\bigtriangleup_2}\}}_{\bigtriangleup_3} \rangle
\right\}\quad
\left\{ \bigotimes_{\bigtriangleup_3 \in \partial \mathcal{M}_1} \langle V^{ \{a_{\bigtriangleup_2}\}}_{\bigtriangleup_3}  \mid  \right\}
\label{B_Cd}
\end{align}
\end{widetext}
with the partition function~\eqref{OMN TQFT} on the 4D generalization of Fig.~\ref{LW evolution}.  The $B_{\v C}$ operator maps states on 24 tetrahedrons($\otimes_{\bigtriangleup_3 \in \partial \mathcal{M}_1} \mid V^{ \{a_{\bigtriangleup_2}\}}_{\bigtriangleup_3} \rangle$ or $\otimes_{\v V \in \v C} \mid V^{\{a\}}_{\v V} \rangle
$ in membrane-net lattice) to states on another 24 tetrahedrons($\otimes_{\bigtriangleup_3 \in \partial \mathcal{M}_2} \mid V^{ \{a'_{\bigtriangleup_2}\}}_{\bigtriangleup_3} \rangle$ or $ \otimes_{\v V \in \v C} \mid V^{\{a'\}}_{\v V} \rangle
$ in membrane-net lattice).
Thus $B_{\v C}$ on membrane-net lattice reads
\begin{widetext}
\begin{align}
B_{\v C} = &\frac{1}{|D_\mathcal{B}|}\sum_{\substack{\{A\}, \{A'\} \in \mathcal{B}\\ \{a\}, \{a'\} \in \mathcal{B}}}
\prod_{\substack{\v E \in \mathcal{K}^{in}\\ \v F \in \mathcal{K}^{in}}}f_{\v F, \v E}
\Gamma_{\{A\}, \{a\}}^{\{A'\}, \{a'\}} 
\left\{ \bigotimes_{\v V \in \v C} \mid V^{\{a'\}}_{\v V} \rangle
\right\}\quad
\left\{ \bigotimes_{\v V \in \v C} \langle V^{\{a\}}_{\v V}  \mid  \right\},
\label{B_C}
\end{align}
\end{widetext}
where $\v V$, $\v E$, and $\v F$ are the vertices, edges, and faces in the cell $\v C$, and $\Gamma_{\{A\}, \{a\}}^{\{A'\}, \{a'\}} =  \prod_{\rm{\bigtriangleup_4\in \mathcal{K}}} \Gamma_{\bigtriangleup_4}^{\epsilon(\bigtriangleup_4)}$.  $\{A\}$ ($\{A'\}$) and $\{a\}$ ($\{a'\}$) are the objects and 1-morphisms of the initial (final) string-net configuration of the cell $\v C$ respectively (see more details in appendix \ref{B_Cmap}).

To this end, we have provided a complete construction of membrane-net model through which all three-dimensional topological orders can be realized on lattice. However, we have not got any string-like and particle-like excitations in the model yet. In fact, it is formidable to find out all membrane operators/string operators commuting with the Hamiltonian. Fortunately, excitations in membrane-net model may be constructed by the 3d tube algebra.

\section{Tube algebra and excitations in three-dimensional topological order}

\subsection{2d Tube algebra and Levin-Wen model}

Tube algebra\cite{ocneanu1994} or Q-algebra is a finite dimensional $\mathbb{C}$-algebra.
It has been shown that there is a one-to-one correspondence between the irreducible central idempotents (ICIs) of tube algebra or the simple modules over Q-algebra and the particle-like excitations in Levin-Wen model \cite{Lan2014,Bultinck2017,Lootens2020}. The basis element in a tube algebra can be presented as
\begin{align}
\label{2d tube Tri}
\mathbf{A}_{ABCD}=
\begin{array}{c}
\begin{tikzpicture}[scale=0.4]
\draw[->-] (-2.5,2.5)--(2.5,2.5) node[midway,below] {$D$};
\draw[->-] (-2.5,2.5)--(-2.5,-2.5) node[midway,left] {$A$};
\draw[->-] (-2.5,-2.5)--(2.5,-2.5) node[midway,below] {$D$};
\draw[->-] (2.5,2.5)--(2.5,-2.5) node[midway,right] {$C$};
\draw[->-] (-2.5,2.5)--(2.5,-2.5) node[midway,below] {$B$};
\end{tikzpicture}
\end{array} 
\end{align}
in triangular lattice.  The boundary labeled by $D$ are glued together, hence $\Ab_{ABCD}$ looks like a cylinder, or $S^1\times I$.  Therefore, the algebra formed by $\Ab_{ABCD}$ are called a tube algebra, while $\Ab_{ABCD}$ is called a tube.  In the honeycomb lattice, $\Ab_{ABCD}$ can be represented as
\begin{align}
\label{2d tube hc}
\mathbf{A}_{ABCD}=
\begin{array}{c}
\begin{tikzpicture}[scale=0.4]
\draw[->-] (-0.7,-0.7)--(-3,-0.7) node[midway,below] {$A$};
\draw[->-] (-0.7,-0.7)--(-0.7,-3) node[midway,left] {$D$};
\draw[->-] (0.7,0.7)--(-0.7,-0.7) node[midway,below] {$B$};
\draw[->-] (3,0.7)--(0.7,0.7) node[midway,below] {$C$};
\draw[->-] (0.7,3)--(0.7,0.7) node[midway,left] {$D$};
\draw [dashed](0.7,3).. controls (2,5) and (2,-5) ..(-0.7,-3);
\end{tikzpicture}
\end{array} 
\end{align}
where the legs labeled by $D$ are connected.

\begin{figure}[htbp]
\centering
\begin{align*}
\begin{array}{c}
\begin{tikzpicture}[scale=0.9]
\draw [->-] (0,0) circle (1);
\draw [->-] (0,0) circle (1.5);
\draw [->-] (0,0) circle (2.5);
\draw[ultra thick,dotted] (0,2.1)--(0,2.5);
\draw[->-] (210:0.4)--(210:1);
\draw[->-] (150:0.4)--(150:1);
\draw[->-] (90:0.4)--(90:1);
\draw[->-] (30:0.4)--(30:1);
\draw[->-] (-90:0.4)--(-90:1) node[midway,above] {$n-legs$};
\draw[->-] (-30:0.4)--(-30:1);
\draw[->-] (180:1)--(180:1.5);
\draw[->-] (120:1)--(120:1.5)  node[midway,right] {$1-loop$};
\draw[->-] (60:1)--(60:1.5) ;
\draw[->-] (0:1)--(0:1.5);
\draw[->-] (-60:1)--(-60:1.5) ;
\draw[->-] (-120:1)--(-120:1.5);
\draw[->-] (210:1.5)--(210:2);
\draw[->-] (150:1.5)--(150:2);
\draw[->-] (90:1.5)--(90:2);
\draw[->-] (30:1.5)--(30:2);
\draw[->-] (-90:1.5)--(-90:2);
\draw[->-] (-30:1.5)--(-30:2);
\draw[->-] (180:1.9)--(180:2.5);
\draw[->-] (120:1.9)--(120:2.5) node[midway,above] {$L-loop$};
\draw[->-] (60:1.9)--(60:2.5);
\draw[->-] (0:1.9)--(0:2.5);
\draw[->-] (-60:1.9)--(-60:2.5);
\draw[->-] (-120:1.9)--(-120:2.5);
\draw[->-] (210:2.5)--(210:3);
\draw[->-] (150:2.5)--(150:3);
\draw[->-] (90:2.5)--(90:3);
\draw[->-] (30:2.5)--(30:3);
\draw[->-] (-90:2.5)--(-90:3) node[midway,right] {$m-legs$};
\draw[->-] (-30:2.5)--(-30:3);
\end{tikzpicture}
\end{array} 
\begin{array}{c}
\begin{tikzpicture}[scale=1]
\draw[ultra thick,->] (2.5,0)--(4.0,0) node[midway,above] {$renormalize$};
\draw [->-] (5.2,0) circle (0.8);
\coordinate[label=left:$B$] (0) at (4.5,-0.5);
\coordinate[label=above:$\mathbf{A}_{ABCD}$] (1) at (5.2,-1.8);
\draw[->-] (5.2,-0.8)--(5.2,-1.3) node[midway,left] {$A$};
\draw[->-] (4.9,0)--(4.4,0) node[midway,above] {$C$};
\draw[->-] (5.21,0.8)--(5.2,0.8) node[midway,above] {$D$};
%
\end{tikzpicture}
\end{array} 
\end{align*}
\caption{A graph of string-net configuration on on cylinder renormalized into tube element. 
Note that here we have drawn an annulus instead of cylinder $S^1\times I$ for convenience due to a cylinder $S^1\times I$ is homeomorphic to an annulus}
\label{RG-String-net}
\end{figure}
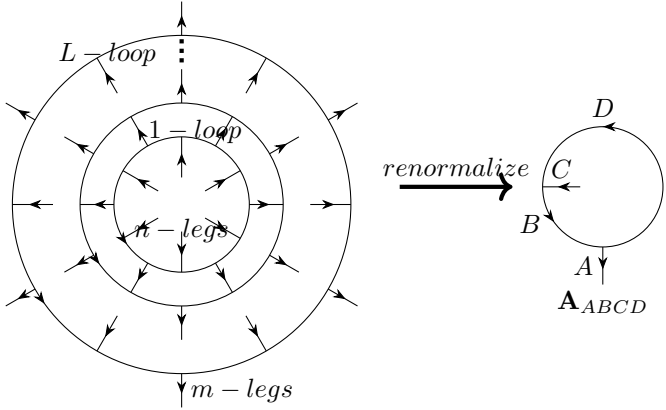

Physically, in the sense of wave function renormalization process, the basis element $\Ab_{ABCD}$ can be viewed as a basis vector in the Hilbert space of the Levin-Wen model on an open manifold $S^1\times I$, or a tube. 
In general, an arbitrary string-net configuration on a tube can be denoted as $\mathbf{v}(L)^m_n$, where $m/n$ is the number of legs of outer/inner boundary, $L$ is the number of large loops around the $S^1$ direction.  The Hilbert space of all the configurations for given $m, n, $ and $L$ are denoted as $\mathcal{H}(L)^{m}_{n}$.
Since Levin-Wen is a fixed point theory, we can freely perform renormalization along the radial direction and tangential direction of the tube(annulus). And the renormalization in these two directions will reduce the number of loops $L$ and boundary legs $m,n$ respectively.  Finally, the string-net configuration $\mathbf{v}(L)^m_n$ and Hilbert space $\mathcal{H}(L)^m_n$ will become the tube $\mathbf{v}(1)^1_1$, which is just $\Ab_{ABCD}$, and the tube space $\mathcal{H}(1)^1_1$ as shown in Fig.~\ref{RG-String-net}.

On the other hand, since the Levin-Wen model is at the fixed point, we can reverse the renormalization process above, and recover the string-net configuration and Hilbert space from the basic tube elements $\Ab_{ABCD}$.  Roughly speaking, an arbitrary string-net sate can be constructed by ``gluing'' the tubes properly.  The process that gluing two tubes and then renormalizing them back to another tube defines the multiplication of tubes, which can be represented as
\begin{align*}
\begin{array}{c}
\begin{tikzpicture}[scale=0.3]
\draw[->-] (-7.5,2.5)--(-2.5,2.5) node[midway,below] {$D$};
\draw[->-] (-7.5,2.5)--(-7.5,-2.5) node[midway,right] {$A$};
\draw[->-] (-7.5,-2.5)--(-2.5,-2.5) node[midway,below] {$D$};
\draw[->-] (-7.5,2.5)--(-2.5,-2.5) node[midway,below] {$B$};
\draw[->-] (-2.5,2.5)--(2.5,2.5) node[midway,below] {$H$};
\draw[->-] (-2.5,2.5)--(-2.5,-2.5) node[midway,left] {$C$};
\draw[->-] (-2.5,-2.5)--(2.5,-2.5) node[midway,below] {$H$};
\draw[->-] (2.5,2.5)--(2.5,-2.5) node[midway,left] {$G$};
\draw[->-] (-2.5,2.5)--(2.5,-2.5) node[midway,below] {$F$};
\end{tikzpicture}
\end{array} 
=  \sum_{IJ} \Omega^{ABCD, CFGH}_{AJGI}
\begin{array}{c}
\begin{tikzpicture}[scale=0.3]
\draw[->-] (-2.5,2.5)--(2.5,2.5) node[midway,below] {$I$};
\draw[->-] (-2.5,2.5)--(-2.5,-2.5) node[midway,right] {$A$};
\draw[->-] (-2.5,-2.5)--(2.5,-2.5) node[midway,below] {$I$};
\draw[->-] (2.5,2.5)--(2.5,-2.5) node[midway,left] {$G$};
\draw[->-] (-2.5,2.5)--(2.5,-2.5) node[midway,below] {$J$};
\end{tikzpicture}
\end{array},
\end{align*}
or
\begin{align}
\Ab_{ABCD}\times \Ab_{EFGH}= \delta_{CE} \sum_{IJ} \Omega^{{ABCD}, {CFGH}}_{{AJGI}} \Ab_{AJGI}.
\end{align}
$\delta_{CE}$ means only the 1-simplexes labeled by same object can be glued. 
Physically, the ``multiplication'' suggests that $\Ab_{ABCD}$ is not only a state but can also be viewed as a linear operator acting on a string-net state (described by another tube) in Levin-Wen model, this is the so-called {\em state-operator correspondence}.

The multiplication of tube algebra maps a vector in $\mathcal{H}(1)^1_1 \otimes \mathcal{H}(1)^1_1$ to a vector in $\mathcal{H}(1)^1_1$.  Therefore, the structure factor $\Omega$ can be calculated with the partition function of a 2+1D TQFT on an open manifold $D^2\times S^1$ with a triangulation shown below
\begin{widetext}
\begin{align}
\label{Z_2tube}
Z\Biggl(
\begin{array}{c}
\begin{tikzpicture}[scale=0.56]
\coordinate[label=below:$\widetilde{2}$] (0) at (0,0,0) ; 
\coordinate[label=below:$\widetilde{1}$] (1) at (4,0,0) ; 
\coordinate[label=below:$\widetilde{0}$] (1') at (3.5,0.5,4) ;  
\coordinate[label=above:$2$] (0t) at (0,4,0); 
\coordinate[label=above:$1$] (1t) at (4,4,0); 
\coordinate[label=above:$0$] (1't) at (3.5,4.5,4); 
\draw[->-,thick,dashed] (1)--(0) node[midway,left] {$H$};
\draw[->-,thick] (1')--(1) node[midway,below] {$D$};
\draw[->-,thick] (0t)--(0) node[midway,right] {$G$};
\draw[->-,thick] (1't)--(1') node[midway,right] {$A$};
\draw[->-,thick] (1t)--(1) node[midway,left] {$C$};
\draw[->-,thick] (1t)--(0t) node[midway,above] {$H$};
\draw[->-,thick] (1't)--(1t) node[midway,above] {$D$};
\draw[->-,thick] (1')--(0) node[midway,below] {$I$};
\draw[->-,thick,dashed] (1t)--(0) node[midway,right] {$F$};
\draw[->-,thick] (1't)--(1) node[midway,below] {$B$};
\draw[->-,thick] (1't)--(0t) node[midway,below] {$I$};
\draw[->-,thick] (1't)--(0) node[midway,above] {$J$};
\end{tikzpicture}
\end{array} 
\Biggr)
= \frac{1}{|D_\mathcal{C}|^\frac{1}{2}}
\sum_{I,J}
\delta_{CE}
F^{DIH}_{GFJ} {F^{DBC}_{HFJ}}^{-1} F^{ADB}_{HIJ}
\Biggl|
\begin{array}{c}
\begin{tikzpicture}[scale=0.4]
\draw[->-] (-2.5,2.5)--(2.5,2.5) node[midway,below] {$I$};
\draw[->-] (-2.5,2.5)--(-2.5,-2.5) node[midway,right] {$A$};
\draw[->-] (-2.5,-2.5)--(2.5,-2.5) node[midway,below] {$I$};
\draw[->-] (2.5,2.5)--(2.5,-2.5) node[midway,left] {$G$};
\draw[->-] (-2.5,2.5)--(2.5,-2.5) node[midway,below] {$J$};
\end{tikzpicture}
\end{array} 
\Biggr> 
\Biggl<
\begin{array}{c}
\begin{tikzpicture}[scale=0.4]
\draw[->-] (-7.5,2.5)--(-2.5,2.5) node[midway,below] {$D$};
\draw[->-] (-7.5,2.5)--(-7.5,-2.5) node[midway,right] {$A$};
\draw[->-] (-7.5,-2.5)--(-2.5,-2.5) node[midway,below] {$D$};
\draw[->-] (-7.5,2.5)--(-2.5,-2.5) node[midway,below] {$B$};
\draw[->-] (-2.5,2.5)--(2.5,2.5) node[midway,below] {$H$};
\draw[->-] (-2.5,2.5)--(-2.5,-2.5) node[midway,left] {$C$};
\draw[->-] (-2.5,-2.5)--(2.5,-2.5) node[midway,below] {$H$};
\draw[->-] (2.5,2.5)--(2.5,-2.5) node[midway,left] {$G$};
\draw[->-] (-2.5,2.5)--(2.5,-2.5) node[midway,below] {$F$};
\end{tikzpicture}
\end{array} 
\Biggl|,
\end{align}
\end{widetext}
where $\frac{1}{|D_\mathcal{C}|^\frac{1}{2}}$ is just a constant for normalization.  Please note that the up surface and lower surface of the prism is glued together.  We have divide the boundary ($S^1 \times S^1$) of the prism into three pieces of $S^1\times I$, each corresponds to a tube element.

The definition of the F symbol in eqn.~\eqref{Z_2tube} is given in the following.  For a tetrahedron with the triangulation
\begin{align*}
\begin{tikzpicture}[scale=0.7]
\coordinate[label=below left:$2$] (0) at (0,0,0) ; 
\coordinate[label=below right:$1$] (1) at (4,0,0) ; 
\coordinate[label=below:$0$] (1') at (3.5,0.5,4) ;  
\coordinate[label=above:$3$] (1't) at (3.5,4.5,4); 
\draw[->-,thick,dashed] (1)--(0) node[midway,above left] {$C$};
\draw[->-,thick] (1')--(1) node[midway,below right] {$A$};
\draw[->-,thick] (1')--(1't) node[midway,right] {$F$};
\draw[->-,thick] (1')--(0) node[midway,below left] {$B$};
\draw[->-,thick] (1)--(1't) node[midway, above right] {$E$};
\draw[->-,thick] (0)--(1't) node[midway,above left] {$D$};
\end{tikzpicture},
\end{align*}
the $F$ symbol assigned to it is defined as
\begin{align}
\label{F basis}
&{F^{A\,B\,C}_{D\,E\,F}}:
V^+(023)\otimes V^+(012) \rightarrow V^+(123) \otimes V^+(013)\nonumber\\
&({F^{A\,B\,C}_{D\,E\,F}})^{-1}:
V^+(123) \otimes V^+(013) \rightarrow  V^+(023)\otimes V^+(012),
\end{align}
where states on left hand side of eqn.~\eqref{F basis} are called in-state and states on right hand side of eqn.~\eqref{F basis} are called out-state. Note that in all following discussion we set the vertex order $0<1<2\cdots <0'<1'<2'\cdots <\widetilde{0}<\widetilde{1}<\widetilde{2} \cdots  <\widetilde{0'}<\widetilde{1'}<\widetilde{2'}\cdots$.

If we reverse all the arrows of a 2-simplex, which equivalents to change its orientation, $V^{+}(012)$ will become $V^{-}(012)$, and vice versa.  Physically, this corresponds to a time reversal transformation. On the other hand, the time reversal transformation also maps an in-state to an out-state. Thus the $F$ symbol can also be written as
\begin{align}
&{F^{A\,B\,C}_{D\,E\,F}}:
V^+(023) \rightarrow V^+(123) \otimes V^+(013)\otimes V^-(012) \nonumber\\
&{F^{A\,B\,C}_{D\,E\,F}}:
V^+(023)\otimes V^+(012) \otimes V^-(013) \rightarrow V^+(123) \nonumber\\
&\cdots
\end{align}
If we reverse all the arrows of a 3-simplex, which also equivalent to change its orientation, the time reversal transformation will turn F to its Hermitian conjugate.
And for $F^{-1}$, since we consider only the unitary case with unitary fusion 1-category, the inversion of $F$ is equal to its Hermitian conjugate. 

With this definition, the product of the three F symbols in eqn.~\eqref{Z_2tube} determines a map
\begin{align}
\label{2d tube map 2}
& V^{+}(01\widetilde{1})\otimes V^-(0\widetilde{0}\widetilde{1}) \otimes V^{+}(12\widetilde{2})\otimes V^-(1\widetilde{1}\widetilde{2}) \rightarrow \nonumber \\ & V^{+}(02\widetilde{2}) \otimes V^-(0\widetilde{0}
\widetilde{2}),
\end{align}
through which state $\Ab_{ABCD} \otimes \Ab_{EFGH}$ is mapped to state $\Ab_{AJGI}$.

It is known that a 2d tube algebra can be decomposed into a direct sum of simple matrix algebras\cite{Biswas1995,MUGER2003}. And  different blocks in this decomposition correspond to different topological sectors labeled by simple anyons in Levin-Wen model \cite{Lan2014, Bultinck2017} (recall that $\Ab_{ABCD}$ itself is a string-net configuration).  These blocks can be obtained by multiplying all the tubes $\Ab_{ABCD}$ with the ICIs, and hence these ICIs can be labeled by the type of the simple anyons in the Levin-Wen model.   In the following, we denote the ICI as $P_i$, where $i$ is a label of simple anyons.  In the tube algebra, the ICIs satisfy $P_i P_j=\delta_{ij}P_i$, $P_i \Ab_{ABCD}=\Ab_{ABCD} P_i$ (see appendix \ref{ICIalg} for detailed algorithm), and can be expanded with the tubes as
\begin{align}
\label{CI}
P_{i}=\frac{1}{D^2_{\mathcal{C}}}\sum_{ABD} \sigma^{i}_{ABD} \Ab_{ABAD},
\end{align}
or represented graphically as
\begin{align}
\label{graph CI}
\begin{array}{c}
\begin{tikzpicture}[scale=0.9]
\draw[thick] (-0.4,-0.4) rectangle (0.4,0.4) node[midway] {$P_i$};
\draw[] (0,0.4) arc (160:-154:1);
\draw[->-] (-0.4,0)--(-0.8,0) node[midway,above] {};
\draw[->-] (0.8,0)--(0.4,0) node[midway,above] {};
\end{tikzpicture}
\end{array}
=\sum_{ABD} \sigma^{i}_{ABD}
\begin{array}{c}
\begin{tikzpicture}[scale=0.36]
\draw[->-] (-0.7,-0.7)--(-3,-0.7) node[midway,above] {$A$};
\draw[->-] (-0.7,-0.7)--(-0.7,-3) node[midway,left] {$D$};
\draw[->-] (0.7,0.7)--(-0.7,-0.7) node[midway,below] {$B$};
\draw[->-] (3,0.7)--(0.7,0.7) node[midway,below] {$A$};
\draw[->-] (0.7,3)--(0.7,0.7) node[midway,left] {$D$};
\draw [dashed](0.7,3).. controls (2,5) and (2,-5) ..(-0.7,-3);
\end{tikzpicture}
\end{array}
\end{align}
In practice, $P_i$'s can be obtained by the so-called idempotent decomposition, i.e. decompose the identity of the tube algebra into the sum of $\{P_i\}$.

An ICI $P_i$ is more than just a mathematical concept.  It has been shown that there is a direct relation between $P_i$ and the string operators that create anyons \cite{Lan2014, Bultinck2017}.  Physically, this can be understood as the following.  In Levin-Wen model, the anyons created by a string operator can move freely in the system.  On the other hand, the Hamiltonian of Levin-Wen model can be written with the tubes $\Ab_{ABCD}$.  Since $P_i$ commutes with all the $\Ab_{ABCD}$ by definition, the excitation created by $P_i$ can move freely in the system.  Notice that $P_i$ is also labeled with the anyon type, it is not surprising to find that $P_i$ is related to the string operator that creates $i$-type anyons.  In fact, the shortest string operator of anyon type $i$ can be represented as
\begin{align}
\label{graph Str}
\begin{array}{c}
\begin{tikzpicture}[scale=1.2]
\draw[ultra thick,->-](0.8,0.5)--(-0.8,-0.5)  node[below] {$i$};
\draw[thick,->-](-0.8,0.5)--(-0.16,0.1)  node[midway,below] {$D$};
\draw[thick,->-](0.16,-0.1)--(0.8,-0.5)  node[midway,below] {$D$};
\end{tikzpicture}
\end{array}
=\sum_{AB}\mathcal{W}^i_{ABAD} 
\begin{array}{c}
\begin{tikzpicture}[scale=1.4]
\draw[thick,->-](-0.7,0.4)--(-0.3,0)  node[midway,above] {$D$};
\draw[thick,->-](0.3,0)--(0.7,-0.4)  node[midway,below] {$D$};
\draw[thick,->-](0.3,0)--(-0.3,0)  node[midway,below] {$B$};
\draw[thick,->-](0.7,0.4)--(0.3,0)  node[midway,right] {$A$};
\draw[thick,->-](-0.3,0)--(-0.7,-0.4)  node[midway,left] {$A$};
\end{tikzpicture}
\end{array}
\end{align}
where $\mathcal{W}^i_{ABAD}= \sigma^{i}_{ABD}/d_D$, and $d_D$ is the quantum dimension of $D$.  By comparing it with eqn.~\eqref{graph CI}, we can find that we can get the string operator by cutting the $D$ loop in $P_i$.

It should be noticed that we have made choices of the basis of F symbols, the basis of tube algebras, the triangulation, and the orientations of the simplexes.  With difference choices, we may have different form of ICIs, string operators, and even the Hamiltonian.  However, the physical observables of the anyons such as S T matrix are independent of these choices. Through the reconstructed string operator from ICIs, the S T matrix can be calculated,
\begin{equation}
\begin{gathered}
T_i=\mathrm{e}^{-\mathrm{i} \theta_{i}}=\frac{1}{d_{i}} \sum_{A} d_{A}^{2} \mathcal{W}(i)_{ A 0 A A^*}, \\
S_{i j}=\frac{1}{D_{\mathcal{Z}(\mathcal{C})}} \sum_{ABD} d_A d_D \mathcal{W}(i)_{ A B^* A D} \mathcal{W}(j)_{ D^* B D^* A^*},
\end{gathered}
\end{equation}
where $d_{i}=\sum_{A\in i}d_A$ and $D_{\mathcal{Z}(\mathcal{C})}$ is quantum dimension of the UMTC $\mathcal{Z}(\mathcal{C})$, which is center of the input fusion 1-category.

\subsection{3d Tube algebra and membrane-net model}

Similarly, the basis element in 3d tube algebra is a basis vector in Hilbert space of membrane-net model, represented as:
\begin{widetext}
\begin{align}
\mathbf{A}^{ABCDEFGHIJ}_{abcdefghijklmn}=
\begin{array}{c}
\begin{tikzpicture}[scale=1.3]
\coordinate[label=below:$0$] (0) at (0,0,0) ; 
\coordinate[label=below:$1$] (1) at (4,0,0) ; 
\coordinate[label=left:$0'$] (0') at (-0.5,0.5,4) ; 1
\coordinate[label=below:$1'$] (1') at (3.5,0.5,4) ;  
\coordinate[label=above:$\widetilde{0}$] (0t) at (0,4,0); 
\coordinate[label=right:$\widetilde{1}$] (1t) at (4,4,0); 
\coordinate[label=left:$\widetilde{0'}$] (0't) at (-0.5,4.5,4); 
\coordinate[label=above:$\widetilde{1'}$] (1't) at (3.5,4.5,4); 
\coordinate[label=above:$ $] (00) at (6,0.5,3); 
\coordinate[label=above:$ $] (01) at (8,0.5,3); 
\coordinate[label=above:$ $] (02) at (6,2.5,3); 
\coordinate[label=above:$ $] (03) at (5.75,0.75,5); 
\draw[thick,->-] (00)--(01) node[midway,above] {$x$};%
\draw[thick,->-] (00)--(02) node[midway,right] {$y$};%
\draw[thick,->-] (00)--(03) node[midway,above] {$z$};%
\coordinate[label=below:$ $] (a) at (6.5/4,5.5/4,3);
\coordinate[label=below:$ $] (b) at (2.5/4,9.5/4,3);
\coordinate[label=below:$ $] (c) at (3/4,13/4,2);
\coordinate[label=below:$ $] (d) at (11/4,5/4,2);
\coordinate[label=below:$ $] (e) at (11.5/4,9.5/4,1);
\coordinate[label=below:$ $] (f) at (7.5/4,12.5/4,1);
\coordinate[label=below:$ $] (cb) at (3/4,13/4-4,2);
\coordinate[label=below:$ $] (fb) at (7.5/4,12.5/4-4,1);
\coordinate[label=below:$ $] (au) at (6.5/4,5.5/4+4,3);
\coordinate[label=below:$ $] (du) at (11/4,5/4+4,2);
\coordinate[label=below:$ $] (cr) at (3/4+4,13/4,2);
\coordinate[label=below:$ $] (br) at (2.5/4+4,9.5/4,3);
\coordinate[label=below:$ $] (dl) at (11/4-4,5/4,2);
\coordinate[label=below:$ $] (el) at (11.5/4-4,9.5/4,1);
\coordinate[label=below:$ $] (eif) at (11.5/4-0.5,9.5/4+0.5,1+4);
\coordinate[label=below:$ $] (fif) at (7.5/4-0.5,12.5/4+0.5,1+4);
\coordinate[label=below:$ $] (abh) at (6.5/4+0.5,5.5/4-0.5,3-4);
\coordinate[label=below:$ $] (bbh) at (2.5/4+0.5,9.5/4-0.5,3-4);
\draw[thick,->-,dashed] (0)--(1) node[midway,below] {$ $};%
\draw[thick,->-,dashed] (0)--(0') node[midway,above] {$C$};%
\draw[thick,->-] (0')--(1') node[midway,below] {$A$};%
\draw[thick,->-] (1)--(1') node[midway,right] {$H$};%
\draw[thick,->-,dashed] (0)--(0t) node[midway,left] {$ $};%
\draw[thick,->-] (0')--(0't) node[midway,left] {$B$};%
\draw[thick,->-] (1')--(1't) node[midway,right] {$G$};%
\draw[thick,->-] (1)--(1t) node[midway,right] {$ $};%
\draw[thick,->-] (0t)--(1t) node[midway,above] {$ $};%
\draw[thick,->-] (0t)--(0't) node[midway,above] {$ $};%
\draw[thick,->-] (0't)--(1't) node[midway,above] {$ $};%
\draw[thick,->-] (1t)--(1't) node[midway,below] {$ $}; %
\draw[thick,->-,dashed] (0)--(1') node[midway,below] {$E$};%
\draw[thick,->-,dashed] (0)--(0't) node[midway,below] {$J$}; %
\draw[thick,dashed,->-] (0)--(1t) node[midway,right] {$ $};%
\draw[thick,->-] (0')--(1't) node[midway,left] {$D$};%
\draw[thick,->-] (1)--(1't) node[midway,below] {$I$};%
\draw[thick,->-] (0t)--(1't) node[midway,below] {$ $};%
\draw[thick,dashed,->-] (0)--(1't) node[midway,below] {$F$};
\draw[thick,dashed,red] (a)--(b) node[red,midway,left] {$i$};
\draw[thick,dashed,red] (b)--(c) node[red,midway,left] {$g$};
\draw[thick,dashed,red] (a)--(d) node[red,midway,below] {$j$};
\draw[thick,dashed,red] (d)--(e) node[red,midway,above] {$h$};
\draw[thick,dashed,red] (e)--(f) node[red,midway,left] {$k$};
\draw[thick,dashed,red] (c)--(f) node[red,midway,below] {$l$};
\draw[thick,dashed,red] (a)--(cb) node[red,midway,left] {$d$};
\draw[thick,dashed,red] (d)--(fb) node[red,midway,left] {$e$};
\draw[thick,dashed,red] (c)--(au) node[red,midway,right] {$ $};
\draw[thick,dashed,red] (f)--(du) node[red,midway,left] {$ $};
\draw[thick,dashed,red] (d)--(br) node[red,midway,right] {$f$};%
\draw[thick,dashed,red] (e)--(cr) node[red,midway,above] {$m$};
\draw[thick,dashed,red] (b)--(dl) node[red,midway,above] {$a$};%
\draw[thick,dashed,red] (c)--(el) node[red,midway,above] {$n$};
\draw[thick,dashed,red] (a)--(eif) node[red,midway,right] {$c$};%
\draw[thick,dashed,red] (b)--(fif) node[red,midway,left] {$b$};%
\draw[thick,dashed,red] (abh)--(e) node[red,midway,right] {$ $};%
\draw[thick,dashed,red] (bbh)--(f) node[red,midway,left] {$ $};%
\end{tikzpicture}
\label{3d tube Tri}
\end{array}
\end{align}
\end{widetext}
where the up surface and the lower surface in $y$ direction of the cube is glued, and the front surface and the back surface in $z$ direction is also glued together, and hence some labels are left implicit. The membrane-net lattice is depicted with red lines for guidance. The two surfaces in $x$ direction generate a pair of excitations in membrane-net model. To avoid confusion, in the following, all the elements of 3d tube algebra we discussed are depicted in tetrahedron lattice and all the excitations we discussed are referred to the excitations in membrane-net lattice. 

Again, we can introduce the multiplication of the tubes
\begin{align}
\mathbf{A}^{\{A_1\}}_{\{a_1\}} \times \mathbf{A}^{\{A_2\}}_{\{a_2\}} = \sum_{\{A_3\}, \{a_3\}}
\Xi^{\{A_1\} \{A_2\},\{A_3\}}_{ \{a_1\} \{a_2\}, \{a_3\}} \mathbf{A}^{\{A_3\}}_{\{a_3\}},
\end{align}
where $\{A_i\}$ is an abbreviation of the label of 10 objects $A_i, B_i, \dots, J_i$, and $\{a_i\}$ is an abbreviation of the label of 14 1-morphisms $a_i, b_i, \dots, n_i$ shown in eqn.~\eqref{3d tube Tri}.  Similar to the $\delta_{CE}$ in the 2d case, there are also some constraints on the structure factor $\Xi$
\begin{align}
    &\quad\quad\quad\quad \Xi^{\{A_1\} \{A_2\},\{A_3\}}_{ \{a_1\} \{a_2\}, \{a_3\}} = \nonumber \\ 
    &\delta^{B_1}_{B_3}\delta^{C_1}_{C_3} \delta^{G_1}_{B_2}
    \delta^{H_1}_{C_2}
    \delta^{G_2}_{G_3}\delta^{H_2}_{H_3}\delta^{I_2}_{I_3} \delta^{I_1}_{J_2} \delta^{J_1}_{J_3}\delta^{a_1}_{a_3}\delta^{f_1}_{a_2}\delta^{f_2}_{f_3}
    \delta^{m_2}_{m_3} \delta^{m_1}_{n_2} \delta^{n_1}_{n_3}\nonumber \\ 
    &\quad\quad \times \Xi^{\{A_1\} \{A_2\}, A_3B_1C_1D_3E_3F_3G_2H_2I_2J_1}_{\{a_1\} \{a_2\}, a_1b_3c_3d_3e_3f_2g_3h_3i_3j_3k_3l_3m_2n_1}
\end{align}

The structure factor can be calculated by a TQFT on $\mathcal{M} = D^2\times S^1 \times S^1$ with a triangulation $\mathcal{K}$ whose boundary $\partial \mathcal{M} = S^1\times S^1\times S^1$ can be divided into three pieces of $S^1\times S^1\times I$
\begin{widetext}
\begin{align}
\label{Z_3tube}
Z(\mathcal{M})
=& \frac{1}{|D_\mathcal{B}|^\frac{1}{2}}
\sum_{\{A_3\}, \{a_3\}} \delta^{B_1}_{B_3}\delta^{C_1}_{C_3} \delta^{G_1}_{B_2}
    \delta^{H_1}_{C_2}
    \delta^{G_2}_{G_3}\delta^{H_2}_{H_3}\delta^{I_2}_{I_3} \delta^{I_1}_{J_2} \delta^{J_1}_{J_3}\delta^{a_1}_{a_3}\delta^{f_1}_{a_2}\delta^{f_2}_{f_3}
    \delta^{m_2}_{m_3} \delta^{m_1}_{n_2} \delta^{n_1}_{n_3} \sum_{\bigtriangleup_2\in {\mathcal{K}}^{in}}
\left(\prod_{\substack{\bigtriangleup_1\in \mathcal{K}^{in}\\\bigtriangleup_2\in \mathcal{K}^{in}}}f_{\bigtriangleup_1, \bigtriangleup_2} \prod_{\rm{\bigtriangleup_4\in \mathcal{K}}} \Gamma_{\bigtriangleup_4} \right) \nonumber\\
&\quad \times \Biggl|
\begin{array}{c}
\begin{tikzpicture}[scale=0.4]
\coordinate[label=below:$0$] (0) at (0,0,0) ; 
\coordinate[label=below:$2$] (1) at (4,0,0) ; 
\coordinate[label=left:$0'$] (0') at (-0.5,0.5,4) ; 1
\coordinate[label=below:$2'$] (1') at (3.5,0.5,4) ;  
\coordinate[label=above:$\widetilde{0}$] (0t) at (0,4,0); 
\coordinate[label=above:$\widetilde{2}$] (1t) at (4,4,0); 
\coordinate[label=left:$\widetilde{0'}$] (0't) at (-0.5,4.5,4); 
\coordinate[label=above:$\widetilde{2'}$] (1't) at (3.5,4.5,4); 
\draw[dashed] (0)--(1) node[midway,below] {$ $};%
\draw[dashed] (0)--(0') node[midway,above] {$ $};%
\draw[] (0')--(1') node[midway,below] {$ $};%
\draw[] (1)--(1') node[midway,right] {$ $};%
\draw[dashed] (0)--(0t) node[midway,left] {$ $};%
\draw[] (0')--(0't) node[midway,left] {$ $};%
\draw[] (1')--(1't) node[midway,right] {$ $};%
\draw[] (1)--(1t) node[midway,right] {$ $};%
\draw[] (0t)--(1t) node[midway,above] {$ $};%
\draw[] (0t)--(0't) node[midway,above] {$ $};%
\draw[] (0't)--(1't) node[midway,above] {$ $};%
\draw[] (1t)--(1't) node[midway,below] {$ $}; %
\draw[dashed] (0)--(1') node[midway,below] {$ $};%
\draw[dashed] (0)--(0't) node[midway,below] {$ $}; %
\draw[dashed] (0)--(1t) node[midway,right] {$ $};%
\draw[] (0')--(1't) node[midway,left] {$ $};%
\draw[] (1)--(1't) node[midway,below] {$ $};%
\draw[] (0t)--(1't) node[midway,below] {$ $};%
\draw[dashed] (0)--(1't) node[midway,below] {$ $};
\end{tikzpicture}
\end{array} 
\Biggr> 
\Biggl<
\begin{array}{c}
\begin{tikzpicture}[scale=0.4]
\coordinate[label=below:$0$] (0) at (0,0,0) ; 
\coordinate[label=below:$1$] (1) at (4,0,0) ; 
\coordinate[label=below:$2$] (2) at (8,0,0) ; 
\coordinate[label=left:$0'$] (0') at (-0.5,0.5,4) ; 1
\coordinate[label=below:$1'$] (1') at (3.5,0.5,4) ;  
\coordinate[label=below:$2'$] (2') at (7.5,0.5,4) ;
\coordinate[label=above:$\widetilde{0}$] (0t) at (0,4,0); 
\coordinate[label=above:$\widetilde{1}$] (1t) at (4,4,0); 
\coordinate[label=above:$\widetilde{2}$] (2t) at (8,4,0);
\coordinate[label=left:$\widetilde{0'}$] (0't) at (-0.5,4.5,4); 
\coordinate[label=above:$\widetilde{1'}$] (1't) at (3.5,4.5,4); 
\coordinate[label=above:$\widetilde{2'}$] (2't) at (7.5,4.5,4); 
\draw[dashed] (0)--(1) node[midway,below] {$ $};%
\draw[dashed] (0)--(0') node[midway,above] {$ $};%
\draw[] (0')--(1') node[midway,below] {$ $};%
\draw[dashed] (0)--(0t) node[midway,left] {$ $};%
\draw[] (0')--(0't) node[midway,left] {$ $};%
\draw[] (1')--(1't) node[midway,right] {$ $};%
\draw[] (0t)--(1t) node[midway,above] {$ $};%
\draw[] (0t)--(0't) node[midway,above] {$ $};%
\draw[] (0't)--(1't) node[midway,above] {$ $};%
\draw[dashed] (0)--(1') node[midway,below] {$ $};%
\draw[dashed] (0)--(0't) node[midway,below] {$ $}; %
\draw[dashed] (0)--(1t) node[midway,right] {$ $};%
\draw[] (0')--(1't) node[midway,left] {$ $};%
\draw[] (0t)--(1't) node[midway,below] {$ $};%
\draw[dashed] (0)--(1't) node[midway,below] {$ $};
\draw[dashed] (1)--(2) node[midway,below] {$ $};%
\draw[dashed] (1)--(1') node[midway,above] {$ $};%
\draw[] (1')--(2') node[midway,below] {$ $};%
\draw[] (2)--(2') node[midway,right] {$ $};%
\draw[dashed] (1)--(1t) node[midway,left] {$ $};%
\draw[] (1')--(1't) node[midway,left] {$ $};%
\draw[] (2')--(2't) node[midway,right] {$ $};%
\draw[] (2)--(2t) node[midway,right] {$ $};%
\draw[] (1t)--(2t) node[midway,above] {$ $};%
\draw[] (1t)--(1't) node[midway,above] {$ $};%
\draw[] (1't)--(2't) node[midway,above] {$ $};%
\draw[] (2t)--(2't) node[midway,below] {$ $}; %
\draw[dashed] (1)--(2') node[midway,below] {$ $};%
\draw[dashed] (1)--(1't) node[midway,below] {$ $}; %
\draw[dashed] (1)--(2t) node[midway,right] {$ $};%
\draw[] (1')--(2't) node[midway,left] {$ $};%
\draw[] (2)--(2't) node[midway,below] {$ $};%
\draw[] (1t)--(2't) node[midway,below] {$ $};%
\draw[dashed] (1)--(2't) node[midway,below] {$ $};
\end{tikzpicture}
\end{array} 
\Biggl|,
\end{align}
\end{widetext}
where objects and 1-morphisms are left implicit. $\bigtriangleup_1\in \mathcal{K}^{in}$ and $\bigtriangleup_2\in \mathcal{K}^{in}$ refer to the object and 1-morphisms inside the manifold $\mathcal{M}$.  Please notice that since $Z(\mathcal{M})$ is partition function of a TQFT, it is independent on the choice of $\mathcal{K}$.  Thus with a special choice of $\mathcal{K}$, $\prod_{\rm{\bigtriangleup_4\in \mathcal{K}}} \Gamma_{\bigtriangleup_4}$ can be written as the product of twelve 10j symbols 
\begin{widetext}
\begin{align}
\label{3d tube map}
\prod_{\rm{\bigtriangleup_4\in \mathcal{K}}} \Gamma_{\bigtriangleup_4}=&\Gamma_{(00'1'2'\widetilde{2'})}  \Gamma_{(00'1'\widetilde{1'}\widetilde{2'})}^{-1} \Gamma_{(00'\widetilde{0'}\widetilde{1'}\widetilde{2'})} \Gamma_{(0\widetilde{0}\widetilde{0'}\widetilde{1'}\widetilde{2'})}^{-1} \Gamma_{(011'2'\widetilde{2'})} ^{-1} \Gamma_{(011'\widetilde{1'}\widetilde{2'})} \Gamma_{(01\widetilde{1}\widetilde{1'}\widetilde{2'})}^{-1} \nonumber\\
&\times \Gamma_{(0\widetilde{0}\widetilde{1}\widetilde{1'}\widetilde{2'})}\Gamma_{(0122'\widetilde{2'})} \Gamma_{(012\widetilde{2}\widetilde{2'})}^{-1} \Gamma_{(01\widetilde{1}\widetilde{2}\widetilde{2'})} \Gamma_{(0\widetilde{0}\widetilde{1}\widetilde{2}\widetilde{2'})}^{-1}, 
\end{align}
\end{widetext}
where the 10j symbols are defined as
\begin{align}
\label{def10j}
\Gamma_{(01234)}:& V^{+}(1234)\otimes V^{+}(0134) \otimes V^{+}(0123) \nonumber \\
& \rightarrow V^{+}(0234) \otimes V^{+}(0124),\nonumber\\
\end{align}
$\Gamma^{-1}$ is the inversion of $\Gamma$, which is also equal to Hermitian conjugate of $\Gamma$ because we consider only the unitary fusion 2-category. 

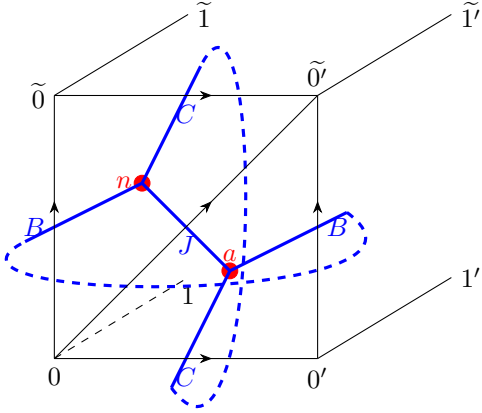
\begin{figure}[htbp]
\centering
\begin{tikzpicture}[scale=0.7]
\coordinate[label=below:$0$] (0) at (0,0,0);
\coordinate[label=below:$0'$] (0') at (5,0,0);
\coordinate[label=left:$\widetilde{0}$] (0t) at (0,5,0);
\coordinate[label=above:$\widetilde{0'}$] (0't) at (5,5,0);
\coordinate[label=below:$1$] (1) at (1,0,-4);
\coordinate[label=right:$1'$] (1') at (6,0,-4);
\coordinate[label=right:$\widetilde{1}$] (1t) at (1,5,-4);
\coordinate[label=right:$\widetilde{1'}$] (1't) at (6,5,-4);
\filldraw[red] (10/3,5/3) circle (0.15) node[above=0.3] {$a$};
\filldraw[red] (5/3,10/3) circle (0.15) node[above=0.2,left=0.3] {$n$};
\draw[blue,very thick] (5/3+10/9,5+5/9)--(5/3,10/3) node[midway,below] {$ $};
\draw[blue,very thick] (5/3,10/3)--(10/3,5/3) node[midway,below] {$ $};
\draw[blue,very thick] (10/3,5/3)--(10/3-10/9,-5/9) node[midway,below] {$ $};
\draw [dashed,blue,very thick] (10/3-10/9,-5/9).. controls (4,-3) and (4,8) ..(5/3+10/9,5+5/9);
\draw[blue,very thick] (-5/9,5/3+5/9)--(5/3,10/3) node[midway,below] {$ $};
\draw[blue,very thick] (10/3,5/3)--(5+5/9,10/3-5/9) node[midway,below] {$ $};
\draw [dashed,blue,very thick] (5+5/9,10/3-5/9).. controls (8,1) and (-3,1) ..(-5/9,5/3+5/9);
\draw[->-] (0)--(0') node[midway,below,blue] {$C$};
\draw[dashed] (0)--(1);
\draw[] (0')--(1');
\draw[] (0t)--(1t);
\draw[] (0't)--(1't);
\draw[->-] (0)--(0t) node[midway,left,blue] {$B$};
\draw[->-] (0)--(0't) node[midway,below,blue] {$J$};
\draw[->-] (0')--(0't) node[midway,right,blue] {$B$};
\draw[->-] (0t)--(0't) node[midway,below,blue] {$C$};
\end{tikzpicture}
\caption{Excitation of 3d topological order created on $00'\widetilde{0}\widetilde{0'}$ plane where string-like excitations are drawn in blue and particle-like excitations are drawn in red.}
\label{3d excitation}
\end{figure}

The structure factor in the multiplication of 3d tube algebra is much more complicated than its 2d analogue. The reason is that the excitation in a 3d topological order is not merely a simple string or a simple particle but a combination of 3 string-like excitations and 2 particle-like excitations (the two particle-like excitations are actually the domain walls between the 3 strings).  To see this, we move to the dual lattice and check the excitations created by the two ends of the 3d tube~\eqref{3d tube Tri} along $x$ direction.  For simplicity, we consider only the $00'\widetilde{0}\widetilde{0'}$ surface, where each $1$-simplex is dual to a membrane and each $2$-simplex is dual to a string in membrane-net model.  If we cut these membranes and strings alone $00'\widetilde{0}\widetilde{0'}$ plane, five strings (from the five membranes, or five 1-simplexes in the tetrahedron lattice) and two particles (from the two strings, or two triangles in the tetrahedron lattice) will remain on the plane.  Since $0\widetilde{0}$ and $0'\widetilde{0'}$ are assigned with same object(membrane) and $00'$ and $\widetilde{0}\widetilde{0'}$ are assigned with same object(membrane), there is only three different strings and two different particles on the plane as depicted in Fig.~\ref{3d excitation}.
This is the structure of excitation in 3d topological order, and we will refer to this kind of excitation in 3d topological order as crossed strings in following discussion.

Similarly, we can illustrate the relation between tube element $\mathbf{A}^{\{A\}}_{\{a\}}$ and the membrane-net configuration. For membrane-net Hilbert space $\mathcal{H}(L)^{m}_{n}$ living on the manifold $S^1\times S^1\times I$, an arbitrary membrane-net configuration is a vector $\mathbf{v}(L)^m_n$, where $m/n$ denotes the number of crossed strings of outer/inner boundary, respectively, $L$ denotes the number of torus. Analogy to 2d case, due to the fixed-point property of membrane-net model, the membrane-net configuration $\mathbf{v}(L)^m_n$ can be renormalized to a 3d tube $\mathbf{v}(1)^1_1$ (or $\mathbf{A}^{\{A\}}_{\{a\}}$). And the process that gluing any two tubes and then renormalizing back to another tube defines the multiplication of tube algebra. With the structure factor, ICIs of 3d tube algebra can be calculated in the same way as its 2d analogue such that  $P_i P_j=\delta_{ij}P_i$, $P_i \mathbf{A}^{\{A\}}_{\{a\}}=\mathbf{A}^{\{A\}}_{\{a\}} P_i$. The ICIs can be found by decomposing the identity of the tube algebra, which is guaranteed by the semi-simpleness of the input fusion 2-category. 

Recall that it has been pointed out that the ground state degeneracy for a two-dimensional topological order on torus ($S^1 \times S^1$) is equal to the number of ICIs \cite{Lan2014,Bultinck2017}.  In the 2d case, different ground states (minimally entangled states) on torus are labeled by different non-contractible strings, which can be regarded as the ICIs connected in $x$-direction \cite{Bultinck2017}.  It is not difficult to verify that different ground states on $T^3$ are also characterized by different non-contractible membranes.  We will show that these membrane structures correspond to connect the 3D ICIs in $x$-direction.

To do this, we only need to verify that a state on $T^3$ characterized by an ICI has lowest energy. Since the membrane-net structure of an ICI intrinsically satisfies flatness condition, eigenvalue of $Q_{\v V}$ and $Q_{\v E}$ must be $1$. And it is easy to check that eigenvalue of $B_{\v C}$ is also $1$ through some straightforward calculation similar to the process in Ref.~\onlinecite{Bei2017} for 2d case.
Furthermore, each ICI has a one-to-one correspondence with irreducible representation of tube algebra mathematically. This fact guarantees different ground states labeled by different ICIs are orthogonal. As a result, the number of ICIs naturally give the ground state degeneracy of membrane-net model on $T^3$.

Alternatively, one can reach same conclusion by thinking about the simple modules over tube algebra $\mathcal{A} P_i$ which are generated by acting all tube elements $\mathbf{A}^{\{A\}}_{\{a\}}$ on each $P_i$. Because each simple module has a one-to-one correspondence with a topological sector, all excited states in same topological sector can be created by acting some tube elements $\mathbf{A}^{\{A\}}_{\{a\}}$ on left side of tube $P_i$. Furthermore, since $P_i$ is in the center of tube algebra $\mathcal{A}$, we have $\mathcal{A} P_i=P_i \mathcal{A}$. Physically, this property guarantees that acting same tube elements from either left or right side of tube $P_i$ yields to same state. 
This result is quite intuitive because based on braiding non-degeneracy, the state which is transparent for all excited states must be the vacuum state.
Consequently, $P_i$ must be the ground state in the corresponding topological sector.

Besides ground state degeneracy, there are another important property can be used to characterize a 3d TO--mutual statistic between excitations. However, unlike the 2d case, there is no general framework to calculate mutual statistics between excitations in 3d, and hence we can not read these information from ICIs directly. We will proposal a general framework to calculate the mutual statistics of excitations in 3d topological orders based on tube algebra in next section.

\subsection{Tube algebra constructed with $2\mathcal{V}ec^\omega_G$}
\label{sec:2vect}

The algebraic structure for a general 3d tube algebra is very complicated.  However, if the input fusion 2-category is $2\mathcal{V}ec^\omega_G$, the $G$-graded 2-vector spaces where $G$ is a group, the algebraic structure of tube algebra can be vastly simplified.  Physically, we will show that all the gauged 3d bosonic symmetry protected states (SPT) states can be realized with $2\mathcal{V}ec^\omega_G$ as the input fusion 2-category.  Thus $2\mathcal{V}ec^\omega_G$ are very good examples to examine our theory.

The 2-category $2\mathcal{V}ec$ is the category of 2-vector spaces.  Its object is a 2-vector space, which is just a formal symbol $V[n]$ with non-negative integers $n$.  $V[1]$ is the simple object of $2\mathcal{V}ec$.  A 1-morphism $M$ in $\hom(V[n], V[m])$ is a 2-matrix, which is an $n\times m$ array and whose element $M_{ij}$ is a vector space.  Then a 2-morphism between $M, N \in \hom(V[n], V[m])$ is an $n\times m$ array of linear maps or matrices $L_{ij} : M_{ij} \rightarrow N_{ij}$. (see Ref.~\onlinecite{johnson2020,Bullivant2019} for more details)

For $2\mathcal{V}ec^{\omega}_G$, its objects are $G$ graded 2-vector spaces, where the simple objects can be labeled by group elements.  
There is no non-trivial $1$-morphism between simple objects labeled by different group elements.
The tensor product of objects are induced by the multiplication of the group $G$.
The 10j symbol of this category is given by group 4-cocycle $\omega$ in $\mathcal{H}^4(G,U(1))$.

In the case that $2\mathcal{V}ec^\omega_G$ is the input category of our membrane-net model, one may simply think that the objects assigned to $1$-simplex are group elements of $G$.  Since there is no non-trivial $1$-morphism between different simple objects, a tube can be fully characterized by objects on its $1$-simplexes.  Furthermore, with the flatness condition, which respects tensor product of objects such that on a 2-simplex $v_0 v_1 v_2$, $g_{v_0 v_1}g_{v_1 v_2}=g_{v_0 v_2}$, a 3d tube can be fully determined by only three objects.  Therefore, we will denote such a tube as $\mathbf{A}^{ABC}$ where $A$ is the object (group element) assigned on $01$, $B$ is the object (group element) assigned on $0\widetilde{0}$ and $C$ is the object (group element) assigned on $00'$ as drawn in eqn.~\eqref{3d tube Tri}.  Please notice that the normalization factor $\prod_{\substack{\bigtriangleup_1\in \mathcal{K}\\\bigtriangleup_2\in \mathcal{K}}}f_{\bigtriangleup_1, \bigtriangleup_2}$ in eqn.~\eqref{Z_3tube} is just $1$. For an arbitrary group (Abelian or Non-Abelian), the multiplication of 3d tube algebra constructed with $2\mathcal{V}ec^\omega_G$ is given by:
\begin{widetext}
\begin{eqnarray}
\label{eq:tubemul2vect}
    \Ab^{A_1B_1C_1} \times \Ab^{A_2B_2C_2}&=&\sum_{A_3}\delta^{A_3}_{A_1A_2}\delta^{B_2}_{A^{-1}_1B_1A_{1}}\delta^{C_2}_{A^{-1}_1C_1A_{1}}\prod_{\rm{\bigtriangleup_4\in \mathcal{K}}}\Gamma^{A_1,B_1,C_1,A_2,A_3}\Ab^{A_3,B_1,C_1}\\ \nonumber
\prod_{\rm{\bigtriangleup_4\in\mathcal{K}}}\Gamma^{A_1,B_1,C_1,A_2,A_3}&=&\omega(C_1,A_1,A_2,A^{-1}_3B_1A_3)\omega^{-1}(C_1,A_1,A^{-1}_1B_1A_1,A_2)\omega(C_1,B_1,A_1,A_2)\omega^{-1}(B_1,C_1,A_1,A_2)\\\nonumber
&&\omega^{-1}(A_1,A^{-1}_1C_1A_1,A_2,A^{-1}_3B_1A_3)\omega(A_1,A^{-1}_1C_1A_1,A^{-1}_1B_1A_1,A_2)\omega^{-1}(A_1,A^{-1}_1B_1A_1,A^{-1}_1C_1A_1,A_2)\\\nonumber
&&\omega(B_1,A_1,A^{-1}_1C_1A_1,A_2)\omega(A_1,A_2,A^{-1}_1C_1A_1,A^{-1}_1B_1A_1)\omega^{-1}(A_1,A_2,A^{-1}_1B_1A_1,A^{-1}_1C_1A_1)\\\nonumber
&&\omega(A_1,A^{-1}_1B_1A_1,A_2,A^{-1}_1C_1A_1)\omega^{-1}(B_1,A_1,A_2,A^{-1}_1C_1A_1).
\end{eqnarray},
\end{widetext}
where all the $A_i$, $B_i$, and $C_i$ are group elements of $G$.

Since it corresponds to assign group element of $G$ to each $1$-simplex, the TQFT we achieved is nothing but the 3+1D Dijkgraaf-Witten theory.  Thus the 3d topological orders constructed with $2Vec^\omega_G$ are just gauged SPT phase. For example, the simplest example of 3d topological order--toric code, is a gauged 3d bosonic trivial $\Zb_2$ SPT phase.  It can be achieved with our membrane-net model by using the spherical fusion 2-category $2Vec_{\mathcal{Z}_2}$ of $\Zb_2$-graded 2-vector spaces as the input. We will study the statistics of topological excitation in 3d toric code in next section.

\section{Statistics of topological excitations in 3d topological order}

One of the most important phenomena of a topological order is the self and mutual statistics of topological excitations in the system. In contrast to the 2d case where only particle-like excitations are available, the 3d topological order also have string-like excitations, and hence support a much richer statistical pattern.  In the literature, various statistics, such as particle-particle braiding, particle-loop braiding, loop-loop braiding, three-loop braiding and so on have been studied with different approaches\cite{Wang2014,Jiang2014,JuvenWang2015,Ye2021}. However, a general framework for the statistics between topological excitations in an exact solvable 3d lattice model is still lacking.

As we discussed in previous section, a topological excitation $i$ can always emerged as the boundary loops of the corresponding ICIs  of 3d tube algebra (a particle-like excitation corresponds to the trivial boundary loop). Therefore, the ICIs of tube algebra in 3d membrane-net model can naturally provide a unified view of the statistics of all excitations in 3d topological order.

Here, we will present a primitive framework on how to calculate some fundamental datas of statistics of the topological excitations by manipulating tubes. The basic idea is the following:
\begin{enumerate}
    \item For a 3d topological order on a closed manifold $M$, its ground state subspace forms a projective representation of the mapping class group $\mathrm{MCG}$($M$) of $M$.  And all the representations together encode all the information on the statistics of topological excitations of the topological order \cite{kong2014braided,Moradi2015}.  In the following, we consider the case that $M$ is the $3$-dimensional Torus $\Tb^3$, which may contain the most important part of the information of the statistics.
    \item On the 2-dimensional torus $\Tb^2$, it is well known that the ICIs $P_i$ allow one to construct the minimally entangled states (MES), namely the ground states with minimal entanglement entropy\cite{ZhangYi2012,Bultinck2017}. These states can be taken as a set of basis in ground state subspace on the torus $\Tb^2$. Therefore, the ground state subspace can be constructed by ICIs. For the higher dimensional torus $\Tb^3$, we argue that it still holds, hence the ground state subspace on $\Tb^3$ can be constructed by ICIs of the 3d tube algebra.
    \item  The mapping class group of $\Tb^3$, $\mathrm{MCG}(\Tb^3)=SL(3,\Zb)$, has only two kinds of nontrivial generators $S^{zxy}_{3d}$ and $T^{zx}_{3d}$\cite{coxeter2013}.  Each of them can transform one membrane-net configuration to another membrane-net configuration.  Thus the representation matrices of these generators can be calculated with the partition function by taking the two membrane-net configurations as boundary configurations.
\end{enumerate}

\begin{figure}
\centering
\includegraphics[width=9cm]{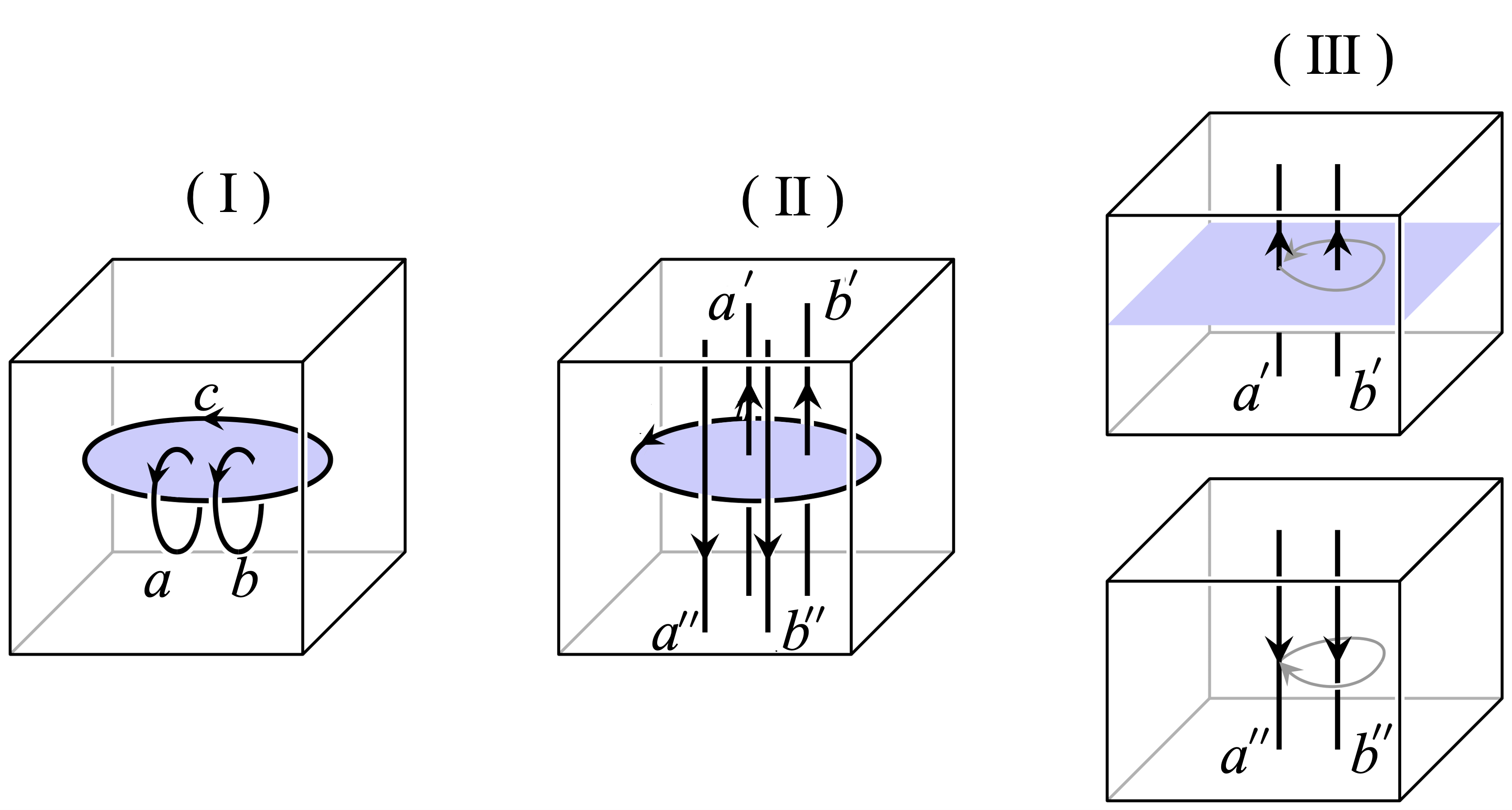}
\caption{Process of three loop braiding. Cited from Ref.~\onlinecite{Wang2014}, To avoid confusion of symbols in different papers, we have replaced some symbols }
\label{3lbp}
\end{figure}

In principle, the statistical information originated from the particle-particle braiding, particle-loop braiding, and loop-loop braiding can be easily read out from the S/T matrices, while the one of the three loop braiding is a little bit tricky.
In previous literature \cite{Wang2014}, the phase of three loop braiding is defined as
\begin{align}
\label{3lb}
\theta^c_{ab}=\theta^c_{a'b'}-\theta^0_{a''b''},
\end{align}
which is achieved from a dimension reduction scenario shown in Fig.~\ref{3lbp}.  The first term on right hand side is statistical phases associated with the braiding of two particle-like excitations $a'$ and $b'$ around each other in the $z$-$x$ plane with a gauge flux $c$ through $y$-hole. The second term comes from braiding two particle-like excitations $a''$ and $b''$ around each other in the $z$-$x$ plane without gauge flux $c$.  Here the additional relative sign comes from the fact that the two pairs are braided in opposite directions.

In our framework, the dimension reduction is just fixing the object $B$ on $00'$ bond (alone $y$ direction) and treating the section in $x$-$z$ plane of the 3d tube effectively as a 2d tube. 
Thus, the three-loop braiding is naturally described by the $S^{zx}$ matrix of the 2d modular transformation in the 2d $z$-$x$ plane.  In general, the matrix element of $S^{zx}$ with {ICIs} basis is not necessarily a phase. However, for Abelian group $G$, we can get a phase from the $S^{zx}$ and the phase of the three-loop braiding is given by
\begin{align}\label{3loop-S}
\theta^c_{ab}=arg(S^{zx}_{\{0, c, a\},\{0, c, b\}})-arg(S^{zx}_{\{0, 0, a\},\{0, 0, b\}}).
\end{align}
where we choose the {ICIs} basis for $S^{zx}$.

In principle, we may provide more information on statistics than the ones discussed above.  In the following,  we will show that our approach will give same results as the ones in previous works\cite{Wang2014,Jiang2014,JuvenWang2015}.  A more complete study will be presented in a different paper.

\subsection{3d toric code}
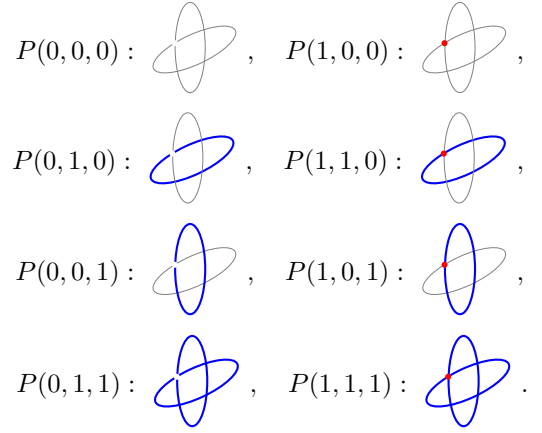
\begin{figure}[htbp]
\begin{eqnarray*}
%
%
P(0,0,0)
:
\begin{array}{c} 
\begin{tikzpicture}
\coordinate[] (c) at (-0.0625,0.5,0.5) ;

\draw[help lines, rotate=25] (0.1,0.2) ellipse (0.6 and 0.2);

\draw[help lines, ,rotate=90] (0.25,0.05) ellipse (0.6 and 0.2);

\fill[white] (c) circle (0.04) node[right,black] {} ;
\draw[] (-0.0625,0.9,0.5) node[right,black] {} ;
\draw[] (-0.0625,0.1,0.5) node[right,black] {} ;

\end{tikzpicture}
\end{array},~~~
%
%
P(1,0,0)
:
\begin{array}{c} 
\begin{tikzpicture}

\draw[help lines, rotate=25] (0.1,0.2) ellipse (0.6 and 0.2);

\draw[help lines, ,rotate=90] (0.25,0.05) ellipse (0.6 and 0.2);

\fill[red] (c) circle (0.04) node[right,black] {} ;
\draw[] (-0.0625,0.9,0.5) node[right,black] {} ;
\draw[] (-0.0625,0.1,0.5) node[right,black] {} ;

\end{tikzpicture}
\end{array},
\\
%
%
P(0,1,0)
:
\begin{array}{c} 
\begin{tikzpicture}

\draw[thick, blue, rotate=25] (0.1,0.2) ellipse (0.6 and 0.2);

\draw[help lines, ,rotate=90] (0.25,0.05) ellipse (0.6 and 0.2);

\fill[white] (c) circle (0.04) node[right,black] {} ;
\draw[] (-0.0625,0.9,0.5) node[right,black] {} ;
\draw[] (-0.0625,0.1,0.5) node[right,black] {} ;

\end{tikzpicture}
\end{array},~~~
%
%
P(1,1,0)
:
\begin{array}{c} 
\begin{tikzpicture}

\draw[thick,blue, rotate=25] (0.1,0.2) ellipse (0.6 and 0.2);

\draw[help lines, rotate=90] (0.25,0.05) ellipse (0.6 and 0.2);

\fill[red] (c) circle (0.04) node[right,black] {} ;
\draw[] (-0.0625,0.9,0.5) node[right,black] {} ;
\draw[] (-0.0625,0.1,0.5) node[right,black] {} ;

\end{tikzpicture}
\end{array},
\\
%
%
P(0,0,1)
:
\begin{array}{c} 
\begin{tikzpicture}

\draw[help lines, rotate=25] (0.1,0.2) ellipse (0.6 and 0.2);

\draw[thick,blue,rotate=90] (0.25,0.05) ellipse (0.6 and 0.2);

\fill[white] (c) circle (0.04) node[right,black] {} ;
\draw[] (-0.0625,0.9,0.5) node[right,black] {} ;
\draw[] (-0.0625,0.1,0.5) node[right,black] {} ;

\end{tikzpicture}
\end{array},~~~
%
%
P(1,0,1)
:
\begin{array}{c} 
\begin{tikzpicture}

\draw[help lines, rotate=25] (0.1,0.2) ellipse (0.6 and 0.2);

\draw[thick,blue,rotate=90] (0.25,0.05) ellipse (0.6 and 0.2);

\fill[red] (c) circle (0.04) node[right,black] {} ;
\draw[] (-0.0625,0.9,0.5) node[right,black] {} ;
\draw[] (-0.0625,0.1,0.5) node[right,black] {} ;

\end{tikzpicture}
\end{array},
\\
%
%
P(0,1,1)
:
\begin{array}{c} 
\begin{tikzpicture}

\draw[thick,blue,rotate=25] (0.1,0.2) ellipse (0.6 and 0.2);

\draw[thick,blue,rotate=90] (0.25,0.05) ellipse (0.6 and 0.2);

\fill[white] (c) circle (0.04) node[right,black] {} ;
\draw[] (-0.0625,0.9,0.5) node[right,black] {} ;
\draw[] (-0.0625,0.1,0.5) node[right,black] {} ;

\end{tikzpicture}
\end{array},~~~
%
%
P(1,1,1)
:
\begin{array}{c} 
\begin{tikzpicture}

\draw[thick,blue,rotate=25] (0.1,0.2) ellipse (0.6 and 0.2);

\draw[thick,blue,rotate=90] (0.25,0.05) ellipse (0.6 and 0.2);

\fill[red] (c) circle (0.04) node[right,black] {} ;
\draw[] (-0.0625,0.9,0.5) node[right,black] {} ;
\draw[] (-0.0625,0.1,0.5) node[right,black] {} ;

\end{tikzpicture}
\end{array}.
\end{eqnarray*}
\caption{Eight {ICIs} $P_i(n_e,\alpha,\beta)$ describes trivial pure particle(white point), non-trivial pure particle $e$(red point), $\alpha$-loop, $\alpha$-loop attached by $e$, $\beta$-loop, $\beta$-loop attached by $e$, $\alpha,\beta$-loop and $\alpha,\beta$-loop attached by $e$, respectively. Non-trivial string is drawn in blue and trivial string is drawn in gray.}
\label{fig: boundary excitation}
\end{figure}

The simplest example of 3d topological orders is the 3d toric code, which can be viewed as a gauged 3d bosonic trivial $\Zb_2$ symmetry protected topological(SPT) phase. Since the group 4-cohomology $\mathcal{H}^4(\Zb_2,U(1))$ is trivial, there is no nontrivial bosonic $\Zb_2$ SPT phase, and hence 3d toric code is the only gauged bosonic $\Zb_2$ SPT in three spatial dimension.  We will show below that as a gauged trivial $\Zb_2$ SPT, the 3d toric code can be achieved with our membrane-net model by using the spherical fusion 2-category $2 \mathcal{V}ec_{\Zb_2}$ of $\Zb_2$-graded 2-vector spaces as the input.

\begin{figure}
$$
\begin{array}{ccc}
\begin{array}{c} 
\begin{tikzpicture}
\coordinate[label=below:$0$] (0) at (0,0,0) ; 
\coordinate[label=below:$0'$]  (0') at (1,0,0) ; 
\coordinate[label=below:$0''$]  (0'') at (2,0,0) ; 
\coordinate[label=left:$1$] (1) at (0.125,-0.125,-1) ;    
\coordinate[label=below:$1'$] (1') at (1.125,-0.125,-1) ;  
\coordinate[label=below:$1''$] (1'') at (2.125,-0.125,-1) ; 
\coordinate[label=above:$\widetilde{0}$] (0t) at (0,1,0) ;  
\coordinate[label=above: $\widetilde{1}$]  (1t) at (0.125,0.875,-1) ; 
\coordinate[label=above:$\widetilde{0'}$]  (0't) at (1,1,0) ; 
\coordinate[label=above:$\widetilde{0''}$]  (0''t) at (2,1,0) ; 
\coordinate[label=above:$\widetilde{1'}$]  (1't) at (1.125,0.875,-1) ;
\coordinate[label=above:$\widetilde{1''}$]  (1''t) at (2.125,0.875,-1) ;

\coordinate[] (c) at (-0.0625,0.5,0.5) ;
\coordinate[label=above:$ $] (00) at (0,0,0); 
\coordinate[label=above:$ $] (01) at (3,0,0); 
\coordinate[label=above:$ $] (02) at (0,2.5,0); 
\coordinate[label=above:$ $] (03) at (0.75,-0.75,-6);

\draw[line width=0.5mm, blue] (0)--(0') node[midway,below] {$ $};%
\draw[dashed,line width=0.5mm, blue] (1)--(1') node[midway,below] {$ $};%
\draw[] (0t)--(0't) node[midway,above] {$ $};%
\draw[] (1t)--(1't) node[midway,above] {$ $};%

\draw[line width=0.5mm,orange] (0)--(0t) node[midway,below] {$ $};%
\draw[] (0')--(0't) node[midway,left] {$ $};%
\draw[dashed] (1)--(1t) node[midway,right] {$ $};%
\draw[] (1')--(1't) node[midway,right] {$ $};%

\draw[line width=0.5mm,green,dashed] (0)--(1) node[midway,below] {$ $};%
\draw[line width=0.5mm, green] (0')--(1') node[midway,above] {$ $};%
\draw[] (0t)--(1t) node[midway,right] {$ $};%
\draw[] (0't)--(1't) node[midway,below] {$ $};%

\draw[help lines] (0')--(0'') node[midway,below] {$ $};%
\draw[dashed, help lines] (1')--(1'') node[midway,below] {$ $};%
\draw[help lines] (0't)--(0''t) node[midway,above] {$ $};%
\draw[help lines] (1't)--(1''t) node[midway,above] {$ $};%

\draw[help lines] (0')--(0't) node[midway,below] {$ $};%
\draw[help lines] (0'')--(0''t) node[midway,left] {$ $};%
\draw[dashed, help lines] (1')--(1't) node[midway,right] {$ $};%
\draw[help lines] (1'')--(1''t) node[midway,right] {$ $};%

\draw[help lines] (0'')--(1'') node[midway,above] {$ $};%
\draw[help lines] (0''t)--(1''t) node[midway,below] {$ $};%

\draw[->,help lines, blue!50] (00)--(01) node[at end,above] {$z$};%
\draw[help lines,->,blue!50] (00)--(02) node[at end,right] {$y$};%
\draw[help lines,->,blue!50] (00)--(03) node[at end,above] {$x$};%

\end{tikzpicture}
\end{array}&
\xrightarrow{T^{zx}_{3d}}&
\begin{array}{c} 
\begin{tikzpicture}
\coordinate[label=below:$0$] (0) at (0,0,0) ; 
\coordinate[label=below:$0'$]  (0') at (1,0,0) ; 
\coordinate[label=below:$0''$]  (0'') at (2,0,0) ; 
\coordinate[label=left:$1$] (1) at (0.125,-0.125,-1) ;    
\coordinate[label=below:$1'$] (1') at (1.125,-0.125,-1) ;  
\coordinate[label=below:$1''$] (1'') at (2.125,-0.125,-1) ; 
\coordinate[label=above:$\widetilde{0}$] (0t) at (0,1,0) ;  
\coordinate[label=above: $\widetilde{1}$]  (1t) at (0.125,0.875,-1) ; 
\coordinate[label=above:$\widetilde{0'}$]  (0't) at (1,1,0) ; 
\coordinate[label=above:$\widetilde{0''}$]  (0''t) at (2,1,0) ; 
\coordinate[label=above:$\widetilde{1'}$]  (1't) at (1.125,0.875,-1) ;
\coordinate[label=above:$\widetilde{1''}$]  (1''t) at (2.125,0.875,-1) ;

\draw[line width=0.5mm,blue] (0)--(0') node[midway,below] {$ $};%
\draw[dashed,help lines] (1)--(1') node[midway,below] {$ $};%
\draw[] (0t)--(0't) node[midway,above] {$ $};%
\draw[help lines] (1t)--(1't) node[midway,above] {$ $};%

\draw[line width=0.5mm,orange] (0)--(0t) node[midway,below] {$ $};%
\draw[] (0')--(0't) node[midway,left] {$ $};%
\draw[dashed, help lines] (1)--(1t) node[midway,right] {$ $};%
\draw[dashed] (1')--(1't) node[midway,right] {$ $};%

\draw[dashed, help lines] (0)--(1) node[midway,below] {$ $};%
\draw[dashed] (0')--(1') node[midway,above] {$ $};%
\draw[help lines] (0t)--(1t) node[midway,right] {$ $};%
\draw[] (0't)--(1't) node[midway,below] {$ $};%

\draw[dashed,line width=0.5mm, cyan] (0)--(1') node[midway,below] {$ $};%
\draw[] (0t)--(1't) node[midway,below] {$ $};%

\draw[help lines] (0')--(0'') node[midway,below] {$ $};%
\draw[line width=0.5mm,blue,dashed] (1')--(1'') node[midway,below] {$ $};%
\draw[help lines] (0't)--(0''t) node[midway,above] {$ $};%
\draw[] (1't)--(1''t) node[midway,above] {$ $};%

\draw[help lines] (0')--(0't) node[midway,below] {$ $};%
\draw[help lines] (0'')--(0''t) node[midway,left] {$ $};%
\draw[dashed] (1')--(1't) node[midway,right] {$ $};%
\draw[] (1'')--(1''t) node[midway,right] {$ $};%

\draw[help lines] (0'')--(1'') node[midway,above] {$ $};%
\draw[help lines] (0''t)--(1''t) node[midway,below] {$ $};%

\draw[line width=0.5mm,cyan] (0')--(1'') node[midway,below] {$ $};%
\draw[] (0't)--(1''t) node[midway,below] {$ $};%

\draw[->,help lines, blue!50] (00)--(01) node[at end,above] {$z$};%
\draw[help lines,->,blue!50] (00)--(02) node[at end,right] {$y$};%
\draw[help lines,->,blue!50] (00)--(03) node[at end,above] {$x$};%

\end{tikzpicture}
\end{array}\\
\begin{array}{c} 
\begin{tikzpicture}
\coordinate[label=below:$0$] (0) at (0,0,0) ; 
\coordinate[label=below:$0'$]  (0') at (1,0,0) ; 
\coordinate[label=above:$1$] (1) at (0,1,0) ;  
\coordinate[label=above:$1'$]  (1') at (1,1,0) ; 
\coordinate[label=below:$0''$] (0'') at (2,0,0) ;  
\coordinate[label=above:$1''$]  (1'') at (2,1,0) ;

\coordinate[label=above:$ $] (00) at (0,0,0); 
\coordinate[label=above:$ $] (01) at (2.5,0,0); 
\coordinate[label=above:$ $] (02) at (0,2.5,0);

\draw[->-,thick, blue] (0)--(0') node[midway,below] {$ $};%
\draw[->-,thick, blue] (1)--(1') node[midway,below] {$ $};%
\draw[->-,thick, green] (0)--(1) node[midway,below] {$ $};%
\draw[->-,thick, green] (0')--(1') node[midway,below] {$ $};%
\draw[->-] (0)--(1') node[midway,below] {$ $};%

\draw[help lines] (0')--(0'') node[midway,below] {$ $};%
\draw[help lines] (1')--(1'') node[midway,below] {$ $};%
\draw[help lines] (0')--(1') node[midway,below] {$ $};%
\draw[help lines] (0'')--(1'') node[midway,below] {$ $};%
\draw[help lines] (0')--(1'') node[midway,below] {$ $};%

\draw[->,help lines, blue!50] (00)--(01) node[at end,above] {$z$};%
\draw[help lines,->,blue!50] (00)--(02) node[at end,right] {$x$};%

\end{tikzpicture}
\end{array}
&
\xrightarrow{T^{zx}_{2d}}
&\begin{array}{c} 
\begin{tikzpicture}
\coordinate[label=below:$0$] (0) at (0,0,0) ; 
\coordinate[label=below:$0'$]  (0') at (1,0,0) ; 
\coordinate[label=above:$1$] (1) at (0,1,0) ;  
\coordinate[label=above:$1'$]  (1') at (1,1,0) ; 
\coordinate[label=below:$0''$] (0'') at (2,0,0) ;  
\coordinate[label=above:$1''$]  (1'') at (2,1,0) ;

\coordinate[label=above:$ $] (00) at (0,0,0); 
\coordinate[label=above:$ $] (01) at (2.5,0,0); 
\coordinate[label=above:$ $] (02) at (0,2.5,0);

\draw[->-,thick, blue] (0)--(0') node[midway,below] {$ $};%
\draw[help lines] (1)--(1') node[midway,below] {$ $};%
\draw[help lines] (0)--(1) node[midway,below] {$ $};%
\draw[->-] (0')--(1') node[midway,below] {$ $};%
\draw[->-,thick, cyan] (0)--(1') node[midway,below] {$ $};%

\draw[help lines] (0')--(0'') node[midway,below] {$ $};%
\draw[->-,thick, blue] (1')--(1'') node[midway,below] {$ $};%
\draw[help lines] (0')--(1') node[midway,below] {$ $};%
\draw[help lines] (0'')--(1'') node[midway,below] {$ $};%
\draw[->-,thick, cyan] (0')--(1'') node[midway,below] {$ $};%

\draw[->,help lines, blue!50] (00)--(01) node[at end,above] {$z$};%
\draw[help lines,->,blue!50] (00)--(02) node[at end,right] {$x$};%

\end{tikzpicture}
\end{array}
\end{array}
$$
\caption{3d Dehn Twist process with tube algebra. $T^{zx}_{3d}$ transformation acting on an arbitrary tube of 3d Toric code in real 3d space(upper) and $z$-$x$ plane projective $T^{zx}_{2d}$ transformation(lower), the different colors label the different edges. The dotted line indicates the line on the backside of the viewpoint.}
\label{fig: Twist_real}
\end{figure}
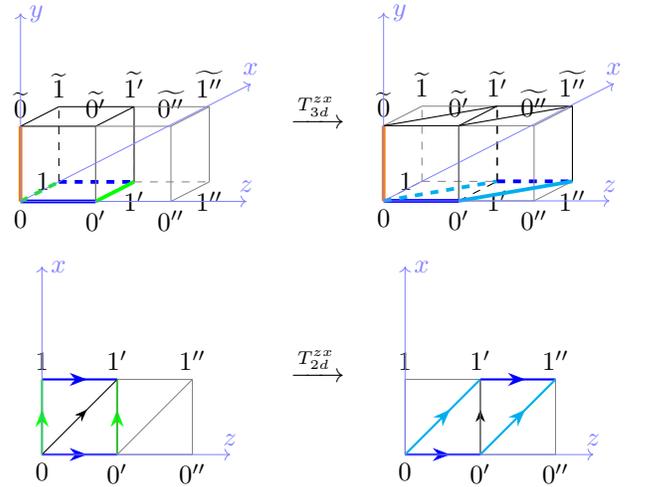

\begin{figure}
$$
\begin{array}{cc}
\begin{array}{c} 
\begin{tikzpicture}

\coordinate[label=below:$0$] (0) at (0,0,0) ; 
\coordinate[label=below:$1$] (1) at (1,0,0) ; 
\coordinate[label=below:$2$] (2) at (2,0,0) ;
\coordinate[label=below:$0'$] (0') at (-0.125,0.125,1) ; 1
\coordinate[label=below:$1'$] (1') at (0.875,0.125,1) ; 
\coordinate[label=below:$2'$] (2') at (1.875,0.125,1) ;   
\coordinate[label=above:$\widetilde{0}$] (0t) at (0,1,0); 
\coordinate[label=above:$\widetilde{1}$] (1t) at (1,1,0); 
\coordinate[label=above:$\widetilde{2}$] (2t) at (2,1,0); 
\coordinate[label=left:$\widetilde{0'}$] (0't) at (-0.125,1.125,1); 
\coordinate[label=above:$\widetilde{1'}$] (1't) at (0.875,1.125,1); 
\coordinate[label=above:$\widetilde{2'}$] (2't) at (1.875,1.125,1);

\coordinate[] (c) at (-0.0625,0.5,0.5) ;
\coordinate[label=above:$ $] (00) at (0,0,0); 
\coordinate[label=above:$ $] (01) at (2.5,0,0); 
\coordinate[label=above:$ $] (02) at (0,2,0); 
\coordinate[label=above:$ $] (03) at (-0.25,0.25,2);

\draw[line width=0.5mm, blue,dashed] (0)--(0') node[midway,below] {$ $};%
\draw[line width=0.5mm, blue,dashed] (1)--(1') node[midway,below] {$ $};%
\draw[] (0t)--(0't) node[midway,above] {$ $};%
\draw[] (1t)--(1't) node[midway,above] {$ $};%
\draw[] (2)--(2') node[midway,above] {$ $};%
\draw[] (2t)--(2't) node[midway,above] {$ $};%

\draw[line width=0.5mm,orange,dashed] (0)--(0t) node[midway,below] {$ $};%
\draw[] (0')--(0't) node[midway,left] {$ $};%
\draw[line width=0.5mm,orange,dashed] (1)--(1t) node[midway,right] {$ $};%
\draw[] (1')--(1't) node[midway,right] {$ $};%
\draw[] (2)--(2t) node[midway,right] {$ $};%
\draw[] (2')--(2't) node[midway,right] {$ $};%

\draw[dashed,green,line width=0.5mm] (0)--(1) node[midway,below] {$ $};%
\draw[] (0')--(1') node[midway,above] {$ $};%
\draw[] (0t)--(1t) node[midway,right] {$ $};%
\draw[] (0't)--(1't) node[midway,below] {$ $};%
\draw[line width=0.5mm, blue,dashed] (1)--(2) node[midway,below] {$ $};%
\draw[] (1t)--(2t) node[midway,below] {$ $};%
\draw[] (1')--(2') node[midway,below] {$ $};%
\draw[] (1't)--(2't) node[midway,below] {$ $};%

\draw[->,help lines, blue!50] (00)--(01) node[at end,above] {$x$};%
\draw[help lines,->,blue!50] (00)--(02) node[at end,right] {$y$};%
\draw[help lines,->,blue!50] (00)--(03) node[at end,above] {$z$};%

\end{tikzpicture}
\end{array}&
\xrightarrow{Z}
\begin{array}{c} 
\begin{tikzpicture}

\coordinate[label=below:$0$] (0) at (0,0,0) ; 
\coordinate[label=below:$2$] (2) at (1,0,0) ; 
\coordinate[label=below:$0'$] (0') at (-0.125,0.125,1) ; 1
\coordinate[label=below:$2'$] (2') at (0.875,0.125,1) ;  
\coordinate[label=above:$\widetilde{0}$] (0t) at (0,1,0); 
\coordinate[label=above:$\widetilde{2}$] (2t) at (1,1,0); 
\coordinate[label=left:$\widetilde{0'}$] (0't) at (-0.125,1.125,1); 
\coordinate[label=above:$\widetilde{2'}$] (2't) at (0.875,1.125,1); 
\coordinate[label=above:$ $] (01) at (1.5,0,0); 

\draw[line width=0.5mm,blue] (0)--(0') node[midway,below] {$ $};%
\draw[dashed] (2)--(2') node[midway,below] {$ $};%
\draw[] (0t)--(0't) node[midway,above] {$ $};%
\draw[] (2t)--(2't) node[midway,above] {$ $};%

\draw[line width=0.5mm,orange,dashed] (0)--(0t) node[midway,below] {$ $};%
\draw[] (0')--(0't) node[midway,left] {$ $};%
\draw[dashed] (2)--(2t) node[midway,right] {$ $};%
\draw[] (2')--(2't) node[midway,right] {$ $};%

\draw[line width=0.5mm,cyan,dashed] (0)--(2) node[midway,below] {$ $};%
\draw[] (0')--(2') node[midway,above] {$ $};%
\draw[] (0t)--(2t) node[midway,right] {$ $};%
\draw[] (0't)--(2't) node[midway,below] {$ $};%

\draw[->,help lines, blue!50] (00)--(01) node[at end,above] {$x$};%
\draw[help lines,->, blue!50] (00)--(02) node[at end,right] {$y$};%
\draw[help lines,->, blue!50] (00)--(03) node[at end,above] {$z$};%

\end{tikzpicture}
\end{array}\\
\begin{array}{c} 
\begin{tikzpicture}
\draw[thick,blue,rotate=90] (0.25,0.05) ellipse (0.6 and 0.2);
\fill[red] (-0.0625,0.5,0.5) circle (0.04) node[right,black] {\small $n_e{=}1$} ;
\draw[] (-0.0625,0.1,0.5) node[right,black] {\small $\beta=1$} ;

\foreach \y  in {-0.2}
\foreach \x  in {1}
{\fill[blue!50, opacity=0.4] (\x-0.2,\y,0)--(\x-0.2,1+\y,0)--(\x,1+\y,0)--(\x,\y,0)--cycle ;
\draw[blue,->-](\x-0.2,1+\y,0)--(\x,1+\y,0);
\draw[blue,->-](\x-0.2,\y,0)--(\x,\y,0);}  

\foreach \y  in {-0.2}
\foreach \x  in {1}
{\fill[blue!50,opacity=0.4] (\x,\y,0)--(\x,1+\y,0) .. controls (\x+0.2,1+\y,1) and (\x+1.9,1+\y,1.5) .. (\x+2,1+\y,0)--(\x+2,\y,0) .. controls (\x+1.9,\y,1.5) and (\x+0.2,\y,1) .. cycle ;
\draw[blue,->-](\x,1+\y,0) .. controls (\x+0.2,1+\y,1) and (\x+1.9,1+\y,1.5) .. (\x+2,1+\y,0);
\draw[blue,->-](\x,\y,0) .. controls (\x+0.2,\y,1) and (\x+1.9,\y,1.5) .. (\x+2,\y,0);}

\foreach \y  in {-0.2}
\foreach \x  in {1}   
{\fill[blue!30,opacity=0.2](\x+2,\y,0)--(\x+2,1+\y,0)..
   controls (\x+1.8,1+\y,-0.8) and (\x+0.8,1+\y,-1).. (\x+1,1+\y,0)--(\x+1,\y,0)..
  controls (\x+0.8,\y,-1) and (\x+1.8,\y,-0.8)..cycle;
\draw[blue,->-](\x+2,1+\y,0)..
   controls (\x+1.8,1+\y,-0.8) and (\x+0.8,1+\y,-1).. (\x+1,1+\y,0);  
 
 \draw[blue,->-](\x+2,\y,0)..
   controls (\x+1.8,\y,-0.8) and (\x+0.8,\y,-1).. (\x+1,\y,0);   }

\foreach \y  in {-0.2}
\foreach \x  in {1}
{\fill[blue!50, opacity=0.4] (\x+1.2,\y,0)--(\x+1.2,1+\y,0)--(\x+1,1+\y,0)--(\x+1,\y,0)--cycle ;
\draw[blue,->-](\x+1,\y,0)--(\x+1.2,\y,0);
\draw[blue,->-](\x+1,\y+1,0)--(\x+1.2,\y+1,0);}
      

\end{tikzpicture}
\end{array}
& 
\xLeftrightarrow{self-2\pi}
\begin{array}{c} 
\begin{tikzpicture}
\draw[thick,blue,rotate=90] (0.25,0.05) ellipse (0.6 and 0.2);
\fill[red] (-0.0625,0.5,0.5) circle (0.04) node[right,black] {\small $n_e{=}1$} ;
\draw[] (-0.0625,0.1,0.5) node[right,black] {\small $\beta=1$} ;
\draw[->,thick](0,1.1,0).. controls (-0.5,1.05,-0.2) and (-0.5,0.8,-0.2) ..  node[midway,above]{$2\pi$}(0.2,0.95,0.2); 
\end{tikzpicture}
\end{array}
\end{array}
$$
\caption{$T^{zx}_{3d}$ transformation of $P(n_e,\alpha,\beta)$ is described through the partition function $Z_{\vert P(n_e,\alpha,\beta) \rangle\langle P(n_e,\alpha,\beta) \Ab^{\beta,\alpha,\beta}\vert}$, equivalent to the action of tube $\Ab^{\beta,\alpha,\beta}$(upper). The physical interpretation of this process is the self-$2\pi$ twisted process for the $P_6$ ICI(lower).}
\label{fig: Twist_tube}
\end{figure}
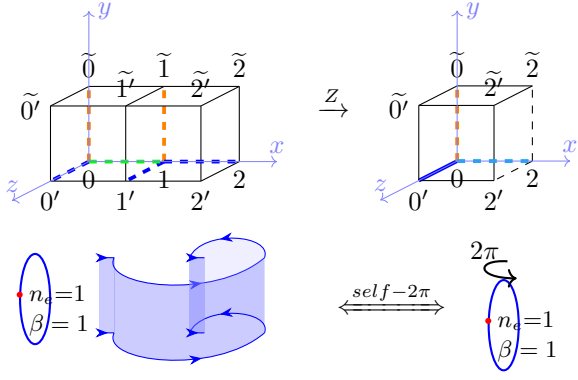

As discussed in Sec. \ref{sec:2vect}, the simple objects of $2\mathcal{V}ec_{\Zb_2}$ can be labeled by $0$ and $1$.  And an element of the tube algebra can be denoted as $\Ab_{\Zb_2}^{A,B,C}$, $A, B, C = 0, 1$.  With the multiplication defined in eqn.~\eqref{eq:tubemul2vect}, we can get the eight one-dimensional ICIs \{$P(n_{e},\alpha,\beta)$\}
\be
\label{eq:Ps}
P(n_e,\alpha,\beta):= \frac{1}{2}\Ab^{0,\alpha,\beta}+\frac{(-1)^{n_e}}{2}\Ab^{1,\alpha,\beta},
\ee
where $\alpha$ and $\beta$ are the group elements (0 or 1) assigned to the edges in the $y$ and $z$ directions of the tube, respectively.  As we discussed before, general topological excitations correspond to the boundaries of the ICIs, which are just linear combinations of the patterns shown in Fig.~\ref{3d excitation}.  Therefore, such an excitation consists of three strings and two possible attached particles.  Please notice that $\alpha$ and $\beta$ in eqn.~\eqref{eq:Ps} is just $B$ and $C$ in Fig. \ref{3d excitation}.

In the 3d toric code model, there are excitations of the trivial string, the $m$-string, the trivial particle and the $e$-particle.  With the help of the statistics data (will be discussed below) and dimension reduction technique, the eight ICIs can be related to the topological excitations as shown in Fig.~\ref{fig: boundary excitation}, where the blue solid lines are m-strings, the gray solid lines are trivial strings, while the red dot are the e-particles.  From the figure, it can be easily identified that the label $n_e$ is just the number of e-particle on the crossing point of the strings, while the $\alpha(\beta) = 0,1$ corresponds to the trivial or m-string along the corresponding direction respectively.

Because of the correspondence between the topological excitations and the ICIs, we can study the statistics of the topological excitations through the changes of the ICIs under the modular transformations $S$'s and $T$'s in $\mathrm{SL}(3, \mathbb{Z})$.  The latter one can be calculated by constructing an appropriate partition function which maps an ICI to the configuration after the transformation.  For example, a topological excitation transformed under the Dehn twist $T^{zx}_{3d}$ can be represented by the lower part of Fig.~\ref{fig: Twist_tube}, and calculated by the multiplication of the tube algebra depicted in the upper part of Fig.~\ref{fig: Twist_tube}.  Thus, we can easily have
\be
T^{zx}_{\{n_{e},\alpha,\beta\},\{n_e',\alpha',\beta'\}}=\text{e}^{\text{i}\pi* n_e\beta}\delta_{\alpha,\alpha'}\delta_{\beta,\beta'}\delta_{n_e,n_{e}'},
\ee
in the \{$P(n_{e},\alpha,\beta)$\} basis. The non-trivial matrix elements of $T^{zx}$ actually arise from the contribution of the particle $e$ wrapping around $\beta$-loop depicted in the lower part of Fig.~\ref{fig: Twist_tube}.

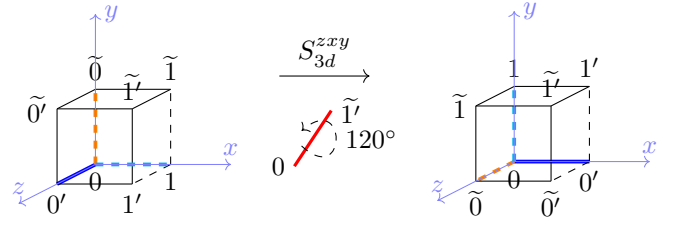
\begin{figure}
\begin{eqnarray*}
\begin{array}{c}
\begin{tikzpicture}
\coordinate[label=below:$0$] (0) at (0,0,0) ; 
\coordinate[label=below:$1$] (1) at (1,0,0) ; 
\coordinate[label=below:$0'$] (0') at (-0.125,0.125,1) ; 1
\coordinate[label=below:$1'$] (1') at (0.875,0.125,1) ;  
\coordinate[label=above:$\widetilde{0}$] (0t) at (0,1,0); 
\coordinate[label=above:$\widetilde{1}$] (1t) at (1,1,0); 
\coordinate[label=left:$\widetilde{0'}$] (0't) at (-0.125,1.125,1); 
\coordinate[label=above:$\widetilde{1'}$] (1't) at (0.875,1.125,1); 
\coordinate[label=above:$ $] (00) at (0,0,0); 
\coordinate[label=above:$ $] (01) at (1.8,0,0); 
\coordinate[label=above:$ $] (02) at (0,2,0); 
\coordinate[label=above:$ $] (03) at (-0.25,0.25,2); 
\draw[line width=0.5mm,blue] (0)--(0') node[midway,below] {$ $};%
\draw[dashed] (1)--(1') node[midway,below] {$ $};%
\draw[] (0t)--(0't) node[midway,above] {$ $};%
\draw[] (1t)--(1't) node[midway,above] {$ $};%
\draw[line width=0.5mm,orange,dashed] (0)--(0t) node[midway,below] {$ $};%
\draw[] (0')--(0't) node[midway,left] {$ $};%
\draw[dashed] (1)--(1t) node[midway,right] {$ $};%
\draw[] (1')--(1't) node[midway,right] {$ $};%
\draw[line width=0.5mm,cyan,dashed] (0)--(1) node[midway,below] {$ $};%
\draw[] (0')--(1') node[midway,above] {$ $};%
\draw[] (0t)--(1t) node[midway,right] {$ $};%
\draw[] (0't)--(1't) node[midway,below] {$ $};%
\draw[->,help lines, blue!50] (00)--(01) node[at end,above] {$x$};%
\draw[help lines,->, blue!50] (00)--(02) node[at end,right] {$y$};%
\draw[help lines,->, blue!50] (00)--(03) node[at end,above] {$z$};%
\end{tikzpicture}
\end{array}
\begin{array}{c}
\begin{tikzpicture}
\coordinate[red,label=left:$0$] (0) at (0,0,0) ; 
\coordinate[red,label=right:$\widetilde{1'}$] (1't) at (0.875,1.125,1); 
\draw[->] (-0.2,1.2)--(1,1.2) node[midway,above] {$S^{zxy}_{3d}$};
\draw[red,very thick] (0)--(1't) node[midway,left] {$ $};%
\draw [dashed,->] (0.575/2,0.925/2,1/2).. controls (0.9,-0.1,1/2) and (0.9,1.2,1/2) ..(0.575/2,1.325/2,1/2) node[midway,right]{$120^{\circ}$};
\end{tikzpicture}
\end{array}
\begin{array}{c}
\begin{tikzpicture}
\coordinate[label=below:$0$] (0) at (0,0,0) ; 
\coordinate[label=below:$0'$] (1) at (1,0,0) ; 
\coordinate[label=below:$\widetilde{0}$] (0') at (-0.125,0.125,1) ; 1
\coordinate[label=below:$\widetilde{0'}$] (1') at (0.875,0.125,1) ;  
\coordinate[label=above:$1$] (0t) at (0,1,0); 
\coordinate[label=above:$1'$] (2t) at (1,1,0); 
\coordinate[label=left:$\widetilde{1}$] (0't) at (-0.125,1.125,1); 
\coordinate[label=above:$\widetilde{1'}$] (2't) at (0.875,1.125,1); 
\coordinate[label=above:$ $] (00) at (0,0,0); 
\coordinate[label=above:$ $] (01) at (1.8,0,0); 
\coordinate[label=above:$ $] (02) at (0,2,0); 
\coordinate[label=above:$ $] (03) at (-0.25,0.25,2); 
\draw[line width=0.5mm,blue] (0)--(2) node[midway,below] {$ $};%
\draw[dashed] (2)--(2') node[midway,below] {$ $};%
\draw[] (0t)--(0't) node[midway,above] {$ $};%
\draw[] (2t)--(2't) node[midway,above] {$ $};%
\draw[line width=0.5mm,orange,dashed] (0)--(0') node[midway,below] {$ $};%
\draw[] (0')--(0't) node[midway,left] {$ $};%
\draw[dashed] (2)--(2t) node[midway,right] {$ $};%
\draw[] (2')--(2't) node[midway,right] {$ $};%
\draw[line width=0.5mm,cyan,dashed] (0)--(0t) node[midway,below] {$ $};
\draw[] (0')--(2') node[midway,above] {$ $};%
\draw[] (0t)--(2t) node[midway,right] {$ $};%
\draw[] (0't)--(2't) node[midway,below] {$ $};%

\draw[->,help lines, blue!50] (00)--(01) node[at end,above] {$x$};%
\draw[help lines,->, blue!50] (00)--(02) node[at end,right] {$y$};%
\draw[help lines,->, blue!50] (00)--(03) node[at end,above] {$z$};%
\end{tikzpicture}
\end{array}
\end{eqnarray*}
    \caption{$S^{zxy}_{3d}$ transformation acting on an arbitrary tube of 3d toric code in real 3d space is equivalent to rotating the tube $120^{\circ}$ counterclockwise around $0$-$\widetilde{1'}$ axis.}
    \label{fig:Szxy}
\end{figure}

However, the computation of a general $S^{zxy}_{3d}$ is very tedious and can not be expressed as a multiplication with some tubes.
Thus we use a different and simpler approach.  As depicted in Fig.~\ref{fig:Szxy}, $S^{zxy}_{3d}$ just rotates the tube without deforming it.  Thus we can easily get the representation matrix of the $S^{zxy}_{3d}$ by acting it on $P_i$ and then expressing the resultant $S^{zxy}_{3d} \cdot P_i$ in the basis of ICIs.  Then the coefficient $S^{zxy}_{P_i,P_j}$ is just the element of the $S$ matrix of the topological excitations in the 3d topological order.  For the 3d toric code, our calculation leads to the following S matrix (see more details in appendix \ref{3DTC})
\be
S^{zxy}_{\{n_{e},\alpha,\beta\},\{n_e',\alpha',\beta'\}}=\frac{1}{2}\text{e}^{\text{i}\pi*(n'_e\beta-n_e\alpha') }\delta_{\alpha,\beta'}.
\ee

So far, we have obtained the matrix representation of the generators $T^{zx}_{3d}$, $S^{zxy}_{3d}$ of $SL(3,\Zb)$ of the 3d toric code.  It has been proposed that there is another kind of very interesting statistics data from the so-called three-loop braidings \cite{Wang2014}.  According to eqn.~\eqref{3loop-S}, the statistics of the three-loop braiding can be calculated with the $S^{zx}$ matrix of the $SL(2, \Zb)$ group.  Since $SL(2,\Zb)$ is a subgroup of $SL(3,\Zb)$, $S^{zx}$ can be calculated with the generators of $SL(3,\Zb)$ through \cite{coxeter2013}
\begin{eqnarray}
S^{zx}_{\{n_{e},\alpha,\beta\},\{n_{e}',\alpha',\beta'\}}&=&\left(\left(T_{3d}^{zx}\right)^{-1} S_{3d}^{zxy}\right)^{3}\left(S_{3d}^{zxy} T_{3d}^{zx}\right)^{2}\nonumber\\
&~&\times  S_{3d}^{zxy}\left(T_{3d}^{zx}\right)^{-1}\nonumber\\
&=&\frac{1}{2}\text{e}^{\text{i}\pi*(n'_e\beta+n_e\beta') }\delta_{\alpha,\alpha'}	
\end{eqnarray}
According to eqn.~\eqref{3loop-S}, $\theta_{\{\alpha,\beta,0\},\{\alpha,\beta',0\}}=\pi*0$, hence there is no non-trivial three-loop braidings in 3d toric code, which is consistent with previous result \cite{Wang2014}.

\subsection{Three loop braiding in 3d $\Zb_2\times\Zb_2-(p,q)$ topological orders}

The simplest example of a 3d TO with non-trivial three-loop braiding is the so-called 3d $\Zb_2\times\Zb_2-(p,q)$ topological orders, which can be viewed as gauged 3d bosonic $\Zb_2\times\Zb_2$ SPT phase.  $p, q = 0, 1$ labels the four different $\Zb_2\times\Zb_2$ SPT phase, which is classified by the nontrivial group 4-cohomology $\mathcal{H}^4(\Zb_2\times \Zb_2,U(1))=\Zb_2\times \Zb_2$.  These 3d $\Zb_2\times\Zb_2-(p,q)$ topological orders can be naturally realized with our membrane-net model constructed with $2\mathcal{V}ec^{\omega_{p, q}}_{\Zb_2 \times \Zb_2}$, where $\omega_{p,q}$ is defined as
\begin{eqnarray}
\omega_{0,0}(a,b,c,d)&=&1 \nonumber\\ 
\omega_{1,0}(a,b,c,d)&=&\text{e}^{\pi/2 a_1b_2(c_2+d_2-\langle c_2+d_2\rangle)} \nonumber \\
\omega_{0,1}(a,b,c,d)&=&\text{e}^{\pi/2 a_2b_1(c_1+d_1-\langle c_1+d_1\rangle)} \nonumber \\
\omega_{1,1}(a,b,c,d)&=&\omega_{1,0}(a,b,c,d)\omega_{0,1}(a,b,c,d),
\end{eqnarray}
where $a=(a_1,a_2), b=(b_1,b_2), c=(c_1,c_2), d=(d_1,d_2)$ are group elements of $\Zb_2\times\Zb_2$.

As discussed in Sec. \ref{sec:2vect}, the object in $2\mathcal{V}ec^{\omega_{p, q}}_{\Zb_2 \times \Zb_2}$ can be labeled by the group element of $\Zb_2\times\Zb_2$.  Thus, we denote the objects with a pair $(i, j)$ with $i, j = 0, 1$.  Then with the similar process, we can obtain the 64 ICIs for the different $\Zb_2\times\Zb_2-(p,q)$ topological orders
\begin{widetext}
\noindent$(p=0,q=0):$
\begin{eqnarray}
P(n, \alpha,\beta)&=&\frac{1}{4}\left(\Ab^{(0,0),\alpha,\beta}+(-1)^{n_1}\Ab^{(1,0),\alpha,\beta}+  (-1)^{n_2}\Ab^{(0,1),\alpha,\beta}+(-1)^{n_1+n_2}\Ab^{(1,1),\alpha,\beta}\right),\nonumber
\end{eqnarray}
\noindent$(p=1,q=0):$
\begin{eqnarray}
P(n_e,\alpha,\beta)&=&\frac{1}{4}\left(\Ab^{(0,0),\alpha,\beta}+(-1)^{n_1}\text{i}^{\text{mod}(\alpha_1\beta_2+\alpha_2\beta_1,2)}\Ab^{(1,0),\alpha,\beta} + (-1)^{n_2}\Ab^{(0,1),\alpha,\beta}\right. \nonumber \\
&+& \left.(-1)^{n_1+n_2}\text{i}^{\text{mod}(\alpha_1\beta_2+\alpha_2\beta_1,2)}\Ab^{(1,1),\alpha,\beta}\right),\nonumber
\end{eqnarray}
\noindent$(p=0,q=1):$
\begin{eqnarray}
P(n_e,\alpha,\beta)&=&\frac{1}{4}\left(\Ab^{(0,0),\alpha,\beta}+(-1)^{n_1}\Ab^{(1,0),\alpha,\beta} + (-1)^{n_2}\text{i}^{\text{mod}(\alpha_1\beta_2+\alpha_2\beta_1,2)}\Ab^{(0,1),\alpha,\beta}\right. \nonumber \\
&+& \left.(-1)^{n_1+n_2}\text{i}^{\text{mod}(\alpha_1\beta_2+\alpha_2\beta_1,2)}\Ab^{(1,1),\alpha,\beta}\right),\nonumber
\end{eqnarray}
\noindent$(p=1,q=1):$
\begin{eqnarray}
P(n_e,\alpha,\beta)&=&\frac{1}{4}\left(\Ab^{(0,0),\alpha,\beta}+(-1)^{n_1}\text{i}^{\text{mod}(\alpha_1\beta_2+\alpha_2\beta_1,2)}\Ab^{(1,0),\alpha,\beta}\right. \nonumber \\
 &+&  \left. (-1)^{n_2}\text{i}^{\text{mod}(\alpha_1\beta_2+\alpha_2\beta_1,2)}\Ab^{(0,1),\alpha,\beta}+(-1)^{n_1+n_2}\Ab^{(1,1),\alpha,\beta}\right),
\end{eqnarray}
\end{widetext}
where $\alpha = (\alpha_1, \alpha_2)$ and $\beta = (\beta_1, \beta_2)$ with $\alpha_i, \beta_i = 0, 1$ are the labels of objects.  $n = (n_1, n_2)$ with $n_i = 0, 1$ is similar to the index $n_e$ in the 3d toric code case.

Our results show that all the classes have non-trivial three-loop braiding phase except the $(0, 0)$ case.  For all the other classes, the braiding between two pure loops $\beta$ and $\beta'$ linked by $\alpha$-loop is determined by
\begin{widetext}
\begin{align}
    S^{zx}_{\{(0,0),\alpha,\beta\},\{(0,0),\alpha',\beta'\}} & =
    \left\{ 
    \begin{matrix}
    \frac{1}{4}\text{e}^{\text{i}\pi/2\left( \left(\beta_2\beta'_1 + \beta_1\beta'_2\right)\alpha_1-2\alpha_2\beta_1\beta'_1\right)}\delta_{\alpha,\alpha'},	 &  (p=1,q=0) \\
   \frac{1}{4}\text{e}^{\text{i}\pi/2\left( \left(\beta_2\beta'_1 + \beta_1\beta'_2\right)\alpha_2-2\alpha_1\beta_2\beta'_2\right)}\delta_{\alpha,\alpha'},	 &  (p=0,q=1) \\
   \frac{1}{4}\text{e}^{\text{i}\pi/2\left( \left(\beta_2\beta'_1 + \beta_1\beta'_2\right)\left(\alpha_1+\alpha_2\right)-2\alpha_1\beta_2\beta'_2-2\alpha_2\beta_1\beta'_1\right)}\delta_{\alpha,\alpha'},	 & (p = 1, q= 1)
    \end{matrix}
    \right.
\end{align}
\end{widetext}
Then according to eqn.~\eqref{3loop-S}, we have that $\theta^{(1,0)}_{(1,0),(0,1)}=\pi/2$ in $(p=1,q=0)$ case, which is consistent with the result of Ref. ~\onlinecite{Wang2014}. 

\section{Discussion and Conclusion}
The above approach to construct the 3d membrane-net model can be generalized to higher dimensional case, where one may construct an n-dimensional lattice model for n-dimensional topological order based on a spherical fusion $(n-1)$-category~\cite{kapustin2010,kong2015,johnsonfreyd2020}. The Hilbert space of the lattice model should be spanned by $(n-1)$-morphism in the spherical fusion $(n-1)$-category. Combination of fusion rules of objects, 1-morphism $\cdots$ and $(n-1)$-morphisms corresponds to flatness condition of the lattice model. And the evolution operator is constructed by state sum of spherical fusion $(n-1)$-category (partition function).
In this way, the lattice structure of general topological order in arbitrary dimension is provided by a natural and simple approach.

Similarly, though we discuss only the 2d tube algebra on $S^1 \times I$ and 3d tube algebra on $S^1 \times S^1 \times I$ in this paper, it is not difficult to construct arbitrary dimensional tube algebra on some different open manifold to study mutual statistics of topological orders in arbitrary dimension. Even for 2d cases, it is known that S, T matrix are not enough to characterize a UMTC and thus more topological invariant is required to characterize a 2d topological order. So we can construct a 2d tube algebra on some different open manifold to study the new topological invariant in 2d. This will be discussed in a different paper.

In a summary, we construct a membrane-net model to systematically study all 3d topological orders with arbitrary gapped boundary. The Hamiltonian of the membrane-net model is derived from the 3+1D partition function, which is taken as the state sum of a spherical fusion 2-category.
Furthermore, membrane-net model should always admit also the canonical boundary described by either $2\mathcal{V}ec_G^\omega$ or an EF 2-category, thus our model provide an explicit approach to demonstrate the Morita equivalence between any fusion 2-category and the canonical ones.
We also provide a universal framework to study mutual statistics of all excitations in the 3d topological order through 3d tube algebra constructed with the input spherical fusion 2-category. We demonstrate by a few examples that our result is consistent with known results. And the approach to construct lattice model and to study mutual statistics can all be generalized for topological orders in arbitrary dimension naturally.

\section{Acknowledgement}
We are grateful to the helpful discussions with Zheng-Cheng Gu, Liang Kong, Laurens Lootens, Yang Qi, Chenjie Wang, Qing-Rui Wang and Zhi-Hao Zhang. This work was supported by NSFC (Grants No. 11861161001), the Science, Technology and Innovation Commission of Shenzhen Municipality (No. ZDSYS20190902092905285), and Center for Computational Science and Engineering at Southern University of Science and Technology. TL acknowledges start-up support via Direct Grant No. 4053501 from The Chinese University of Hong Kong.

\appendix

\section{spherical fusion 2-category} \label{F2CAT}

In this section, we will give a brief introduction for certain properties of spherical fusion 2-category that relates to our construction. To offer a physically intuitive picture, mathematical strictness of these properties will be partially given up.
For more detailed and strict discussion of spherical fusion 2-category, please see Ref.~\onlinecite{douglas2018}.

The data of a spherical fusion 2-category $\mathcal{B}$ consists of objects, 1-morphisms and 2-morphisms.
A strict definition of $\mathcal{B}$ is quite complicated.  Thus we presented the structure of a spherical fusion 2-category with the graphical calculus in the plane, which is the so called ``string diagrams'' in mathematics, for better understanding.
In string diagrams, the objects are depicted as regions in the plane and a 1-morphism between two objects $A$ and $B$ is depicted as a line separating the region as shown below
\begin{align}
\label{graph of obj and 1-mor}
\begin{array}{c}
\begin{tikzpicture}[scale=1.8]
\coordinate[label=right:$ $] (0) at (0,0) ;
\coordinate[label=above:$ $] (1) at (0,1) ;
\coordinate[label=left:$ $] (2) at (1,0) ;
\coordinate[label=below:$ $] (3) at (1,1) ;
\draw[help lines, very thick] (0)--(1) node[midway,right=17pt] {$A$};%
\draw[help lines, very thick] (0)--(2) node[midway,left] {$ $};%
\draw[help lines, very thick] (3)--(1) node[midway,left] {$ $};%
\draw[help lines, very thick] (3)--(2) node[midway,left] {$ $};%
\coordinate[label=right:$ $] (0') at (2,0) ;
\coordinate[label=above:$ $] (1') at (2,1) ;
\coordinate[label=left:$ $] (2')at (3,0) ;
\coordinate[label=below:$ $] (3') at (3,1) ;
\draw[help lines, very thick] (0')--(1') node[midway,right=5pt] {$A$};%
\draw[help lines, very thick] (0')--(2') node[midway,left] {$ $};%
\draw[help lines, very thick] (3')--(1') node[midway,left] {$ $};%
\draw[help lines, very thick] (3')--(2') node[midway,left=5pt] {$B$};%
\draw[very thick] (2.5,1)--(2.5,0) node[below=0pt] {$f$};
\end{tikzpicture}
\end{array}
\end{align}
The collection of all 1-morphisms from object $A$ to $B$, together with all 2-morphisms between these 1-morphisms are denoted by $\hom_{\mathcal{B}}(A, B)$, and called the hom category from object $A$ to object $B$.
The tensor product between objects $A \otimes B$ and 1-morphisms $f \otimes g$ are represented by the following diagram:
\begin{align}
\label{tensor of obj and 1-mor}
\begin{array}{c}
\begin{tikzpicture}[scale=1.8]
\coordinate[label=right:$ $] (0) at (0,0) ;
\coordinate[label=above:$ $] (1) at (0,1) ;
\coordinate[label=left:$ $] (2) at (1,0) ;
\coordinate[label=below:$ $] (3) at (1,1) ;
\draw[help lines, very thick] (0)--(1) node[midway,right=-2pt] {$A$};%
\draw[help lines, very thick] (0)--(2) node[midway,left] {$ $};%
\draw[help lines, very thick] (3)--(1) node[midway,left] {$ $};%
\draw[help lines, very thick] (3)--(2) node[midway,left] {$ $};%
\coordinate[label=right:$ $] (0') at (0.2,0.2) ;
\coordinate[label=above:$ $] (1') at (0.2,1.2) ;
\coordinate[label=left:$ $] (2') at (1.2,0.2) ;
\coordinate[label=below:$ $] (3') at (1.2,1.2) ;
\draw[help lines, very thick, dashed] (0')--(0.2,1) node[midway,right=17pt] {$ $};%
\draw[help lines, very thick] (0.2,1)--(1') node[midway,right=17pt] {$ $};%
\draw[help lines, very thick, dashed] (0')--(1,0.2) node[midway,left] {$ $};%
\draw[help lines, very thick] (1,0.2)--(2') node[midway,left] {$ $};%
\draw[help lines, very thick] (3')--(1') node[midway,left] {$ $};%
\draw[help lines, very thick] (3')--(2') node[midway,left=-2pt] {$B$};%
\coordinate[label=right:$ $] (0) at (2,0) ;
\coordinate[label=above:$ $] (1) at (2,1) ;
\coordinate[label=left:$ $] (2)at (3,0) ;
\coordinate[label=below:$ $] (3) at (3,1) ;
\draw[help lines, very thick] (1)--(0) node[right=5pt,above=-1pt] {$A$};%
\draw[help lines, very thick] (0)--(2) node[midway,left] {$ $};%
\draw[help lines, very thick] (3)--(1) node[midway,left] {$ $};%
\draw[help lines, very thick] (3)--(2) node[left=5pt,above=-1pt] {$B$};%
\draw[very thick] (2.5,1)--(2.5,0) node[below=0pt] {$f$};
\coordinate[label=right:$ $] (0') at (2.2,0.2) ;
\coordinate[label=above:$ $] (1') at (2.2,1.2) ;
\coordinate[label=left:$ $] (2') at (3.2,0.2) ;
\coordinate[label=below:$ $] (3') at (3.2,1.2) ;
\draw[help lines, very thick, dashed] (0')--(2.2,1) node[] {$ $};%
\draw[help lines, very thick] (2.2,1)--(1') node[right=5pt,below=-1pt] {$C$};%
\draw[help lines, very thick, dashed] (0')--(3,0.2) node[midway,left] {$ $};%
\draw[help lines, very thick] (3,0.2)--(2') node[midway,left] {$ $};%
\draw[help lines, very thick] (3')--(1') node[midway,left] {$ $};%
\draw[help lines, very thick] (2')--(3') node[left=5pt,below=-1pt] {$D$};%
\draw[very thick, dashed] (2.7,0.2)--(2.7,1) node[below=0pt] {$ $};
\draw[very thick] (2.7,1)--(2.7,1.2) node[above=0pt] {$g$};
\end{tikzpicture}
\end{array}
\end{align}

A 2-morphism $\eta$ from 1-morphisms $f$ to $g$ in $\hom_{\mathcal{B}}(A, B)$ is depicted as a node separating the line labeled by $f$ from the line labeled by $g$:
\begin{align}
\begin{array}{c}
\begin{tikzpicture}[scale=1.8]
\coordinate[label=right:$ $] (0) at (0,0) ;
\coordinate[label=above:$ $] (1) at (0,1) ;
\coordinate[label=left:$ $] (2) at (1,0) ;
\coordinate[label=below:$ $] (3) at (1,1) ;
\draw[help lines, very thick] (0)--(1) node[midway,right=3pt] {$A$};%
\draw[help lines, very thick] (0)--(2) node[midway,left] {$ $};%
\draw[help lines, very thick] (3)--(1) node[midway,left] {$ $};%
\draw[help lines, very thick] (3)--(2) node[midway,left=3pt] {$B$};%
\draw[very thick] (0.5,0.5)--(0.5,0) node[below=0pt] {$f$};
\draw[very thick] (0.5,0.5)--(0.5,1) node[above=0pt] {$g$};
\filldraw[] (0.5,0.5) circle (0.03) node[left=0.0] {$\eta$};
\end{tikzpicture}
\end{array}
\end{align}
Again we denote by $\hom_{\mathcal{B}}(f, g)$ the collection of all 2-morphisms from $f$ to $g$.  For a spherical fusion 2-category, $\hom_{\mathcal{B}}(f, g)$ is just a vector space.  The above graphical representation is very similar to our membrane-net model, where we assign objects in $\mathcal{B}$ on faces, 1-morphisms in $\mathcal{B}$ on edges, and 2-morphisms in $\mathcal{B}$ on the vertices in the membrane-net lattice.

\subsection{Fusion theory}

The Fusion rule of a spherical fusion 2-category is related to the flatness condition of membrane-net model.
A more complicated 2-morphism, which is also the main element of the fusion theory--the associator state space $V^+(ijkl)$ is represented by the following diagram:

\begin{align}
\begin{array}{c}
\begin{tikzpicture}[scale=1.3]
\label{gradia2}
\draw[help lines, very thick] (0,0)--(0.6,0) node[midway,below] {$A$};
\draw [help lines, very thick](0.6,0).. controls (1,0.2) and (1.4,0.4) ..(2.2,0.4) node[above right] {$F$};
\draw [help lines, very thick](0.6,0).. controls (0.8,-0.15) and (1.0,-0.2) ..(1.2,-0.2) node[below left] {$B$};
\draw [help lines, very thick](1.2,-0.2).. controls (1.4,-0.35) and (1.6,-0.4) ..(1.8,-0.4) node[above right] {$D$};
\draw [help lines, very thick](1.2,-0.2).. controls (1.4,-0.05) and (1.8,0) ..(2,0) node[above right] {$E$};
\draw[help lines, very thick] (0,0)--(0,1.5) node[midway,below] {$ $};
\draw[help lines, very thick] (0,1.5)--(0.6,1.5)
node[midway,below] {$ $};
\draw[help lines, very thick] (1.8,-0.4)--(1.8,1.1);
\draw[help lines, very thick] (2,0)--(2,1.5);
\draw[help lines, very thick] (2.2,0.4)--(2.2,1.9);
\draw [help lines, very thick](0.6,1.5).. controls (1,1.3) and (1.4,1.1) ..(1.8,1.1) node[above left] {$ $};
\draw [help lines, very thick](0.6,1.5).. controls (0.8,1.65) and (1.0,1.7) ..(1.2,1.7) node[above left] {$C$};
\draw [help lines, very thick](1.2,1.7).. controls (1.4,1.55) and (1.8,1.5) ..(2,1.5) node[above left] {$ $};
\draw [help lines, very thick](1.2,1.7).. controls (1.4,1.85) and (1.8,1.9) ..(2.2,1.9) node[above left] {$ $};
\draw [very thick](0.6,0).. controls (0.7,0.5) and (0.8,0.7) ..(0.9,0.75) node[midway,left] {$i$};
\draw [very thick](1.2,-0.2).. controls (1.1,0.4) and (1.0,0.7) ..(0.9,0.75) node[midway,right] {$j$};
\draw [very thick](0.6,1.5).. controls (0.7,1) and (0.8,0.8) ..(0.9,0.75) node[midway,left] {$k$};
\draw [very thick](1.2,1.7).. controls (1.1,1.1) and (1.0,0.8) ..(0.9,0.75) node[midway,right] {$l$};
\filldraw[] (0.9,0.75) circle (0.04) node[above=0.2] {$\eta$};
\end{tikzpicture}
\end{array}
:
\begin{array}{c}
\begin{tikzpicture}[scale=1.6]
\draw[help lines, very thick] (0.2,0.3)--(0,0) node[below] {$D$};
\draw[help lines, very thick] (0.2,0.3)--(0.4,0) node[below] {$E$};
\draw[help lines, very thick] (0.2,0.3)--(0.4,0.6) node[midway, left] {$B$};
\draw[help lines, very thick] (0.4,0.6)--(0.8,0.0) node[below] {$F$};
\draw[help lines, very thick] (0.4,0.6)--(0.4,1) node[left] {$A$};
\filldraw[] (0.2,0.3) circle (0.03) node[right] {$j$};
\filldraw[] (0.4,0.6) circle (0.03) node[right] {$i$};
\end{tikzpicture}
\end{array}
\rightarrow
\begin{array}{c}
\begin{tikzpicture}[scale=1.6]
\draw[help lines, very thick] (0.4,0.6)--(0,0) node[below] {$D$};
\draw[help lines, very thick] (0.6,0.3)--(0.4,0) node[below] {$E$};
\draw[help lines, very thick] (0.6,0.3)--(0.4,0.6) node[midway, right] {$C$};
\draw[help lines, very thick] (0.6,0.3)--(0.8,0.0) node[below] {$F$};
\draw[help lines, very thick] (0.4,0.6)--(0.4,1) node[left] {$A$};
\filldraw[] (0.6,0.3) circle (0.03) node[left] {$l$};
\filldraw[] (0.4,0.6) circle (0.03) node[left] {$k$};
\end{tikzpicture}
\end{array}
\end{align}
where $A, B, \dots, F$ are objects. $i \in \hom_{\mathcal{B}}(A, B \otimes F)$, $j \in \hom_{\mathcal{B}}(B, D \otimes E)$, $k \in \hom_{\mathcal{B}}(A, D \otimes C)$, and $l \in \hom_{\mathcal{B}}(C, E \otimes F)$ are 1-morphisms.  $\eta \in \hom_{\mathcal{B}}(i \otimes j, k \otimes l)$ is a 2-morphism.

For clarity, we draw the source and target of the 2-morphism explicitly such that: $i\otimes j \stackrel{\eta}{\rightarrow} k\otimes l$.
For any combination of 1-morphisms $i, j, k, l \in \mathcal{B}$, there is a fixed finite-dimensional vector space $V^+(ijkl)$. And the numbers $N^{ij}_{kl}=dim(V^+(ijkl))$ are called fusion multiplicities which play an important role in fusion rules of 1-morphisms: 
\begin{align}
\label{2catf}
i\otimes j=\sum_{kl} N^{ij}_{kl} \quad k\otimes l,
\end{align}
The choice of a special 1-morphism $\mathbbm{1}$ (for example, let $l$ equal to the trivial 1-morphism $\mathbbm{1}$) is also considered part of the fusion rules.
It should be noticed that though we only write down 1-morphisms and leave objects implicitly in eqn.~\eqref{2catf}, objects must be involved in the fusion of 1-morphisms since 1-morphisms are boundaries (domain walls) between objects as shown in \eqref{gradia2}.

And there are also fusion rules for objects $A, B, C \in \mathcal{B}$:
\begin{align}
A\otimes B=\sum_{C} N^{AB}_{C} \quad C.
\end{align}
However, unlike the fusion rules of objects in fusion 1-category, we can not read $N^{AB}_{C}$ directly from hom spaces. This is because hom spaces in a fusion 2-category are no longer simple vector spaces. Usually, $N^{AB}_{C}$ is contained in the definition for a given spherical fusion 2-category.

Physically, in membrane-net model, $Q_{\v E}$ and $Q_{\v V}$ operators tell us that any membrane-net configuration allowed must satisfy the fusion rules of objects and 1-morphisms, which actually correspond to flatness condition of membrane-net model.

\subsection{Spherical condition and Quantum dimension}

In Levin-Wen model, spherical condition of the input fusion 1-category $\mathcal{C}$ guarantees that shrinking a closed string in wave function will give a number
\begin{align}
\Phi
\Biggl(
\begin{array}{c}
\begin{tikzpicture}[scale=0.4]
\filldraw[draw=gray!80,fill=gray!20,rounded corners] (0,0) rectangle (2,3);
\draw[->-] (3.5,1.5) circle (1) node[above] {$A$};
\end{tikzpicture}
\end{array} 
\Biggr)
=
d_A \Phi
\Biggl(
\begin{array}{c}
\begin{tikzpicture}[scale=0.4]
\filldraw[draw=gray!80,fill=gray!20,rounded corners] (0,0) rectangle (2,3);
\end{tikzpicture}
\end{array} 
\Biggr).
\end{align}
Here, $d_A$ is quantum dimension of anyon created by string $A$, which corresponds to an object in $\mathcal{C}$.

Physically, spherical condition means that if string-net model is placed on a $2$-sphere, closed string diagrams in ground state wave function should be able to move freely on a $2$-sphere. As a result, quantum dimension gained by shrinking a closed string directly from one side should be equal to the one gained by pulling the string around the back of the $2$-sphere and then shrinking it form another side. From a physical perspective, this condition is quite natural.

Similarly, if membrane-net model is placed on a $3$-sphere, it is natural to believe that a $2$-sphere (object) with a loop (1-morphism) on it should be able to move freely on a $3$-sphere. Besides the two different ways to shrink a loop on a $2$-sphere as we mention above, there are also two different ways to shrink the $2$-sphere (object), namely, from outside and inside of the $2$-sphere, on a $3$-sphere. So there are four different ways to shrink a $2$-sphere with a loop on it on a $3$-sphere. Intuitively, spherical condition of the fusion 2-category guarantees that these four processes will give a same number--quantum dimension of 1-morphism. Graphically, spherical condition can thus be simply represented as:
\begin{align}
\Phi
\Biggl(
\begin{array}{c}
\begin{tikzpicture}[scale=0.4]
\filldraw[draw=gray!80,fill=gray!20,rounded corners] (0,0) rectangle (2,3);
\draw[help lines] (5,1.5) circle (2.5) node[above=18pt] {$A$};
\draw[help lines] (7.5,1.5) arc (0:-180:2.5 and 0.8);
\draw[help lines,dashed] (7.5,1.5) arc (0:180:2.5 and 0.8);
\end{tikzpicture}
\end{array} 
\Biggr)
=
d_A \Phi
\Biggl(
\begin{array}{c}
\begin{tikzpicture}[scale=0.4]
\filldraw[draw=gray!80,fill=gray!20,rounded corners] (0,0) rectangle (2,3);
\end{tikzpicture}
\end{array} 
\Biggr);\nonumber\\
\Phi
\Biggl(
\begin{array}{c}
\begin{tikzpicture}[scale=0.4]
\filldraw[draw=gray!80,fill=gray!20,rounded corners] (0,0) rectangle (2,3);
\draw[help lines] (5,1.5) circle (2.5) node[above=18pt] {$A$};
\draw[help lines] (7.5,1.5) arc (0:-180:2.5 and 0.8);
\draw[help lines,dashed] (7.5,1.5) arc (0:180:2.5 and 0.8) node[right=20pt] {$B$};
\draw[->-,thick] (5,1) ellipse  (0.8 and 1.4) node[left=8pt] {$f$};
\end{tikzpicture}
\end{array} 
\Biggr)
=
d_f \Phi
\Biggl(
\begin{array}{c}
\begin{tikzpicture}[scale=0.4]
\filldraw[draw=gray!80,fill=gray!20,rounded corners] (0,0) rectangle (2,3);
\end{tikzpicture}
\end{array} 
\Biggr).
\end{align}
Here $A$ and $B$ are objects in $\mathcal{B}$ and $f \in hom_{\mathcal{B}}(A,B)$ is a 1-morphism in $\mathcal{B}$. Physically, $d_f$ is quantum dimension of particle-like excitation created by string $f$ which is domain wall between two membranes $A$ and $B$. $d_A$ is quantum dimension of string-like excitation created by membrane $A$.
When $A$ and $B$ are all trivial object, $d_f$ is quantum dimension of real particles which are created by strings corresponding to 1-morphisms between trivial object.
And if the loop on $2$-sphere $A$ is trivial, $d_f$ will reduce to the quantum dimension of string-like excitations $d_A$.

\begin{figure*}
\centering
\begin{tikzpicture} [scale=0.68]
\coordinate[label=right:$0$] (0) at (2,0);
\coordinate[label=right:$1$] (1) at (1,1.73);
\coordinate[label=left:$2$] (2) at (-1,1.73);
\coordinate[label=left:$3$] (3) at (-2,0);
\coordinate[label=left:$4$] (4) at (-1,-1.73);
\coordinate[label=right:$5$] (5) at (1,-1.73);
\draw[thick] (0)--(1);
\draw[thick] (1)--(2);
\draw[thick] (2)--(3);
\draw[thick] (3)--(4);
\draw[thick] (4)--(5);
\draw[thick] (5)--(0);
\draw[thick] (0)--(2);
\draw[thick] (0)--(3);
\draw[thick] (0)--(4);
\draw[thick] (1)--(3);
\draw[thick] (1)--(4);
\draw[thick] (1)--(5);
\draw[thick] (2)--(4);
\draw[thick] (2)--(5);
\draw[thick] (3)--(5);
\draw[thick,->] (2,-1.5)--(4,-3) node[left=10pt] {$\Gamma_{(12345)}$};
\coordinate[label=right:$0$] (0'''') at (8,-4.5);
\coordinate[label=right:$1$] (1'''') at (7,-4.5+1.73);
\coordinate[label=left:$2$] (2'''') at (5,-4.5+1.73);
\coordinate[label=left:$3$] (3'''') at (4,-4.5); 
\coordinate[label=left:$4$] (4'''') at (5,-4.5-1.73);
\coordinate[label=right:$5$] (5'''') at (7,-4.5-1.73);
\draw[thick] (0'''')--(1'''');
\draw[thick] (1'''')--(2'''');
\draw[thick] (2'''')--(3'''');
\draw[thick] (3'''')--(4'''');
\draw[thick] (4'''')--(5'''');
\draw[thick] (5'''')--(0'''');
\draw[thick] (0'''')--(2'''');
\draw[thick] (0'''')--(3'''');
\draw[thick] (0'''')--(4'''');
\draw[thick] (1'''')--(3'''');
\draw[thick] (1'''')--(4'''');
\draw[thick] (1'''')--(5'''');
\draw[thick] (2'''')--(5'''');
\draw[thick] (3'''')--(5'''');
\draw[thick,->] (8.5,-4.5)--(11,-4.5) node[midway,below] {$\Gamma_{(01345)}$};
\coordinate[label=right:$0$] (0''''') at (8+7.5,-4.5);
\coordinate[label=right:$1$] (1''''') at (7+7.5,-4.5+1.73);
\coordinate[label=left:$2$] (2''''') at (5+7.5,-4.5+1.73);
\coordinate[label=left:$3$] (3''''') at (4+7.5,-4.5); 
\coordinate[label=left:$4$] (4''''') at (5+7.5,-4.5-1.73);
\coordinate[label=right:$5$] (5''''') at (7+7.5,-4.5-1.73);
\draw[thick] (0''''')--(1''''');
\draw[thick] (1''''')--(2''''');
\draw[thick] (2''''')--(3''''');
\draw[thick] (3''''')--(4''''');
\draw[thick] (4''''')--(5''''');
\draw[thick] (5''''')--(0''''');
\draw[thick] (0''''')--(2''''');
\draw[thick] (0''''')--(3''''');
\draw[thick] (0''''')--(4''''');
\draw[thick] (1''''')--(3''''');
\draw[thick] (1''''')--(5''''');
\draw[thick] (2''''')--(5''''');
\draw[thick] (3''''')--(5''''');
\draw[thick,->] (2,1.5)--(4,3) node[left=10pt] {$\Gamma_{(01234)}$};
\coordinate[label=right:$0$] (0') at (8,4.5);
\coordinate[label=right:$1$] (1') at (7,4.5+1.73);
\coordinate[label=left:$2$] (2') at (5,4.5+1.73);
\coordinate[label=left:$3$] (3') at (4,4.5); 
\coordinate[label=left:$4$] (4') at (5,4.5-1.73);
\coordinate[label=right:$5$] (5') at (7,4.5-1.73);
\draw[thick] (0')--(1');
\draw[thick] (1')--(2');
\draw[thick] (2')--(3');
\draw[thick] (3')--(4');
\draw[thick] (4')--(5');
\draw[thick] (5')--(0');
\draw[thick] (0')--(2');
\draw[thick] (0')--(3');
\draw[thick] (0')--(4');
\draw[thick] (1')--(4');
\draw[thick] (1')--(5');
\draw[thick] (2')--(4');
\draw[thick] (2')--(5');
\draw[thick] (3')--(5');
\draw[thick,->] (8.5,4.5)--(11,4.5) node[midway,above] {$\Gamma_{(01245)}$};
\coordinate[label=right:$0$] (0'') at (8+7.5,4.5);
\coordinate[label=right:$1$] (1'') at (7+7.5,4.5+1.73);
\coordinate[label=left:$2$] (2'') at (5+7.5,4.5+1.73);
\coordinate[label=left:$3$] (3'') at (4+7.5,4.5); 
\coordinate[label=left:$4$] (4'') at (5+7.5,4.5-1.73);
\coordinate[label=right:$5$] (5'') at (7+7.5,4.5-1.73);
\draw[thick] (0'')--(1'');
\draw[thick] (1'')--(2'');
\draw[thick] (2'')--(3'');
\draw[thick] (3'')--(4'');
\draw[thick] (4'')--(5'');
\draw[thick] (5'')--(0'');
\draw[thick] (0'')--(2'');
\draw[thick] (0'')--(3'');
\draw[thick] (0'')--(4'');
\draw[thick] (1'')--(5'');
\draw[thick] (2'')--(4'');
\draw[thick] (2'')--(5'');
\draw[thick] (3'')--(5'');
\draw[thick,->] (15,3)--(17.5,1.5) node[left=12pt] {$\Gamma_{(02345)}$};
\coordinate[label=right:$0$] (0''') at (8+7.5+6,0);
\coordinate[label=right:$1$] (1''') at (7+7.5+6,1.73);
\coordinate[label=left:$2$] (2''') at (5+7+6.5,1.73);
\coordinate[label=left:$3$] (3''') at (4+7.5+6,0); 
\coordinate[label=left:$4$] (4''') at (5+7.5+6,-1.73);
\coordinate[label=right:$5$] (5''') at (7+7.5+6,-1.73);
\draw[thick] (0''')--(1''');
\draw[thick] (1''')--(2''');
\draw[thick] (2''')--(3''');
\draw[thick] (3''')--(4''');
\draw[thick] (4''')--(5''');
\draw[thick] (5''')--(0''');
\draw[thick] (0''')--(2''');
\draw[thick] (0''')--(3''');
\draw[thick] (0''')--(4''');
\draw[thick] (1''')--(5''');
\draw[thick] (2''')--(5''');
\draw[thick] (3''')--(5''');
\draw[thick,->] (15,-3)--(17.5,-1.5) node[left=12pt] {$\Gamma_{(01235)}$};
\end{tikzpicture}
\caption{Two different paths of map ( bistellar move) from vector spaces $V^+(0123)\otimes V^+(0134)\otimes V^+(1234)\otimes V^+(0145)\otimes V^+(1245)\otimes V^+(2345)$ to vector spaces $V^+(0125)\otimes V^+(0235)\otimes V^+(0345)$.}
\label{bistmove}
\end{figure*}
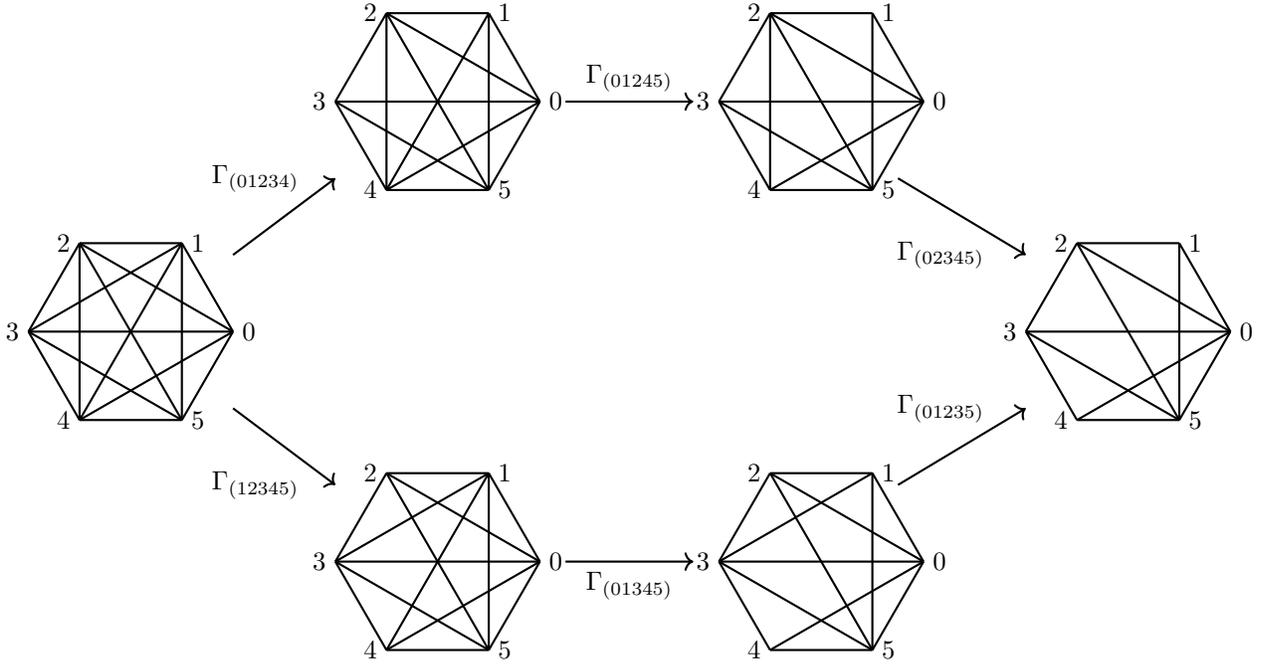

\subsection{Invariance under bistellar moves}

In two dimensional case, each F symbol corresponds to a map from two 2-simplexes to another two 2-simplexes.  For two different complexes glued by three 2-simplexes, different paths of map between these two complexes should be equivalent.  Such a constraint gives the famous pentagon equation, and the the F symbols (6j symbols) should satisfy the pentagon equation.

There are similar constraints in three dimensional cases. Each 10j symbol with positive orientation determines a map from three 3-simplexes to two 3-simplexes according to eqn.~\eqref{def10j}. For a complex glued by six 3-simplexes and another complex glued by three 3-simplexes, different paths of map between these two complexes, drawn in Fig.~\ref{bistmove}, should also be equivalent. And this condition implies 10j symbol must satisfy:
\begin{align}
&\sum_{a_{(024)}\in \mathcal{B}} d_{a_{(024)}} \Gamma_{(01234)} \Gamma_{(01245)} \Gamma_{(02345)}= \nonumber\\& \sum_{a_{(135)}\in \mathcal{B}} d_{a_{(135)}} \Gamma_{(01235)} \Gamma_{(01345)} \Gamma_{(12345)}.
\label{10jmove}
\end{align}

Actually, one can arbitrarily chooses two complexes out of six complexes in Fig.~\ref{bistmove} and requires that two different paths of map between these two complexes are equivalent. This will give another constrain for 10j symbols similar to eqn.~\eqref{10jmove}. Obviously, 10j symbols must satisfy all these constrains.

\section{Algorithm for computing irreducible central idempotents} \label{ICIalg}
In this section, we provide a general algorithm to calculate the irreducible(or minimal or primitive) central idempotents of a given semi-simple algebra $\Ac$. 
Let's first introduce an important theorem, the Wedderburn-Artin theorem, which states that every semi-simple algebra isomorphic to a direct sum of matrix algebras.  Thus for a tube algebra $\Ac$ with dimension $n$, we have $\Ac=\mathbb{M}_{n_{1}} \oplus \mathbb{M}_{n_{2}} \oplus \cdots \oplus$ $\mathbb{M}_{n_{3}} \oplus \cdots \oplus \mathbb{M}_{n_{K}}$, where $n_i$ is dimension of matrix algebra $\mathbb{M}_{n_{i}}$ and $n_{1}^2+n_{2}^2+\dots +n_{K}^2=n^2$. In the language of matrix algebra, we may think the element of $\Ac$ is a $n$ $\times$ $n$ block matrix with $K$ blocks and the dimension of $i$th block is $n_i$. Therefor block diagonalizing matrix representation of $\Ac$ is a straightforward approach to decompose $\Ac$. The approach can be further improved by so-called central idempotents decomposition. As mentioned before, decomposing a semi-simple algebra can be performed by decomposing its identity as the sum of irreducible central idempotents
\begin{equation}\label{DeICIs}
\mathbf{1}=\sum^K_{i=1} P_{i}, \quad P_{i} P_{j}=\delta_{i j} P_{i},
\end{equation}
where $P_i$ is irreducible, i.e. it cannot be further decomposed. Combined eqn.~\eqref{DeICIs} and matrix representation of $\Ac$, in the canonical basis, we know that the $n_i$ $\times$ $n_i$ identity matrix $I_{n_i}$ in the $i$th matrix algebra corresponds to the irreducible central idempotent $P_i$ of $\Ac$ and the direct sum of these identity matrices \{$I_{n_1}$,...,$I_{n_K}$\} of matrix algebras is equal to the identity matrix $I_{\Ac}$ of $\Ac$. 

Based on the matrix representation of ICIs above, we can construct the following program to calculate the ICIs. The input of the program are the structure factors $\Omega^{c}_{ab}$ of the tube algebra $\Ac$ with a natural vector space basis \{$\Ab_1$,...,$\Ab_n$ \} which means $\Ab_a\Ab_b=\sum^{n}_{c=1}\Omega^c_{ab}\Ab_c$. The output are the coefficients $\sigma^{i}_{a}$ in this basis of the irreducible central idempotents $P_i=\sum^{n}_{a=1} \sigma^i_a\Ab_a$. The first step is to determine the center $Z(\Ac)$ of this algebra. For an element $x=\sum^{n}_{b=1} x_b\Ab_b$, $x \in Z(\Ac)$ if and only if $\Ab_ax=x\Ab_a$ for all $a$. It is equivalent to requiring $\sum^{n}_{b=1}(\Omega^{c}_{ab}-\Omega^{c}_{ba})x_b=0$  holds for all $c$ and $a$. This step can be achieved by calculating the null vectors of coefficient matrix $Z_{(a-1)n+c,b}=\Omega^{c}_{ab}-\Omega^{c}_{ba}$.

The second step is to construct the matrix representation of the center $Z(\Ac)$. Through the first step, we can obtain a set of null vectors \{$\mathbf{z}_1$,...,$\mathbf{z}_m$\} as the basis of the center. We have $\mathbf{z}_l=\sum^{n}_{b=1}C^{l}_b \Ab_b$. For convenience, we choose the regular representation $\mathbf{Rep}$ of
 the center to construct its matrix representation.
\begin{equation}
\text { $\mathbf{Rep}$: } Z(\Ac) \rightarrow \operatorname{Aut}(\Ac), \mathbf{z}_l \mapsto \operatorname{\mathbf{Rep}}^{l}
\end{equation}
With the natural basis \{$\Ab_1$,...,$\Ab_n$ \}, 
\begin{eqnarray}
\Ab_a \mathbf{z}_l&=&\sum^{n}_{b=1}C^{l}_{b} \Ab_a\Ab_b=\sum^{n}_{c=1}\sum^{n}_{b=1}C^{l}_{b}\Omega^c_{ab}\Ab_c \nonumber \\
&=&\sum^{n}_{c=1}\mathbf{Rep}^{l}_{ac}\Ab_c
\end{eqnarray}
$\mathbf{Rep}^{l}_{ac}$ is a $n$ $\times$ $n$ matrix representation of basis $\mathbf{z}_l$. Since for any $P_i$, it always belongs to $Z(\Ac)$, $P_i$ can be expressed as the linear
combinations of basis \{$\mathbf{z}_1$,...,$\mathbf{z}_m$\}, $P_i=\sum^m_{l=1}f^i_l\mathbf{z}_{l}$. So far we get the matrix representation of $P_i$: $\mathbf{Rep}^{P_i}_{ac}=\sum^m_{l=1}f^i_l \mathbf{Rep}^l_{ac}$.

Now let's use the eqn~\eqref{DeICIs}. It tells us that up to an unitary transformation, $\mathbf{Rep}^{P_i}_{ac}$ should be equal to be a block-diagonal matrix $E_i$ with $K$ blocks and only the $i$th block is a $n_i$ $\times$ $n_i$ identity matrix $I_{n_i}$ of the $i$th matrix algebra, the other blocks are zero matrices, $U^{\dagger}\mathbf{Rep}^{P_i}_{ac}U=E_i$. Furthermore, we have $I_{\Ac}=\sum^K_{i=1}E_i=\sum^K_{i=1}U^{\dagger}\mathbf{Rep}^{P_i}_{ac}U= \sum^K_{i=1}\mathbf{Rep}^{P_i}_{ac}$. By direct diagonalization of the representation matrix $\mathbf{Rep}^{l}_{ac}$, we can obtain the unitary transformation $U$ satisfying $U^{\dagger}\mathbf{Rep}^{l}_{ac}U=diag(d^l_1,...,d^l_n)=D^l$ and arrive the canonical basis. So the problem is converted to find a set of coefficients $f^i_l$ to satisfy $\sum^m_{l=1}f^i_lD^l=E_i$. The problem can be solved by Moore-Penrose pseudoinverse of the following $n$ $\times$ $m$ matrix
\begin{equation}
\left(\begin{array}{ccc}
d^1_{1} & \dots &d^m_1  \\
\vdots & \ddots & \vdots \\
d^1_n&\dots  & d^{m}_{n}
\end{array}\right)
\end{equation}
Once we have determined $f^i_l$, then the output $\sigma^i_a=\sum^{m}_{l=1} f^i_lC^l_a$ is also known.

\section{3d Toric code}\label{3DTC}
\subsection{Graphical representation of \{ICIs\} of 3d Toric code}
For the 3d Toric Code, there are eight ICIs,
\begin{equation}
\begin{gathered}
P_1(0,0,0)=\frac{1}{2}\mathbf{A}^{000}+\frac{1}{2}\mathbf{A}^{100}=\frac{1}{2}
\begin{array}{c}
\begin{tikzpicture}[scale=0.5]
\coordinate[] (0) at (0,0,0) ; 
\coordinate[] (1) at (1,0,0) ; 
\coordinate[](0') at (-0.125,0.125,1) ; 1
\coordinate[] (1') at (0.875,0.125,1) ;  
\coordinate[] (0t) at (0,1,0); 
\coordinate[] (1t) at (1,1,0); 
\coordinate[] (0't) at (-0.125,1.125,1); 
\coordinate[](1't) at (0.875,1.125,1); 

\draw[dashed,help lines] (0)--(0t) node[midway,below] {$ $};%
\draw[help lines] (0')--(0't) node[midway,left] {$ $};%
\draw[help lines] (1)--(1t) node[midway,right] {$ $};%
\draw[help lines] (1')--(1't) node[midway,right] {$ $};%

\draw[dashed,help lines] (0)--(0') node[midway,below] {$ $};%
\draw[help lines] (0t)--(0't) node[midway,above] {$ $};%
\draw[help lines] (1)--(1') node[midway,right] {$ $};%
\draw[help lines] (1t)--(1't) node[midway,below] {$ $}; %

\draw[dashed,help lines] (0)--(1) node[midway,below] {$ $};%
\draw[help lines] (0')--(1') node[midway,below] {$ $};%
\draw[help lines] (0t)--(1t) node[midway,above] {$ $};%
\draw[help lines] (0't)--(1't) node[midway,above] {$ $};%

\end{tikzpicture}
\end{array}
+
\frac{1}{2}
\begin{array}{c}
\begin{tikzpicture}[scale=0.5]
\coordinate[] (0) at (0,0,0) ; 
\coordinate[] (1) at (1,0,0) ; 
\coordinate[] (0') at (-0.125,0.125,1) ; 1
\coordinate[] (1') at (0.875,0.125,1) ;  
\coordinate[] (0t) at (0,1,0); 
\coordinate[] (1t) at (1,1,0); 
\coordinate[] (0't) at (-0.125,1.125,1); 
\coordinate[] (1't) at (0.875,1.125,1); 

\draw[dashed,help lines] (0)--(0t) node[midway,below] {$ $};%
\draw[help lines] (0')--(0't) node[midway,left] {$ $};%
\draw[help lines] (1)--(1t) node[midway,right] {$ $};%
\draw[help lines] (1')--(1't) node[midway,right] {$ $};%

\draw[dashed,help lines] (0)--(0') node[midway,below] {$ $};%
\draw[help lines] (0t)--(0't) node[midway,above] {$ $};%
\draw[help lines] (1)--(1') node[midway,right] {$ $};%
\draw[help lines] (1t)--(1't) node[midway,below] {$ $}; %

\draw[dashed,thick,blue] (0)--(1) node[midway,below] {$ $};%
\draw[thick,blue] (0')--(1') node[midway,below] {$ $};%
\draw[thick,blue] (0t)--(1t) node[midway,above] {$ $};%
\draw[thick,blue] (0't)--(1't) node[midway,above] {$ $};%

\end{tikzpicture}
\end{array}
\\
%
%
%
P_2(1,0,0)=\frac{1}{2}\mathbf{A}^{000}-\frac{1}{2}\mathbf{A}^{100}=\frac{1}{2}
\begin{array}{c}
\begin{tikzpicture}[scale=0.5]
\coordinate[] (0) at (0,0,0) ; 
\coordinate[] (1) at (1,0,0) ; 
\coordinate[](0') at (-0.125,0.125,1) ; 1
\coordinate[] (1') at (0.875,0.125,1) ;  
\coordinate[] (0t) at (0,1,0); 
\coordinate[] (1t) at (1,1,0); 
\coordinate[] (0't) at (-0.125,1.125,1); 
\coordinate[](1't) at (0.875,1.125,1); 

\draw[dashed,help lines] (0)--(0t) node[midway,below] {$ $};%
\draw[help lines] (0')--(0't) node[midway,left] {$ $};%
\draw[help lines] (1)--(1t) node[midway,right] {$ $};%
\draw[help lines] (1')--(1't) node[midway,right] {$ $};%

\draw[dashed,help lines] (0)--(0') node[midway,below] {$ $};%
\draw[help lines] (0t)--(0't) node[midway,above] {$ $};%
\draw[help lines] (1)--(1') node[midway,right] {$ $};%
\draw[help lines] (1t)--(1't) node[midway,below] {$ $}; %

\draw[dashed,help lines] (0)--(1) node[midway,below] {$ $};%
\draw[help lines] (0')--(1') node[midway,below] {$ $};%
\draw[help lines] (0t)--(1t) node[midway,above] {$ $};%
\draw[help lines] (0't)--(1't) node[midway,above] {$ $};%

\end{tikzpicture}
\end{array}
-
\frac{1}{2}
\begin{array}{c}
\begin{tikzpicture}[scale=0.5]
\coordinate[] (0) at (0,0,0) ; 
\coordinate[] (1) at (1,0,0) ; 
\coordinate[] (0') at (-0.125,0.125,1) ; 1
\coordinate[] (1') at (0.875,0.125,1) ;  
\coordinate[] (0t) at (0,1,0); 
\coordinate[] (1t) at (1,1,0); 
\coordinate[] (0't) at (-0.125,1.125,1); 
\coordinate[] (1't) at (0.875,1.125,1); 

\draw[dashed,help lines] (0)--(0t) node[midway,below] {$ $};%
\draw[help lines] (0')--(0't) node[midway,left] {$ $};%
\draw[help lines] (1)--(1t) node[midway,right] {$ $};%
\draw[help lines] (1')--(1't) node[midway,right] {$ $};%

\draw[dashed,help lines] (0)--(0') node[midway,below] {$ $};%
\draw[help lines] (0t)--(0't) node[midway,above] {$ $};%
\draw[help lines] (1)--(1') node[midway,right] {$ $};%
\draw[help lines] (1t)--(1't) node[midway,below] {$ $}; %

\draw[dashed,thick,blue] (0)--(1) node[midway,below] {$ $};%
\draw[thick,blue] (0')--(1') node[midway,below] {$ $};%
\draw[thick,blue] (0t)--(1t) node[midway,above] {$ $};%
\draw[thick,blue] (0't)--(1't) node[midway,above] {$ $};%

\end{tikzpicture}
\end{array}
\\
%
%
%
%
P_3(0,1,0)=\frac{1}{2}\mathbf{A}^{010}+\frac{1}{2}\mathbf{A}^{110}=\frac{1}{2}
\begin{array}{c}
\begin{tikzpicture}[scale=0.5]
\coordinate[] (0) at (0,0,0) ; 
\coordinate[] (1) at (1,0,0) ; 
\coordinate[](0') at (-0.125,0.125,1) ; 1
\coordinate[] (1') at (0.875,0.125,1) ;  
\coordinate[] (0t) at (0,1,0); 
\coordinate[] (1t) at (1,1,0); 
\coordinate[] (0't) at (-0.125,1.125,1); 
\coordinate[](1't) at (0.875,1.125,1);

\draw[dashed,help lines] (0)--(1) node[midway,below] {$ $};%
\draw[help lines] (0')--(1') node[midway,below] {$ $};%
\draw[help lines] (0t)--(1t) node[midway,above] {$ $};%
\draw[help lines] (0't)--(1't) node[midway,above] {$ $};%

\draw[dashed, thick, blue] (0)--(0t) node[midway,below] {$ $};%
\draw[thick,blue] (0')--(0't) node[midway,left] {$ $};%
\draw[thick,blue] (1)--(1t) node[midway,right] {$ $};%
\draw[thick,blue] (1')--(1't) node[midway,right] {$ $};%

\draw[dashed,help lines] (0)--(0') node[midway,below] {$ $};%
\draw[help lines] (0t)--(0't) node[midway,above] {$ $};%
\draw[help lines] (1)--(1') node[midway,right] {$ $};%
\draw[help lines] (1t)--(1't) node[midway,below] {$ $}; %

\end{tikzpicture}
\end{array}
+
\frac{1}{2}
\begin{array}{c}
\begin{tikzpicture}[scale=0.5]
\coordinate[] (0) at (0,0,0) ; 
\coordinate[] (1) at (1,0,0) ; 
\coordinate[] (0') at (-0.125,0.125,1) ; 1
\coordinate[] (1') at (0.875,0.125,1) ;  
\coordinate[] (0t) at (0,1,0); 
\coordinate[] (1t) at (1,1,0); 
\coordinate[] (0't) at (-0.125,1.125,1); 
\coordinate[] (1't) at (0.875,1.125,1);

\draw[dashed,thick,blue] (0)--(1) node[midway,below] {$ $};%
\draw[thick,blue] (0')--(1') node[midway,below] {$ $};%
\draw[thick,blue] (0t)--(1t) node[midway,above] {$ $};%
\draw[thick,blue] (0't)--(1't) node[midway,above] {$ $};%

\draw[dashed, thick, blue] (0)--(0t) node[midway,below] {$ $};%
\draw[thick, blue] (0')--(0't) node[midway,left] {$ $};%
\draw[thick, blue] (1)--(1t) node[midway,right] {$ $};%
\draw[thick, blue] (1')--(1't) node[midway,right] {$ $};%

\draw[dashed,help lines] (0)--(0') node[midway,below] {$ $};%
\draw[help lines] (0t)--(0't) node[midway,above] {$ $};%
\draw[help lines] (1)--(1') node[midway,right] {$ $};%
\draw[help lines] (1t)--(1't) node[midway,below] {$ $}; %

\end{tikzpicture}
\end{array}
\\
%
%
%
%
P_4(1,1,0)=\frac{1}{2}\mathbf{A}^{010}-\frac{1}{2}\mathbf{A}^{110}=\frac{1}{2}
\begin{array}{c}
\begin{tikzpicture}[scale=0.5]
\coordinate[] (0) at (0,0,0) ; 
\coordinate[] (1) at (1,0,0) ; 
\coordinate[](0') at (-0.125,0.125,1) ; 1
\coordinate[] (1') at (0.875,0.125,1) ;  
\coordinate[] (0t) at (0,1,0); 
\coordinate[] (1t) at (1,1,0); 
\coordinate[] (0't) at (-0.125,1.125,1); 
\coordinate[](1't) at (0.875,1.125,1);

\draw[dashed,help lines] (0)--(1) node[midway,below] {$ $};%
\draw[help lines] (0')--(1') node[midway,below] {$ $};%
\draw[help lines] (0t)--(1t) node[midway,above] {$ $};%
\draw[help lines] (0't)--(1't) node[midway,above] {$ $};%

\draw[dashed, thick, blue] (0)--(0t) node[midway,below] {$ $};%
\draw[thick,blue] (0')--(0't) node[midway,left] {$ $};%
\draw[thick,blue] (1)--(1t) node[midway,right] {$ $};%
\draw[thick,blue] (1')--(1't) node[midway,right] {$ $};%

\draw[dashed,help lines] (0)--(0') node[midway,below] {$ $};%
\draw[help lines] (0t)--(0't) node[midway,above] {$ $};%
\draw[help lines] (1)--(1') node[midway,right] {$ $};%
\draw[help lines] (1t)--(1't) node[midway,below] {$ $}; %

\end{tikzpicture}
\end{array}
-
\frac{1}{2}
\begin{array}{c}
\begin{tikzpicture}[scale=0.5]
\coordinate[] (0) at (0,0,0) ; 
\coordinate[] (1) at (1,0,0) ; 
\coordinate[] (0') at (-0.125,0.125,1) ; 1
\coordinate[] (1') at (0.875,0.125,1) ;  
\coordinate[] (0t) at (0,1,0); 
\coordinate[] (1t) at (1,1,0); 
\coordinate[] (0't) at (-0.125,1.125,1); 
\coordinate[] (1't) at (0.875,1.125,1);

\draw[dashed,thick,blue] (0)--(1) node[midway,below] {$ $};%
\draw[thick,blue] (0')--(1') node[midway,below] {$ $};%
\draw[thick,blue] (0t)--(1t) node[midway,above] {$ $};%
\draw[thick,blue] (0't)--(1't) node[midway,above] {$ $};%

\draw[dashed, thick, blue] (0)--(0t) node[midway,below] {$ $};%
\draw[thick, blue] (0')--(0't) node[midway,left] {$ $};%
\draw[thick, blue] (1)--(1t) node[midway,right] {$ $};%
\draw[thick, blue] (1')--(1't) node[midway,right] {$ $};%

\draw[dashed,help lines] (0)--(0') node[midway,below] {$ $};%
\draw[help lines] (0t)--(0't) node[midway,above] {$ $};%
\draw[help lines] (1)--(1') node[midway,right] {$ $};%
\draw[help lines] (1t)--(1't) node[midway,below] {$ $}; %

\end{tikzpicture}
\end{array}\\
%
%
%
P_5(0,0,1)=\frac{1}{2}\mathbf{A}^{001}+\frac{1}{2}\mathbf{A}^{101}=\frac{1}{2}
\begin{array}{c}
\begin{tikzpicture}[scale=0.5]
\coordinate[] (0) at (0,0,0) ; 
\coordinate[] (1) at (1,0,0) ; 
\coordinate[](0') at (-0.125,0.125,1) ; 1
\coordinate[] (1') at (0.875,0.125,1) ;  
\coordinate[] (0t) at (0,1,0); 
\coordinate[] (1t) at (1,1,0); 
\coordinate[] (0't) at (-0.125,1.125,1); 
\coordinate[](1't) at (0.875,1.125,1);

\draw[dashed, help lines] (0)--(1) node[midway,below] {$ $};%
\draw[help lines] (0')--(1') node[midway,below] {$ $};%
\draw[help lines] (0t)--(1t) node[midway,above] {$ $};%
\draw[help lines] (0't)--(1't) node[midway,above] {$ $};%

\draw[dashed, help lines] (0)--(0t) node[midway,below] {$ $};%
\draw[help lines] (0')--(0't) node[midway,left] {$ $};%
\draw[help lines] (1)--(1t) node[midway,right] {$ $};%
\draw[help lines] (1')--(1't) node[midway,right] {$ $};%

\draw[dashed, thick, blue] (0)--(0') node[midway,below] {$ $};%
\draw[thick, blue] (0t)--(0't) node[midway,above] {$ $};%
\draw[thick, blue] (1)--(1') node[midway,right] {$ $};%
\draw[thick, blue] (1t)--(1't) node[midway,below] {$ $}; %

\end{tikzpicture}
\end{array}
+
\frac{1}{2}
\begin{array}{c}
\begin{tikzpicture}[scale=0.5]
\coordinate[] (0) at (0,0,0) ; 
\coordinate[] (1) at (1,0,0) ; 
\coordinate[] (0') at (-0.125,0.125,1) ; 1
\coordinate[] (1') at (0.875,0.125,1) ;  
\coordinate[] (0t) at (0,1,0); 
\coordinate[] (1t) at (1,1,0); 
\coordinate[] (0't) at (-0.125,1.125,1); 
\coordinate[] (1't) at (0.875,1.125,1);

\draw[dashed,thick,blue] (0)--(1) node[midway,below] {$ $};%
\draw[thick,blue] (0')--(1') node[midway,below] {$ $};%
\draw[thick,blue] (0t)--(1t) node[midway,above] {$ $};%
\draw[thick,blue] (0't)--(1't) node[midway,above] {$ $};%

\draw[dashed, help lines] (0)--(0t) node[midway,below] {$ $};%
\draw[help lines] (0')--(0't) node[midway,left] {$ $};%
\draw[help lines] (1)--(1t) node[midway,right] {$ $};%
\draw[help lines] (1')--(1't) node[midway,right] {$ $};%

\draw[dashed, thick, blue] (0)--(0') node[midway,below] {$ $};%
\draw[thick, blue] (0t)--(0't) node[midway,above] {$ $};%
\draw[thick, blue] (1)--(1') node[midway,right] {$ $};%
\draw[thick, blue] (1t)--(1't) node[midway,below] {$ $}; %

\end{tikzpicture}
\end{array}
\\
%
%
%
%
P_6(1,0,1)=\frac{1}{2}\mathbf{A}^{001}-\frac{1}{2}\mathbf{A}^{101}=\frac{1}{2}
\begin{array}{c}
\begin{tikzpicture}[scale=0.5]
\coordinate[] (0) at (0,0,0) ; 
\coordinate[] (1) at (1,0,0) ; 
\coordinate[](0') at (-0.125,0.125,1) ; 1
\coordinate[] (1') at (0.875,0.125,1) ;  
\coordinate[] (0t) at (0,1,0); 
\coordinate[] (1t) at (1,1,0); 
\coordinate[] (0't) at (-0.125,1.125,1); 
\coordinate[](1't) at (0.875,1.125,1);

\draw[dashed, help lines] (0)--(1) node[midway,below] {$ $};%
\draw[help lines] (0')--(1') node[midway,below] {$ $};%
\draw[help lines] (0t)--(1t) node[midway,above] {$ $};%
\draw[help lines] (0't)--(1't) node[midway,above] {$ $};%

\draw[dashed, help lines] (0)--(0t) node[midway,below] {$ $};%
\draw[help lines] (0')--(0't) node[midway,left] {$ $};%
\draw[help lines] (1)--(1t) node[midway,right] {$ $};%
\draw[help lines] (1')--(1't) node[midway,right] {$ $};%

\draw[dashed, thick, blue] (0)--(0') node[midway,below] {$ $};%
\draw[thick, blue] (0t)--(0't) node[midway,above] {$ $};%
\draw[thick, blue] (1)--(1') node[midway,right] {$ $};%
\draw[thick, blue] (1t)--(1't) node[midway,below] {$ $}; %

\end{tikzpicture}
\end{array}
-
\frac{1}{2}
\begin{array}{c}
\begin{tikzpicture}[scale=0.5]
\coordinate[] (0) at (0,0,0) ; 
\coordinate[] (1) at (1,0,0) ; 
\coordinate[] (0') at (-0.125,0.125,1) ; 1
\coordinate[] (1') at (0.875,0.125,1) ;  
\coordinate[] (0t) at (0,1,0); 
\coordinate[] (1t) at (1,1,0); 
\coordinate[] (0't) at (-0.125,1.125,1); 
\coordinate[] (1't) at (0.875,1.125,1);

\draw[dashed,thick,blue] (0)--(1) node[midway,below] {$ $};%
\draw[thick,blue] (0')--(1') node[midway,below] {$ $};%
\draw[thick,blue] (0t)--(1t) node[midway,above] {$ $};%
\draw[thick,blue] (0't)--(1't) node[midway,above] {$ $};%

\draw[dashed, help lines] (0)--(0t) node[midway,below] {$ $};%
\draw[help lines] (0')--(0't) node[midway,left] {$ $};%
\draw[help lines] (1)--(1t) node[midway,right] {$ $};%
\draw[help lines] (1')--(1't) node[midway,right] {$ $};%

\draw[dashed, thick, blue] (0)--(0') node[midway,below] {$ $};%
\draw[thick, blue] (0t)--(0't) node[midway,above] {$ $};%
\draw[thick, blue] (1)--(1') node[midway,right] {$ $};%
\draw[thick, blue] (1t)--(1't) node[midway,below] {$ $}; %

\end{tikzpicture}
\end{array}
\\
%
%
%
%
P_7(0,1,1)=\frac{1}{2}\mathbf{A}^{011}+\frac{1}{2}\mathbf{A}^{111}=\frac{1}{2}
\begin{array}{c}
\begin{tikzpicture}[scale=0.5]
\coordinate[] (0) at (0,0,0) ; 
\coordinate[] (1) at (1,0,0) ; 
\coordinate[](0') at (-0.125,0.125,1) ; 1
\coordinate[] (1') at (0.875,0.125,1) ;  
\coordinate[] (0t) at (0,1,0); 
\coordinate[] (1t) at (1,1,0); 
\coordinate[] (0't) at (-0.125,1.125,1); 
\coordinate[](1't) at (0.875,1.125,1);

\draw[dashed, help lines] (0)--(1) node[midway,below] {$ $};%
\draw[help lines] (0')--(1') node[midway,below] {$ $};%
\draw[help lines] (0t)--(1t) node[midway,above] {$ $};%
\draw[help lines] (0't)--(1't) node[midway,above] {$ $};%

\draw[dashed, thick, blue] (0)--(0t) node[midway,below] {$ $};%
\draw[thick, blue] (0')--(0't) node[midway,left] {$ $};%
\draw[thick, blue] (1)--(1t) node[midway,right] {$ $};%
\draw[thick, blue] (1')--(1't) node[midway,right] {$ $};%

\draw[dashed, thick, blue] (0)--(0') node[midway,below] {$ $};%
\draw[thick, blue] (0t)--(0't) node[midway,above] {$ $};%
\draw[thick, blue] (1)--(1') node[midway,right] {$ $};%
\draw[thick, blue] (1t)--(1't) node[midway,below] {$ $}; %

\end{tikzpicture}
\end{array}
+
\frac{1}{2}
\begin{array}{c}
\begin{tikzpicture}[scale=0.5]
\coordinate[] (0) at (0,0,0) ; 
\coordinate[] (1) at (1,0,0) ; 
\coordinate[] (0') at (-0.125,0.125,1) ; 1
\coordinate[] (1') at (0.875,0.125,1) ;  
\coordinate[] (0t) at (0,1,0); 
\coordinate[] (1t) at (1,1,0); 
\coordinate[] (0't) at (-0.125,1.125,1); 
\coordinate[] (1't) at (0.875,1.125,1);

\draw[dashed,thick,blue] (0)--(1) node[midway,below] {$ $};%
\draw[thick,blue] (0')--(1') node[midway,below] {$ $};%
\draw[thick,blue] (0t)--(1t) node[midway,above] {$ $};%
\draw[thick,blue] (0't)--(1't) node[midway,above] {$ $};%

\draw[dashed, thick, blue] (0)--(0t) node[midway,below] {$ $};%
\draw[thick, blue] (0')--(0't) node[midway,left] {$ $};%
\draw[thick, blue] (1)--(1t) node[midway,right] {$ $};%
\draw[thick, blue] (1')--(1't) node[midway,right] {$ $};%

\draw[dashed, thick, blue] (0)--(0') node[midway,below] {$ $};%
\draw[thick, blue] (0t)--(0't) node[midway,above] {$ $};%
\draw[thick, blue] (1)--(1') node[midway,right] {$ $};%
\draw[thick, blue] (1t)--(1't) node[midway,below] {$ $}; %

\end{tikzpicture}
\end{array}
\\
%
%
%
%
P_8(1,1,1)=\frac{1}{2}\mathbf{A}^{011}-\frac{1}{2}\mathbf{A}^{111}=\frac{1}{2}
\begin{array}{c}
\begin{tikzpicture}[scale=0.5]
\coordinate[] (0) at (0,0,0) ; 
\coordinate[] (1) at (1,0,0) ; 
\coordinate[](0') at (-0.125,0.125,1) ; 1
\coordinate[] (1') at (0.875,0.125,1) ;  
\coordinate[] (0t) at (0,1,0); 
\coordinate[] (1t) at (1,1,0); 
\coordinate[] (0't) at (-0.125,1.125,1); 
\coordinate[](1't) at (0.875,1.125,1);

\draw[dashed, help lines] (0)--(1) node[midway,below] {$ $};%
\draw[help lines] (0')--(1') node[midway,below] {$ $};%
\draw[help lines] (0t)--(1t) node[midway,above] {$ $};%
\draw[help lines] (0't)--(1't) node[midway,above] {$ $};%

\draw[dashed, thick, blue] (0)--(0t) node[midway,below] {$ $};%
\draw[thick, blue] (0')--(0't) node[midway,left] {$ $};%
\draw[thick, blue] (1)--(1t) node[midway,right] {$ $};%
\draw[thick, blue] (1')--(1't) node[midway,right] {$ $};%

\draw[dashed, thick, blue] (0)--(0') node[midway,below] {$ $};%
\draw[thick, blue] (0t)--(0't) node[midway,above] {$ $};%
\draw[thick, blue] (1)--(1') node[midway,right] {$ $};%
\draw[thick, blue] (1t)--(1't) node[midway,below] {$ $}; %

\end{tikzpicture}
\end{array}
-
\frac{1}{2}
\begin{array}{c}
\begin{tikzpicture}[scale=0.5]
\coordinate[] (0) at (0,0,0) ; 
\coordinate[] (1) at (1,0,0) ; 
\coordinate[] (0') at (-0.125,0.125,1) ; 1
\coordinate[] (1') at (0.875,0.125,1) ;  
\coordinate[] (0t) at (0,1,0); 
\coordinate[] (1t) at (1,1,0); 
\coordinate[] (0't) at (-0.125,1.125,1); 
\coordinate[] (1't) at (0.875,1.125,1);

\draw[dashed,thick,blue] (0)--(1) node[midway,below] {$ $};%
\draw[thick,blue] (0')--(1') node[midway,below] {$ $};%
\draw[thick,blue] (0t)--(1t) node[midway,above] {$ $};%
\draw[thick,blue] (0't)--(1't) node[midway,above] {$ $};%

\draw[dashed, thick, blue] (0)--(0t) node[midway,below] {$ $};%
\draw[thick, blue] (0')--(0't) node[midway,left] {$ $};%
\draw[thick, blue] (1)--(1t) node[midway,right] {$ $};%
\draw[thick, blue] (1')--(1't) node[midway,right] {$ $};%

\draw[dashed, thick, blue] (0)--(0') node[midway,below] {$ $};%
\draw[thick, blue] (0t)--(0't) node[midway,above] {$ $};%
\draw[thick, blue] (1)--(1') node[midway,right] {$ $};%
\draw[thick, blue] (1t)--(1't) node[midway,below] {$ $}; %

\end{tikzpicture}
\end{array}
\end{gathered}
\end{equation}
where the edge in blue labels the non-trivial element $1$, the edge without special colors indicates trivial label. The dotted lines just indicate the line on the backside of the viewpoint. These $1st$-$8th$ ICIs above correspond to pure trivial particle, pure non-trivial particle $e$, $\alpha$-loop, $\alpha$-loop attached by $e$, $\beta$-loop, $\beta$-loop attached by $e$, $\alpha,\beta$-loop and $\alpha,\beta$-loop attached by $e$, respectively.

\subsection{$S^{zxy}_{3d}$ acting on \{ICIs\} of 3d Toric code}

\begin{align}
\label{S_xyz}
\hat{S}^{zxy}_{3d} P_1=\frac{1}{2} ( P_1+P_2 ) + \frac{1}{2} ( P_3+P_4 ) \nonumber\\
\hat{S}^{zxy}_{3d} P_2=\frac{1}{2} ( P_1+P_2 ) - \frac{1}{2} ( P_3+P_4 ) \nonumber\\
\hat{S}^{zxy}_{3d} P_3=\frac{1}{2} ( P_5+P_6 ) + \frac{1}{2} ( P_7+P_8 ) \nonumber\\
\hat{S}^{zxy}_{3d} P_4=\frac{1}{2} ( P_5+P_6 ) - \frac{1}{2} ( P_7+P_8 ) \nonumber\\
\hat{S}^{zxy}_{3d} P_5=\frac{1}{2} ( P_1-P_2 ) + \frac{1}{2} ( P_3-P_4 ) \nonumber\\
\hat{S}^{zxy}_{3d} P_6=\frac{1}{2} ( P_1-P_2 ) - \frac{1}{2} ( P_3-P_4 ) \nonumber\\
\hat{S}^{zxy}_{3d} P_7=\frac{1}{2} ( P_5-P_6 ) + \frac{1}{2} ( P_7-P_8 ) \nonumber\\
\hat{S}^{zxy}_{3d} P_8=\frac{1}{2} ( P_5-P_6 ) - \frac{1}{2} ( P_7-P_8 )
\end{align}

\section{$B_{\v C}$ operator in membrane-net model}
\label{B_Cmap}
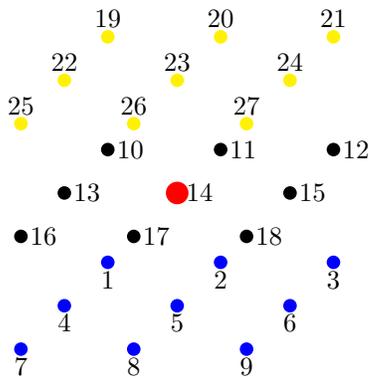
\begin{figure}[htbp]
\centering
\begin{tikzpicture} [scale=1.5]
\fill[blue](0,0,0) circle (0.06) node[below,black] {$1$};
\fill[blue](1,0,0) circle (0.06) node[below,black] {$2$};
\fill[blue](2,0,0) circle (0.06) node[below,black] {$3$};
\fill[blue](0,0,1) circle (0.06) node[below,black] {$4$};
\fill[blue](1,0,1) circle (0.06) node[below,black] {$5$};
\fill[blue](2,0,1) circle (0.06) node[below,black] {$6$};
\fill[blue](0,0,2) circle (0.06) node[below,black] {$7$};
\fill[blue](1,0,2) circle (0.06) node[below,black] {$8$};
\fill[blue](2,0,2) circle (0.06) node[below,black] {$9$};
\fill(0,1,0) circle (0.06) node[right,black] {$10$};
\fill(1,1,0) circle (0.06) node[right,black] {$11$};
\fill(2,1,0) circle (0.06) node[right,black] {$12$};
\fill(0,1,1) circle (0.06) node[right,black] {$13$};
\fill[red](1,1,1) circle (0.1) node[right,black] {$14$};
\fill(2,1,1) circle (0.06) node[right,black] {$15$};
\fill(0,1,2) circle (0.06) node[right,black] {$16$};
\fill(1,1,2) circle (0.06) node[right,black] {$17$};
\fill(2,1,2) circle (0.06) node[right,black] {$18$};
\fill[yellow](0,2,0) circle (0.06) node[above,black] {$19$};
\fill[yellow](1,2,0) circle (0.06) node[above,black] {$20$};
\fill[yellow](2,2,0) circle (0.06) node[above,black] {$21$};
\fill[yellow](0,2,1) circle (0.06) node[above,black] {$22$};
\fill[yellow](1,2,1) circle (0.06) node[above,black] {$23$};
\fill[yellow](2,2,1) circle (0.06) node[above,black] {$24$};
\fill[yellow](0,2,2) circle (0.06) node[above,black] {$25$};
\fill[yellow](1,2,2) circle (0.06) node[above,black] {$26$};
\fill[yellow](2,2,2) circle (0.06) node[above,black] {$27$};
\end{tikzpicture}
\caption{A vertex of tetrahedron lattice labeled by $14$ (in red) shared by 8 cubes. Each cube (for example, 1,2,4,5-10,11,13,14) has same internal structure with the cube drawn in Fig.\ref{3d dual}. Different layers of vertices are labeled by different color.}
\label{vertex14}
\end{figure}

A $B_{\v C}$ operator acting on a vertex of tetrahedron lattice (cell of membrane-net lattice) labeled by $14$ shared by 8 cubes drawn in Fig.\ref{vertex14} induce a map: 

\begin{widetext}
\begin{align}
&V^{+}(1,2,5,14)\otimes V^{+}(1,2,11,14)\otimes V^{+}(1,10,11,14)\otimes V^{+}(1,4,5,14)\otimes V^{+}(1,4,13,14)\otimes V^{+}(1,10,13,14)\otimes \nonumber\\ &V^{+}(2,5,14,15)\otimes V^{+}(2,11,14,15)\otimes V^{+}(4,5,14,17)\otimes V^{+}(4,13,14,17)\otimes V^{+}(5,14,15,18)\otimes \nonumber\\&V^{+}(5,14,17,18)\otimes 
V^{+}(10,13,14,23)\otimes V^{+}(10,11,14,23)\otimes V^{+}(11,14,15,24)\otimes V^{+}(11,14,23,24)\otimes \nonumber\\&V^{+}(13,14,17,26)\otimes  V^{+}(13,14,23,26)\otimes
V^{+}(14,15,18,27)\otimes V^{+}(14,15,24,27)\otimes V^{+}(14,23,24,27)\otimes \nonumber\\ & V^{+}(14,17,18,27)\otimes V^{+}(14,17,26,27)\otimes V^{+}(14,23,26,27)\otimes \longrightarrow \nonumber \\
&V^{+}(1,2,5,14')\otimes V^{+}(1,2,11,14')\otimes V^{+}(1,10,11,14')\otimes V^{+}(1,4,5,14')\otimes V^{+}(1,4,13,14')\otimes V^{+}(1,10,13,14')\otimes \nonumber\\ &V^{+}(2,5,14',15)\otimes V^{+}(2,11,14',15)\otimes V^{+}(4,5,14',17)\otimes V^{+}(4,13,14',17)\otimes V^{+}(5,14',15,18)\otimes \nonumber\\&V^{+}(5,14',17,18)\otimes 
V^{+}(10,13,14',23)\otimes V^{+}(10,11,14',23)\otimes V^{+}(11,14',15,24)\otimes V^{+}(11,14',23,24)\otimes \nonumber\\&V^{+}(13,14',17,26)\otimes  V^{+}(13,14',23,26)\otimes
V^{+}(14',15,18,27)\otimes V^{+}(14',15,24,27)\otimes V^{+}(14',23,24,27)\otimes \nonumber\\ & V^{+}(14',17,18,27)\otimes V^{+}(14',17,26,27)\otimes V^{+}(14',23,26,27),
\end{align}
\end{widetext}

through a path:

\begin{widetext}
\begin{align}
&\Gamma_{(1,2,5,14,14')}:V^{+}(1,2,5,14)\otimes V^{+}(1,2,14,14') \rightarrow V^{+}(1,2,5,14')\otimes V^{+}(1,5,14,14') \otimes V^{+}(2,5,14,14');\nonumber\\
&\Gamma_{(1,2,11,14,14')}:V^{+}(1,2,11,14)\otimes V^{+}(1,11,14,14')\rightarrow V^{+}(1,2,11,14')\otimes V^{+}(1,2,14,14')\otimes V^{+}(2,11,14,14');\nonumber\\
&\Gamma_{(1,10,11,14,14')}:V^{+}(1,10,11,14) \otimes V^{+}(1,10,14,14') \otimes V^{+}(10,11,14,14')\rightarrow V^{+}(1,10,11,14') \otimes V^{+}(1,11,14,14');\nonumber\\
&\Gamma_{(1,4,5,14,14')}:V^{+}(1,4,5,14) \otimes V^{+}(1,5,14,14') \rightarrow V^{+}(1,4,5,14')\otimes V^{+}(1,4,14,14') \otimes V^{+}(4,5,14,14');\nonumber\\
&\Gamma_{(1,4,13,14,14')}:V^{+}(1,4,13,14) \otimes V^{+}(1,4,14,14') \otimes V^{+}(4,13,14,14') \rightarrow V^{+}(1,4,13,14') \otimes V^{+}(1,13,14,14');\nonumber\\
&\Gamma_{(1,10,13,14,14')}:V^{+}(1,10,13,14) \otimes V^{+}(1,13,14,14') \rightarrow V^{+}(1,10,13,14') \otimes V^{+}(1,10,14,14') \otimes V^{+}(10,13,14,14');\nonumber\\
&\Gamma_{(2,5,14,15,14')}:V^{+}(2,5,14,15) \otimes V^{+}(2,5,14,14') \otimes V^{+}(2,14,15,14') \rightarrow V^{+}(2,5,15,14') \otimes V^{+}(5,14,15,14');\nonumber\\
&\Gamma_{(2,11,14,15,14')}:V^{+}(2,11,14,15) \otimes V^{+}(2,11,14,14')\rightarrow V^{+}(2,11,15,14') \otimes V^{+}(2,14,15,14') \otimes V^{+}(11,14,15,14');\nonumber\\
&\Gamma_{(4,5,14,17,14')}:V^{+}(4,5,14,17) \otimes V^{+}(4,5,14,14') \otimes V^{+}(4,14,17,14')\rightarrow V^{+}(4,5,17,14') \otimes V^{+}(5,14,17,14');\nonumber\\
&\Gamma_{(4,13,14,17,14')}:V^{+}(4,13,14,17) \otimes V^{+}(13,14,17,14') \rightarrow V^{+}(4,13,17,14') \otimes V^{+}(4,13,14,14') \otimes V^{+}(4,14,17,14');\nonumber\\
&\Gamma_{(5,14,15,18,14')}:V^{+}(5,14,15,18) \otimes V^{+}(5,14,15,14') \otimes V^{+}(5,14,18,14') \rightarrow V^{+}(5,15,18,14') \otimes V^{+}(14,15,18,14');\nonumber\\
&\Gamma_{(5,14,17,18,14')}:V^{+}(5,14,17,18) \otimes V^{+}(5,14,17,14') \rightarrow V^{+}(5,17,18,14') \otimes V^{+}(5,14,18,14') \otimes V^{+}(14,17,18,14');\nonumber\\
&\Gamma_{(10,13,14,23,14')}:V^{+}(10,13,14,23) \otimes V^{+}(10,13,14,14') \otimes V^{+}(10,14,23,14') \rightarrow V^{+}(10,13,23,14') \otimes V^{+}(13,14,23,14');\nonumber\\
&\Gamma_{(10,11,14,23,14')}:V^{+}(10,11,14,23) \otimes V^{+}(11,14,23,14')\rightarrow V^{+}(10,14,23,14') \otimes V^{+}(10,11,23,14') \otimes V^{+}(10,11,14,14');\nonumber\\
&\Gamma_{(11,14,15,24,14')}:V^{+}(11,14,15,24) \otimes V^{+}(11,14,24,14') \otimes V^{+}(11,14,15,14')\rightarrow V^{+}(11,15,24,14') \otimes V^{+}(14,15,24,14');\nonumber\\
&\Gamma_{(11,14,23,24,14')}:V^{+}(11,14,23,24) \otimes V^{+}(14,23,24,14')\rightarrow V^{+}(11,14,23,14') \otimes V^{+}(11,14,24,14') \otimes V^{+}(11,23,24,14');\nonumber\\
&\Gamma_{(13,14,17,26,14')}:V^{+}(13,14,17,26) \otimes V^{+}(13,14,26,14') \otimes V^{+}(14,17,26,14')\rightarrow V^{+}(13,17,26,14') \otimes (13,14,17,14');\nonumber\\
&\Gamma_{(13,14,23,26,14')}:V^{+}(13,14,23,26) \otimes V^{+}(13,14,23,14') \rightarrow V^{+}(13,14,26,14') \otimes V^{+}(13,23,26,14') \otimes V^{+}(14,23,26,14');\nonumber\\
&\Gamma_{(14,15,18,27,14')}:V^{+}(14,15,18,27) \otimes V^{+}(14,15,18,14') \otimes V^{+}(14,18,27,14')\rightarrow V^{+}(15,18,27,14') \otimes V^{+}(14,15,27,14');\nonumber\\
&\Gamma_{(14,15,24,27,14')}:V^{+}(14,15,24,27) \otimes V^{+}(14,15,27,14') \otimes V^{+}(14,15,24,14')\rightarrow V^{+}(15,24,27,14') \otimes V^{+}(14,24,27,14');\nonumber\\
&\Gamma_{(14,23,24,27,14')}:V^{+}(14,23,24,27) \otimes V^{+}(14,24,27,14')\rightarrow V^{+}(14,23,27,14') \otimes V^{+}(14,23,24,14') \otimes V^{+}(23,24,27,14');\nonumber\\
&\Gamma_{(14,17,18,27,14')}:V^{+}(14,17,18,27) \otimes V^{+}(14,17,18,14') \rightarrow V^{+}(14,18,27,14') \otimes V^{+}(17,18,27,14') \otimes V^{+}(14,17,27,14');\nonumber\\
&\Gamma_{(14,17,26,27,14')}:V^{+}(14,17,26,27) \otimes V^{+}(14,17,27,14') \otimes V^{+}(14,26,27,14')\rightarrow V^{+}(14,17,26,14') \otimes V^{+}(17,26,27,14');\nonumber\\
&\Gamma_{(14,23,26,27,14')}:V^{+}(14,23,26,27) \otimes V^{+}(14,23,26,14') \otimes V^{+}(14,23,26,14') \rightarrow V^{+}(14,26,27,14') \otimes V^{+}(23,26,27,14');
\end{align}
\end{widetext}
which is similar to the evolution as drawn in Fig.\ref{LW evolution}. Please note that here we redefine the map for each 10j symbol instead of applying definition \eqref{def10j}.

\bibliography{bib}
\end{document}